\documentclass[twocolumn,aps,nofootinbib,superscriptaddress]{revtex4-2}
\usepackage{amsmath}
\usepackage{bm}
\usepackage{graphicx}
\usepackage{mathrsfs}
\usepackage{multirow}
\usepackage[normalem]{ulem}
\usepackage{graphicx}
\usepackage{footnote}
\usepackage{booktabs, ctable}
\usepackage{xcolor, color, framed}
\usepackage[colorlinks, linkcolor=red,anchorcolor=green,citecolor=blue]{hyperref}

\graphicspath{{./figs/}}

\begin{document}
\title{Correlations between heavy mesons and the creation of the charmonia, bottomonia, and $B_c$ mesons in high energy $pp$ collisions}

\author{Jiaxing Zhao}
\affiliation{SUBATECH, Nantes University, IMT Atlantique, IN2P3/CNRS, 4 rue Alfred Kastler, 44307 Nantes cedex 3, France}
\affiliation{Helmholtz Research Academy Hessen for FAIR (HFHF),GSI Helmholtz Center for Heavy Ion Research. Campus Frankfurt, 60438 Frankfurt, Germany}
\affiliation{Institute for Theoretical Physics, Johann Wolfgang Goethe Universit\"{a}t, Frankfurt am Main, Germany}
\author{Joerg Aichelin}
\affiliation{SUBATECH, Nantes University, IMT Atlantique, IN2P3/CNRS, 4 rue Alfred Kastler, 44307 Nantes cedex 3, France}
\author{Pol Bernard Gossiaux}
\affiliation{SUBATECH, Nantes University, IMT Atlantique, IN2P3/CNRS, 4 rue Alfred Kastler, 44307 Nantes cedex 3, France}
\author{Klaus Werner}
\affiliation{SUBATECH, Nantes University, IMT Atlantique, IN2P3/CNRS, 4 rue Alfred Kastler, 44307 Nantes cedex 3, France}

\begin{abstract}
The different QCD processes, which can produce a heavy quark-antiquark ($Q\bar Q$) pair, induce different correlations between the heavy quarks. Employing the EPOS4HQ event generator we study the consequences of these correlations and compare the calculation with experimental results on open and hidden heavy flavour mesons, measured in proton-proton (pp) collisions at RHIC and LHC energies. We find that the measured correlations between heavy mesons are a direct image of the different production mechanisms, which contribute also in a different way to the transverse momentum distribution of open and hidden heavy flavour mesons. The latter are calculated in a Wigner density approach which also enables to reproduce quantitatively the
measured $B_c$ spectra. This agreement allows conclusions on the spatial distribution of the heavy quark creation processes. 
\end{abstract}
\date{\today}

\maketitle

%%%%%%%%%%%%%%%%%%%%%%%
\section{Introduction}
%%%%%%%%%%%%%%%%%%%%%%%
In the last decade heavy flavour (HF) hadrons have turned out  to be one of the most promising probes to study the properties of the expanding quark gluon plasma (QGP) produced in ultra-relativistic hadronic collisions. Open heavy flavour hadrons have been the object of investigation of quite a number of transport approaches~\cite{vanHees:2005wb,He:2011qa,Minissale:2020bif,Cao:2015hia,Cao:2016gvr,Cao:2018ews,Cao:2019iqs,Gossiaux:2009mk,Song:2015sfa,Song:2015ykw,He:2019vgs,Li:2020zbk,Beraudo:2022dpz,Zhao:2024ecc}, which succeeded to describe the experimentally observed multiplicity of the different mesons and baryons as well as two key observables, the transverse momentum, $p_T$, dependence of the elliptic flow, $v_2$, and  the ratio between the yield measured in heavy ion collisions and in proton-proton (pp) collisions, scaled by the number of initial hard scatterings, $R_{AA}$~\cite{STAR:2017kkh,STAR:2018zdy,ALICE:2021rxa,ALICE:2020iug,CMS:2020bnz,CMS:2017qjw} . These transport approaches have even revealed that the enhanced multiplicity of charmed baryons as well as the unexpected elliptic flow of charmed mesons can be well understood if one assumes that also in systems as small as pp a QGP is created. 

Correlations between open heavy flavour hadrons as well as hidden heavy flavour mesons have been much less in the focus of theoretical studies although several experimental observation came as a surprise and pose challenges to transport approaches. The heavy quark azimuthal correlation function, measured by LHCb in forward rapidity \cite{LHCb:2012aiv}, shows a very pronounced structure. The  observed values of $R_{AA}$ of $J/\psi$ at low $p_T$ is very different at RHIC \cite{STAR:2019fge} as compared to LHC energies \cite{ALICE:2013osk} and the observed $v_2$ of $J/\psi$ is finite at LHC \cite{ALICE:2020pvw} whereas that observed at RHIC is compatible with zero~\cite{STAR:2012jzy}. For LHC energies this points towards an interaction of the quarkonium (or of the heavy quarks, which form it later) with the expanding QGP because initially single charm quarks are isotropically distributed in azimuthal direction.   
In this work, we will demonstrate that correlations between open heavy flavour mesons and the production of quarkonia are two sides of the same coin because both are sensitive to the azimuthal angle between the $Q$ and $\bar Q$ at production. The production of heavy quark  $(Q\bar Q)$ pairs can be described by perturbative QCD, due to $m_Q\gg \Lambda_{\rm QCD}$, with a heavy quark mass $m_Q$ and the QCD cutoff $\Lambda_{\rm QCD}$ and the single particle distribution has been calculated in FONLL ~\cite{Cacciari:1998it,Cacciari:2005rk}, based on QCD perturbation theory. The results agree quite well with the experimentally observed single particle $p_T$ distribution.  

Azimuthal correlations between $Q\bar Q$ pairs have been studied in a couple of event generators \cite{Norrbin:2000zc,Vogt:2018oje,Souza:2015dgh,Maciula:2014oya}. The comparison with experimental data shows that there is considerable improvement necessary if one would like to obtain them from the measured correlations information of the underlying production processes.

Also the formation of a colorless quarkonium states from a given heavy quark pair $Q\bar Q$ contains information on the azimuthal correlations between the heavy quarks. It is a non-perturbative process and has been studied via many approaches, such as the Color-Evaporation Model (CEM)~\cite{Fritzsch:1977ay,Amundson:1995em,Cheung:2018tvq}, the Color-Singlet Model (CSM)~\cite{Chang:1979nn,Baier:1981uk} and the Color-Octet Model (COM)~\cite{Bodwin:1992qr}, the latter two being encompassed in effective theories~\cite{Bodwin:1994jh,Butenschoen:2012px,Gong:2012ug,Chao:2012iv}. 

Recently, we have advanced a new approach, which is based on the quantum density matrix and which has turned out to give a very good description of the quarkonium $p_T$ spectra in pp, not only for the $J/\psi$ but also for the excited states~\cite{Song:2017phm,Villar:2022sbv,Song:2023zma}. 
For this study the underlying momentum distribution of the $Q\bar Q$ is given by the EPOS4 event generator, in which different processes contribute to the production of a $Q\bar Q$ pair. The purpose of this paper is on the one side to extend this analysis to hidden bottom mesons and to the formation of $B_c$, on the other side to show how the initial correlations between the heavy quarks show up in correlations between two open heavy flavour hadrons and in hidden heavy flavour observables in order to establish how these correlations can be assessed experimentally.

For our study~\footnote{based on version EPOS4.0.4.q4} we use the EPOS4 event generator \cite{Werner:2023zvo,Werner:2023fne,werner:2023-epos4-smatrix,werner:2023-epos4-micro}, which has been shown to reproduce a multitude of light flavour observables. In its extension to heavy quarks, EPOS4HQ, it reproduces as well nicely the available open heavy flavour data in pp \cite{Zhao:2023ucp} as well as in heavy-ion collisions \cite{Zhao:2024ecc} at RHIC and LHC energies.

The paper is organized as follows: In section~\ref{sec.II}, the elementary $Q\bar Q$ production in EPOS4 is analyzed while the model for quarkonium production and its predictions are provided in section~\ref{sec.III}. Section~\ref{sec.IV} contains our conclusions.

%%%%%%%%%%%%%%%%%%%%%%%
\section{$Q \bar Q$ production in EPOS4HQ}
\label{sec.II}
\subsection{Production mechanisms of $Q\bar Q$ pairs}
\begin{figure}[!htb]
\includegraphics[width=0.45\textwidth]{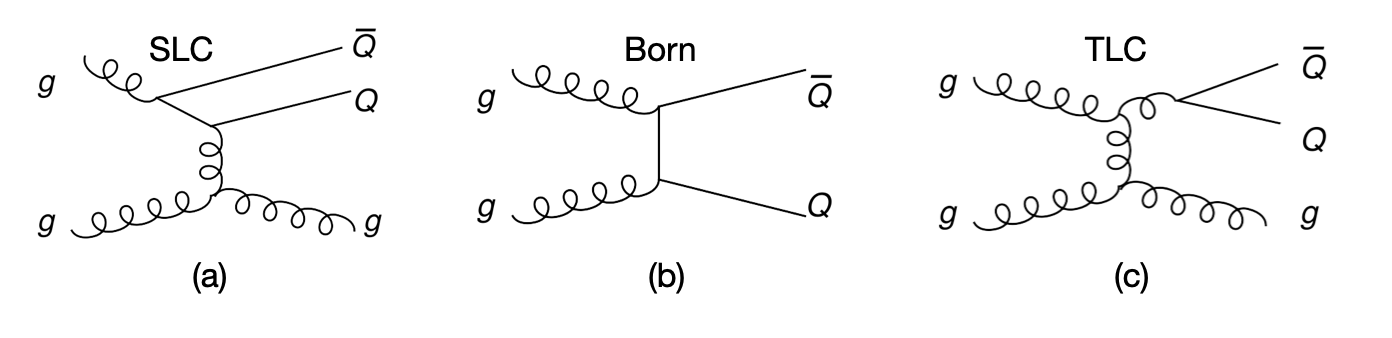}
\caption{The three production mechanisms for heavy quark pairs. From left to right: flavor excitation, flavor creation, and gluon splitting. For details we refer to the text.}
\label{fig.3pro}
\end{figure}

In EPOS4, $Q\bar Q$ pairs can be created in three leading order or next to leading order processes, which are displayed in Fig.~\ref{fig.3pro}. They can be created from a time-like gluon, which disintegrates into a $Q\bar Q$ pair after the gluon has been produced in the hard collision between the partons from projectile and target. This process is called ``gluon splitting'' and is shown in the right figure. It can as well be produced by gluons, which are emitted during the space-like cascade (SLC) of the partons before the hard scattering process. In the hard process between the projectile and the target partons, mediated by a gluon exchange, the $Q\bar Q$ pair gets then on the mass shell. This process, shown in the left figure, is called ``flavor excitation''. A $Q\bar Q$ pair can also be produced in the hard collision between the projectile and target partons, displayed on the middle figure. This leading order process is called ``flavor creation''. In addition to these basic processes, higher order terms contribute as well, due to the realization of full space-like and time-like cascades (TLC), see Ref.~\cite{Werner:2023fne}.
%-------------------------------------------------------
%-------------------------------------------------------

In this paper we employ EPOS4HQ, which was introduced in \cite{Zhao:2024ecc,Zhao:2023ucp}. In distinction to EPOS4, in EPOS4HQ heavy quarks can interact with the QGP, which is formed if the critical energy density of particles produced in hard collisions, is higher than 0.57 GeV/fm$^3$ \cite{werner:2023-epos4-micro}. The formation of a QGP has a significant influence on some observables, like on the elliptic flow and on the chemistry of charmed baryons but leaves other HF-observables, such as momentum spectra and correlations, practically unchanged, see Ref.~\cite{Zhao:2023ucp}.
%\kw{not very clear if in the following we talk about EPOS4 or EPOS4HQ. I guess correlations in pp, the difference is very small, right?  I think a paragraph should be added here to discuss this, and one should quote the correlation results shown in  \cite{Werner:2023zvo}}

\subsection{$Q \bar Q$ azimuthal correlations}
Already from the kinematics one can expect that the different processes yield quite different correlations between the $Q$ and the $\bar Q$. For flavour creation one expects in the center of mass of the heavy quark pair a back to back emission, whereas for gluon splitting the kinematics is determined by the gluon and one expects a small angle between the $Q$ and $\bar Q$. The importance of the different processes depends strongly on the energy and on the kinematic region where the correlations are studied. It is also different for bottom and charm quarks.

%%%%%%%%%%%%%%%%%%%%%%%
\begin{figure}[!htb]
\includegraphics[width=0.45\textwidth]{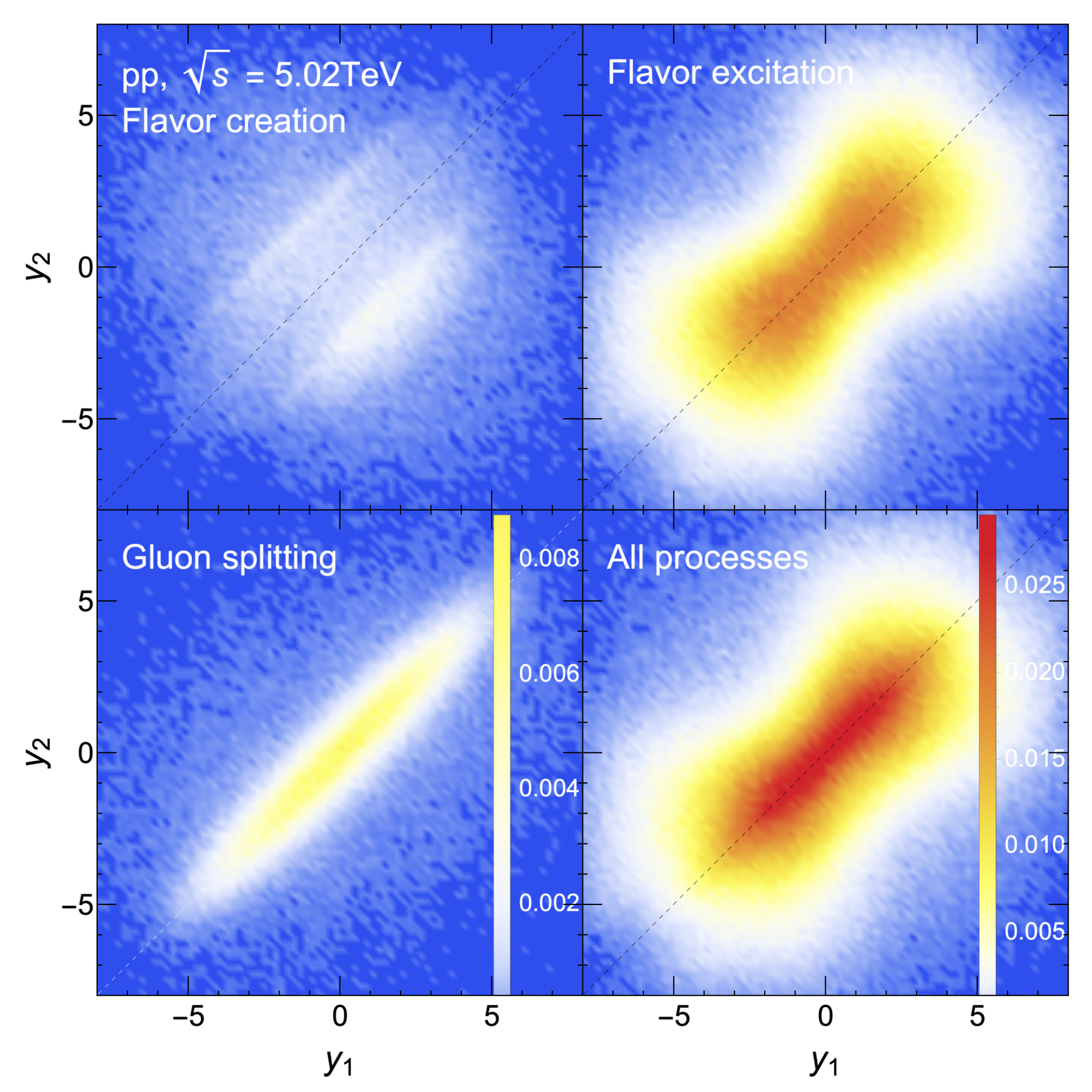}
\caption{The correlation between the rapidities of $c$ and $\bar c$ from  different processes in pp collisions at $\sqrt{s}=5.02 ~ \rm TeV$. }
\label{fig.y1y2}
\end{figure}
%-------------------------------------------------------
\begin{figure}[!htb]
\includegraphics[width=0.45\textwidth]{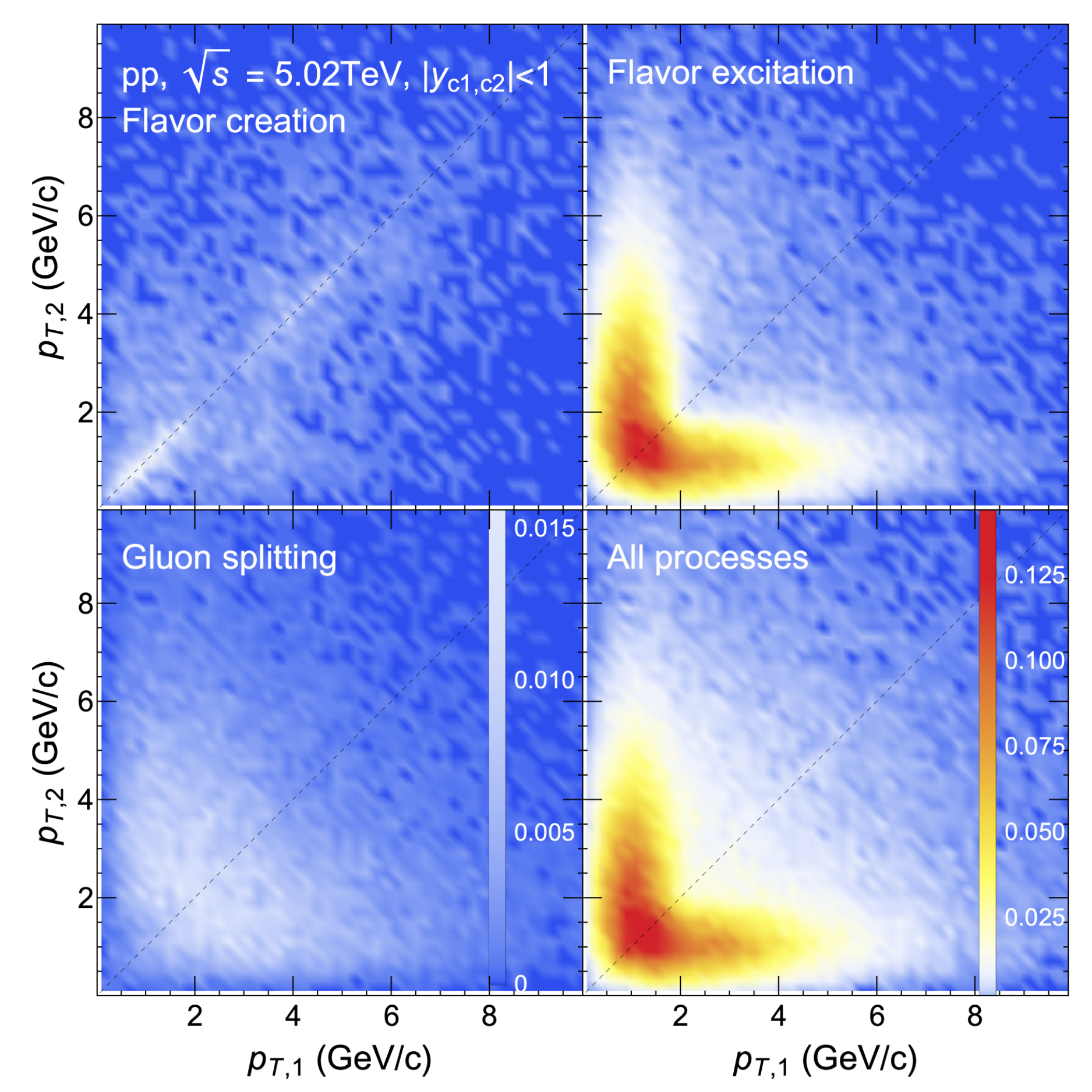}
\caption{The correlation between $p_T$ of $c$ and $\bar c$ from different processes in pp collisions at $\sqrt{s}=5.02 ~ \rm TeV$ and central rapidity $|y|<1$.}
\label{fig.pt1pt2}
\end{figure}
%-------------------------------------------------------
\begin{figure}[!htb]
\includegraphics[width=0.4\textwidth]{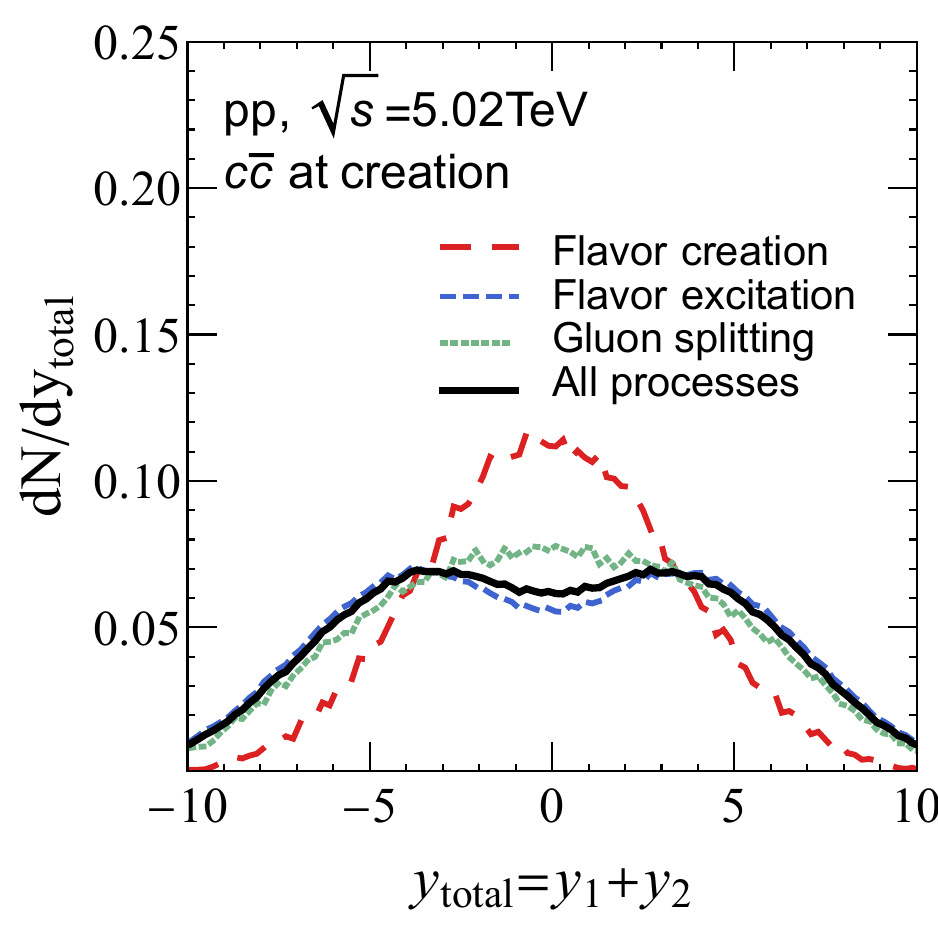}\\
\includegraphics[width=0.4\textwidth]{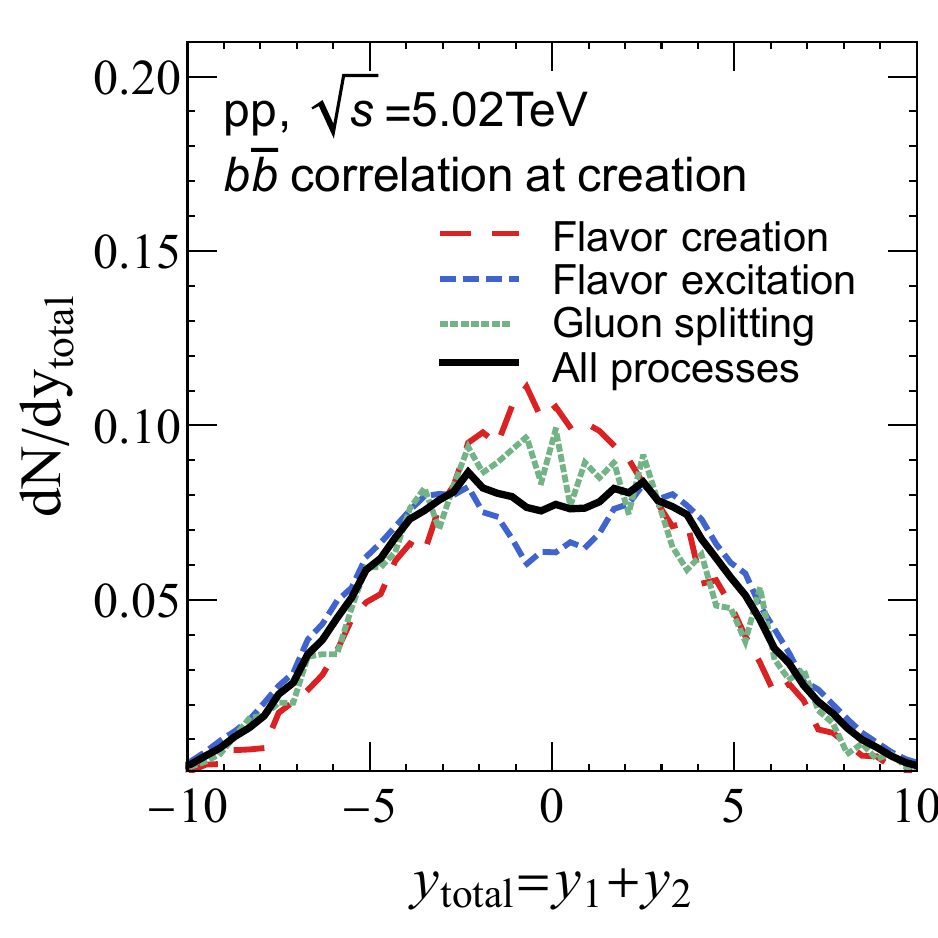}
\caption{The normalized distribution of the total rapidity of $c$ and $\bar c$ (upper), $b$ and $\bar b$ (lower) for different processes in pp collisions at $\sqrt{s}=5.02~\rm TeV$. }
\label{fig.ytotal}
\end{figure}
%-------------------------------------------------------
\begin{figure}[!htb]
\includegraphics[width=0.45\textwidth]{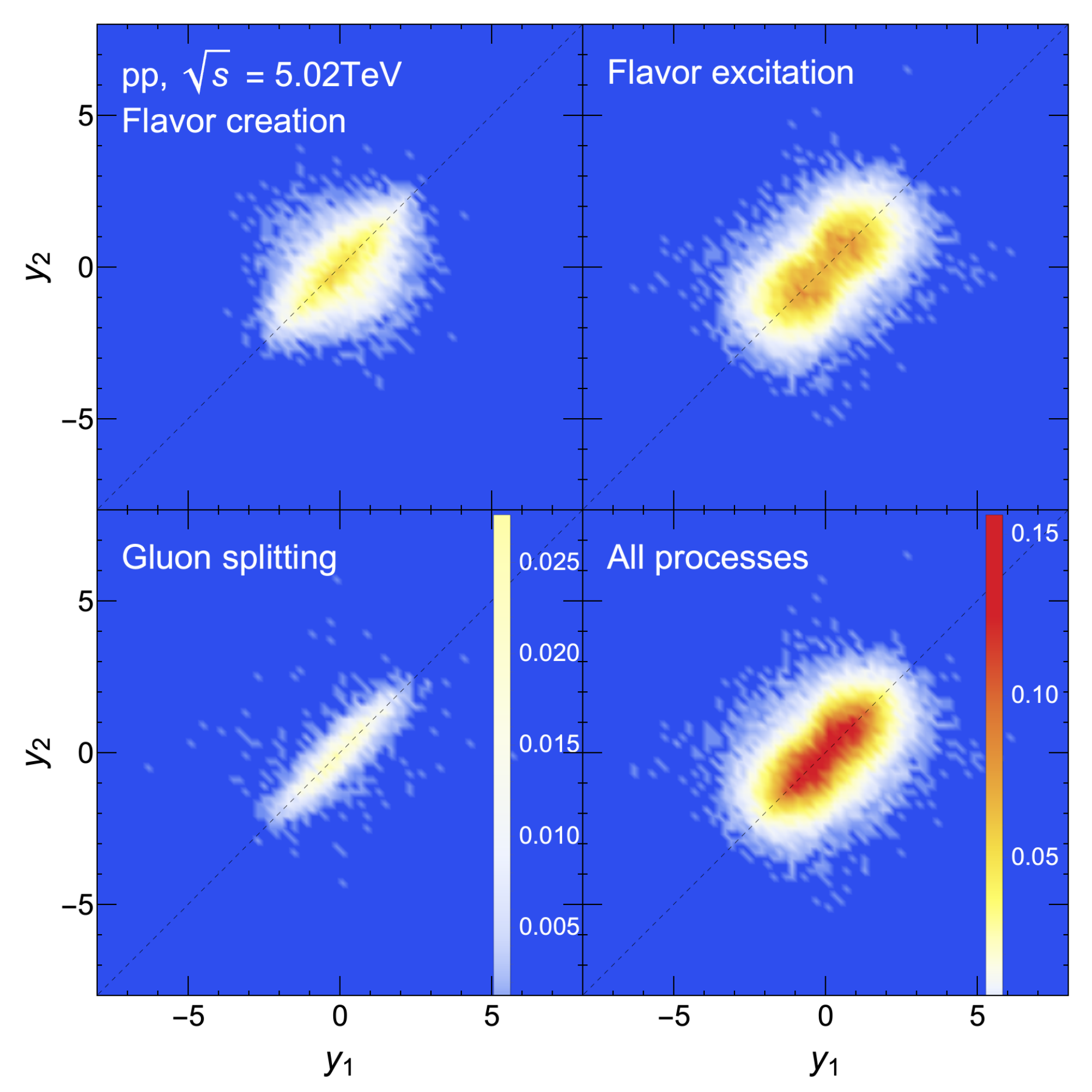}\\
\caption{Same as Fig.~\ref{fig.y1y2} for $b\bar b$ pairs.}
\label{fig.y1y2b}
\end{figure}
%-------------------------------------------------------
%-------------------------------------------------------
\begin{figure}[!htb]
\includegraphics[width=0.45\textwidth]{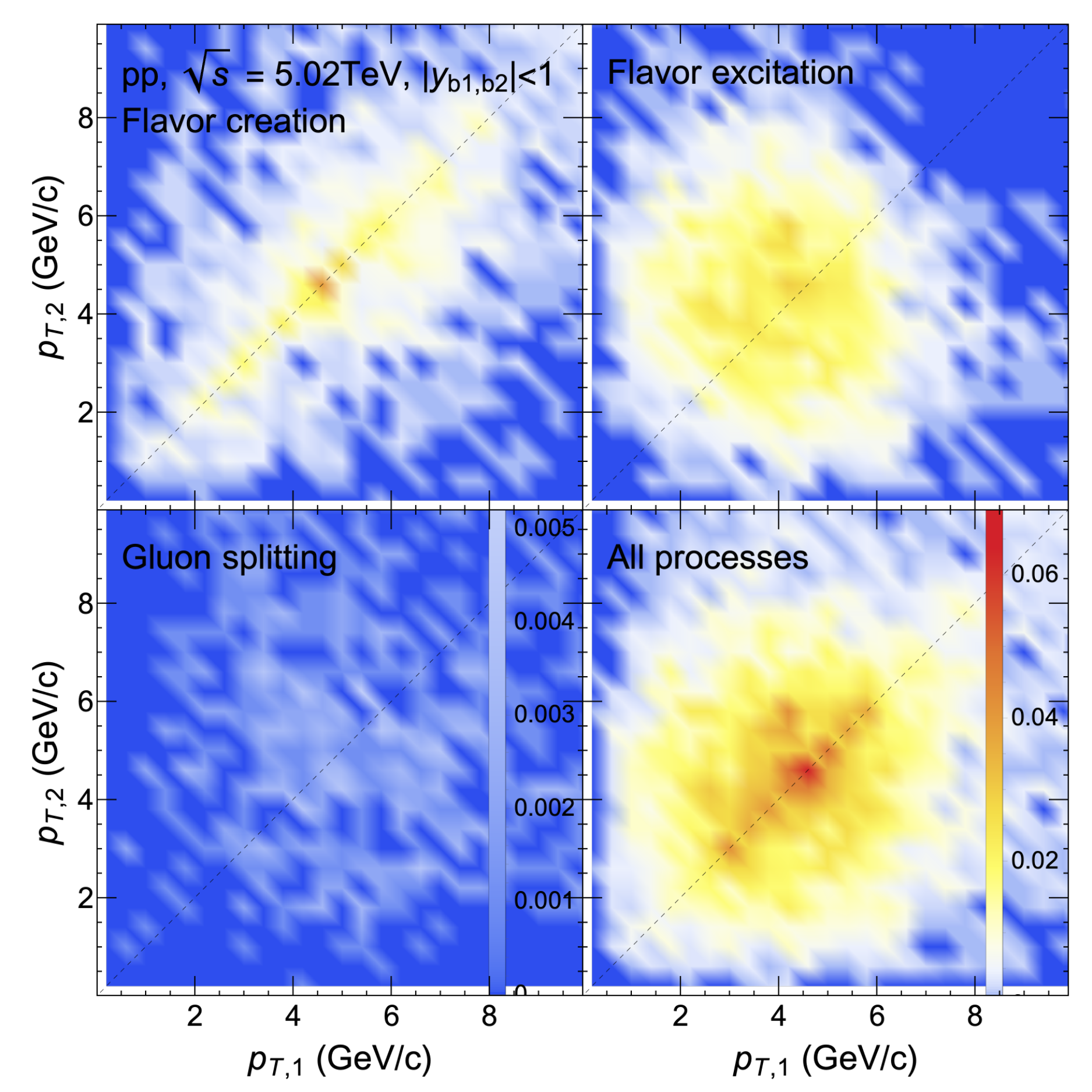}\\
\caption{Same as Fig.~\ref{fig.pt1pt2} for $b\bar b$ pairs.}
\label{fig.pt1pt2b}
\end{figure}
%-
The correlations between a $c$ and a $\bar c$ quark  at creation, produced in pp collisions with $\sqrt{s}$ = 5.02 TeV, are shown in Fig.~\ref{fig.y1y2} and Fig.~\ref{fig.pt1pt2} in a density plot for the different creation processes.
In Fig.~\ref{fig.y1y2} we display the correlation between the rapidities, in Fig.~\ref{fig.pt1pt2} that between the transverse momenta of the charm quark and the antiquark. We see, first of all, that there exist strong correlations between the $c$ and $\bar c$ created at the same vertex. In the flavour creation process the heavy quarks are centered around two bands. The maximum of these bands
is at $y_1=-y_2=\pm 1.5$. $y_1=-y_2$ we expect from back to back emission in the pp center of mass system. Because initially the quark-antiquark pair or gluon pair, which enters the hard process and which creates the $c\bar c$ pair has negligible transverse momentum, in the flavor creation process the heavy quarks have an opposite transverse momentum. We observe indeed a very strong correlation in transverse momentum space between the two heavy quarks.  The rapidity distribution of heavy quarks produced by flavor excitation peaks at midrapidity. The correlation between the heavy quarks is, however, not very strong since one of the heavy quarks gets a kick in the hard process (Fig.~\ref{fig.3pro}, left). In this kick also transverse momentum is transferred to this heavy quarks. Therefore one of the heavy quarks has a considerable larger transverse momentum than the other as seen in the transverse momentum correlation. In the gluon-splitting process the rapidities are strongly aligned, reflecting the rapidity of the gluon, which disintegrates into the pair (Fig.~\ref{fig.3pro}, right). The transverse momenta are little correlated as a recoiling parton, opposed to the gluon, carries away a significant transverse momentum. As we will discuss later, flavour excitation is the dominant mechanism for $c\bar c$ production at low $p_T$ and therefore, if we add up all processes in the bottom right figures, we see that the correlation pattern is governed by the flavour excitation processes.

The normalized distribution of the sum of the rapidities of the $c \bar c$ pair is displayed in Fig.~\ref{fig.ytotal}. The distribution due to flavour creation and gluon splitting are centered around midrapidity. In the flavour–excitation process, the rapidity distribution of heavy quarks shows a maximum at a finite $y_{total}$ because one of the quarks receives a finite momentum kick in the hard interaction.

Figs.~\ref{fig.y1y2b} and~\ref{fig.pt1pt2b} show the $y$ and $p_T$ correlations for $b\bar b$ pairs. For $b$ quarks the flavor creation process yield quarks, which are centered at equal rapidity, because only in this kinematics sufficient energy for the producing of b quarks is available. In the flavor excitation process the kick of one of the quarks due to the hard process is accompanied also by a rapidity transfer. Gluon splitting is for $b$ quarks much less important because the gluon energy for this processes is less frequently available.  Also for $b$ quarks, created by flavor creation, the transverse momenta of the two quarks have the same magnitude. The average $p_T$ of the $b$-quarks, created by flavour excitation, is larger than that for the $c$ quarks and centered around $p_{T1}\approx p_{T2}$.

\begin{figure}[!htb]
\includegraphics[width=0.23\textwidth]{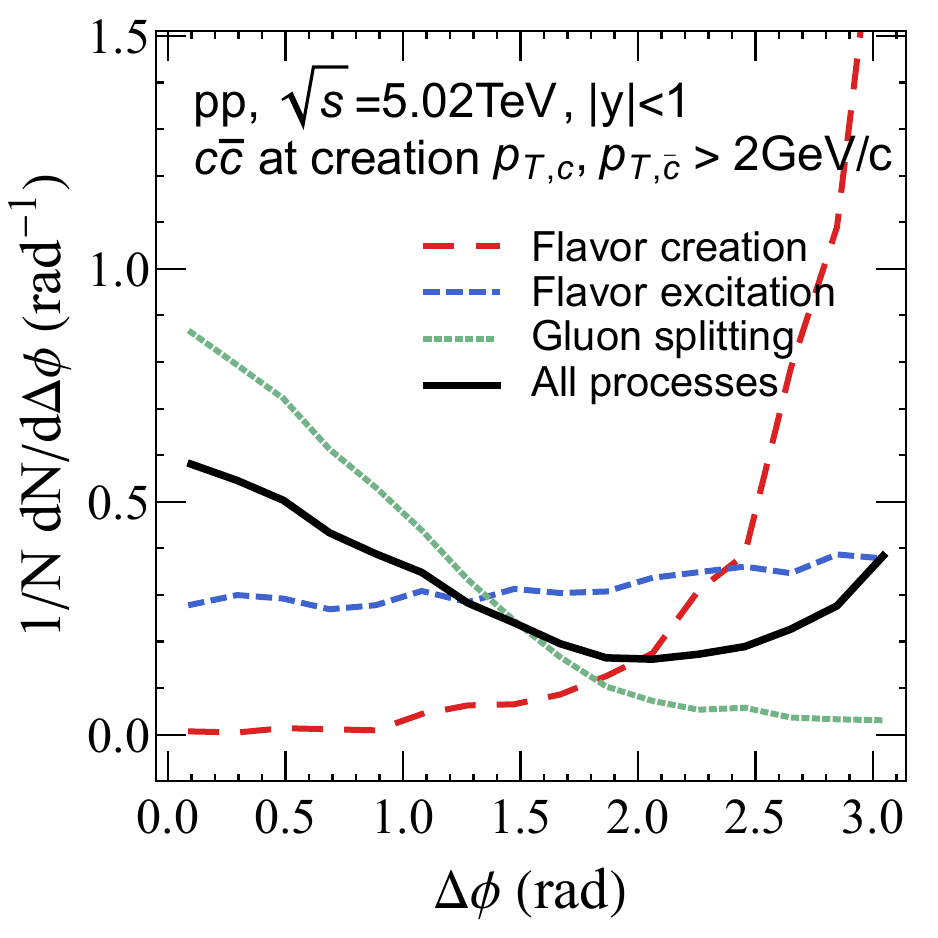}
\includegraphics[width=0.23\textwidth]{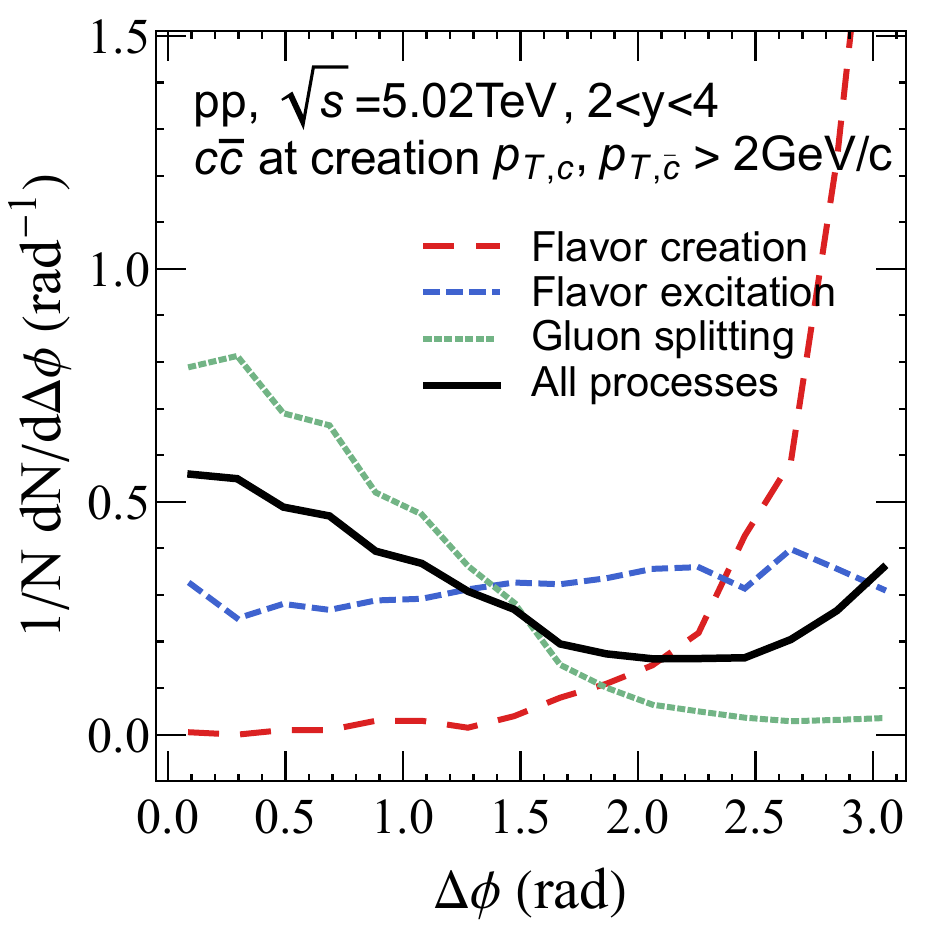}\\
\includegraphics[width=0.23\textwidth]{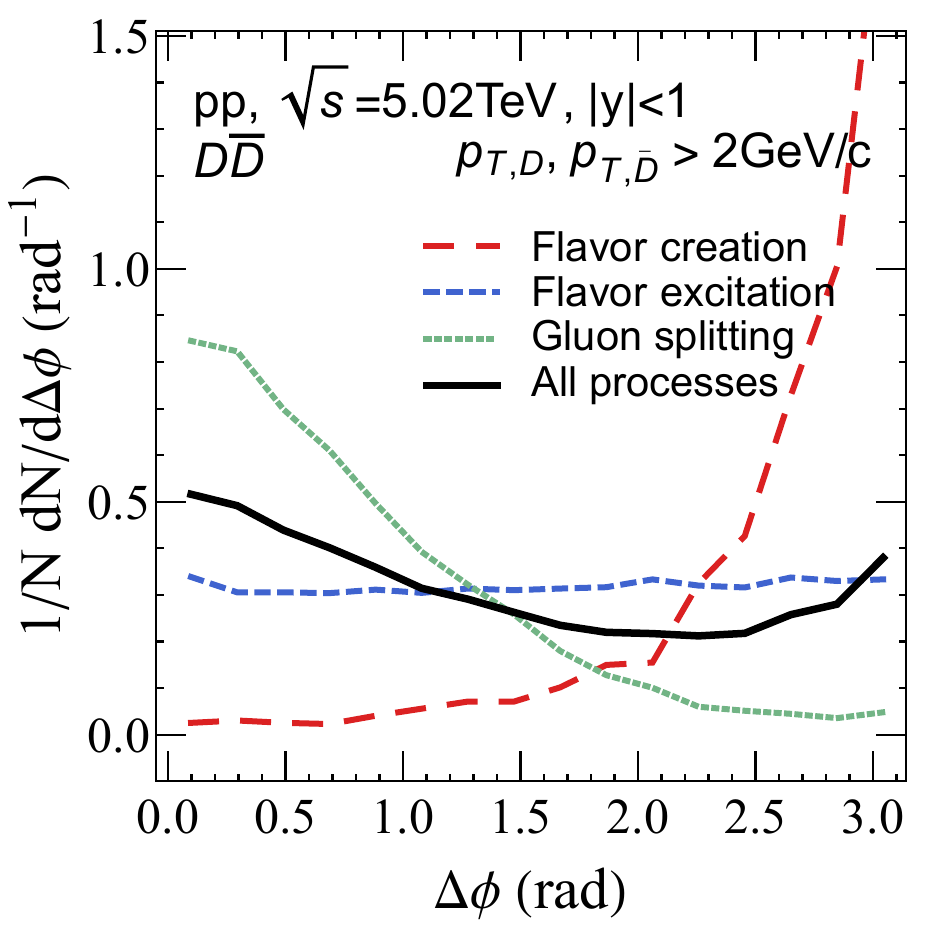}
\includegraphics[width=0.23\textwidth]{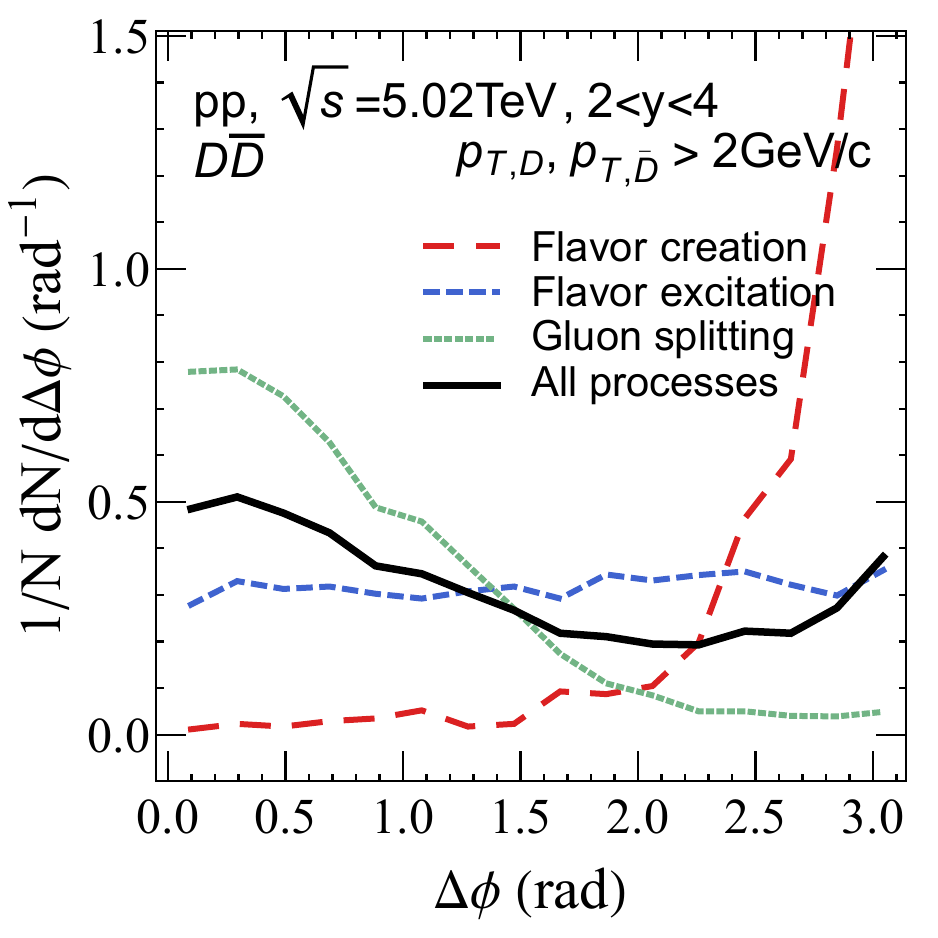}
\caption{Correlation of $c\bar c$ quarks with $p_T>$ 2GeV from the same vertex at creation and of $D^0\bar D^0$ mesons with $p_T>$ 2GeV in pp collisions at $\sqrt{s}=5.02~ \rm TeV$ and central rapidity (left) and forward rapidity (right).}
\label{fig.correlations}
\end{figure}

\begin{figure}[!htb]
\includegraphics[width=0.23\textwidth]{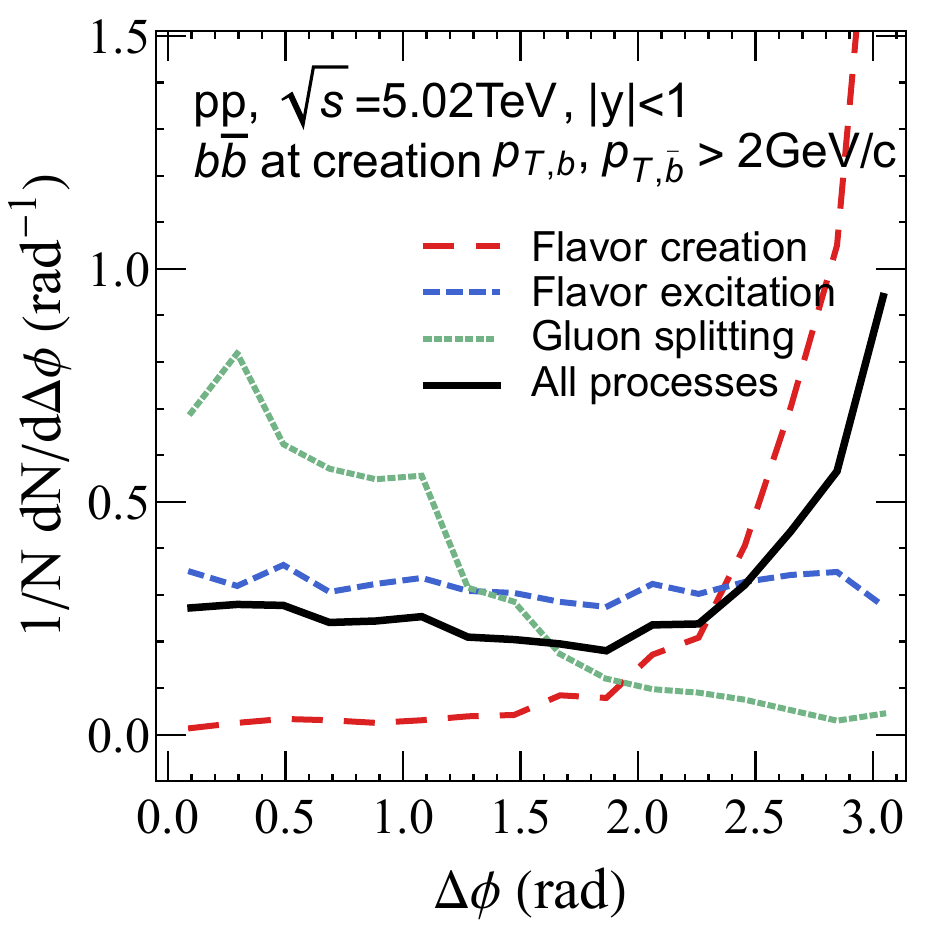}
\includegraphics[width=0.23\textwidth]{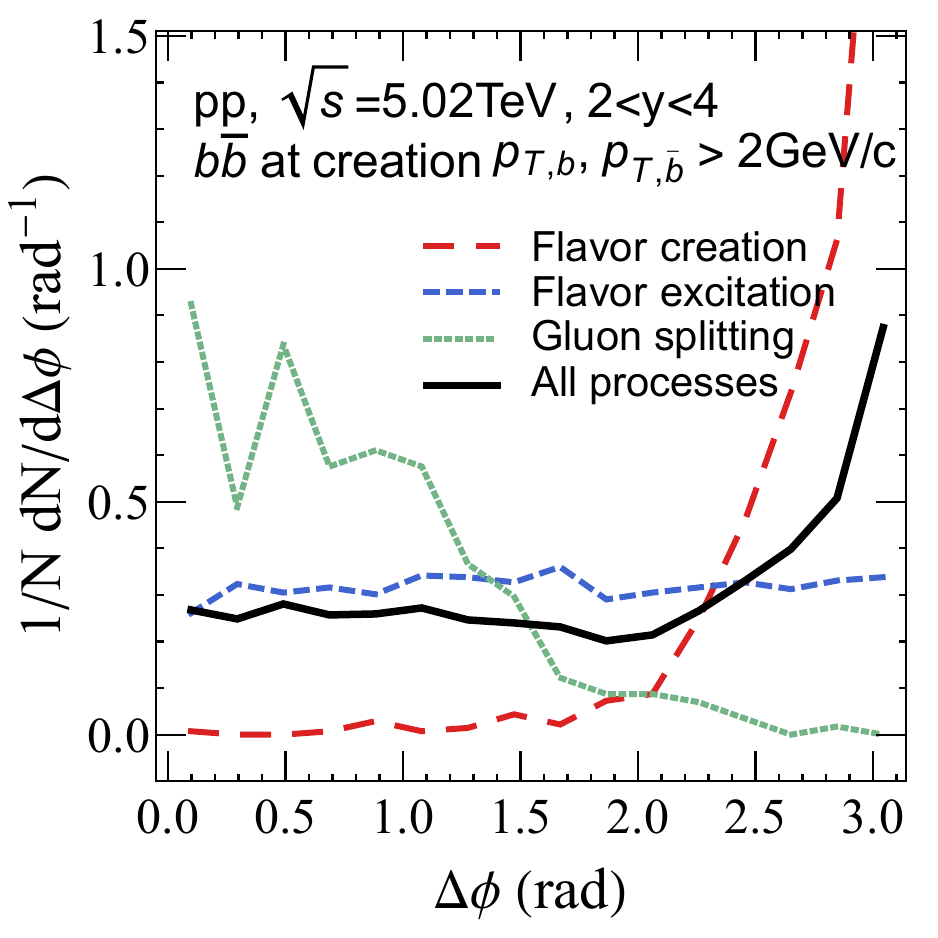}\\
\includegraphics[width=0.23\textwidth]{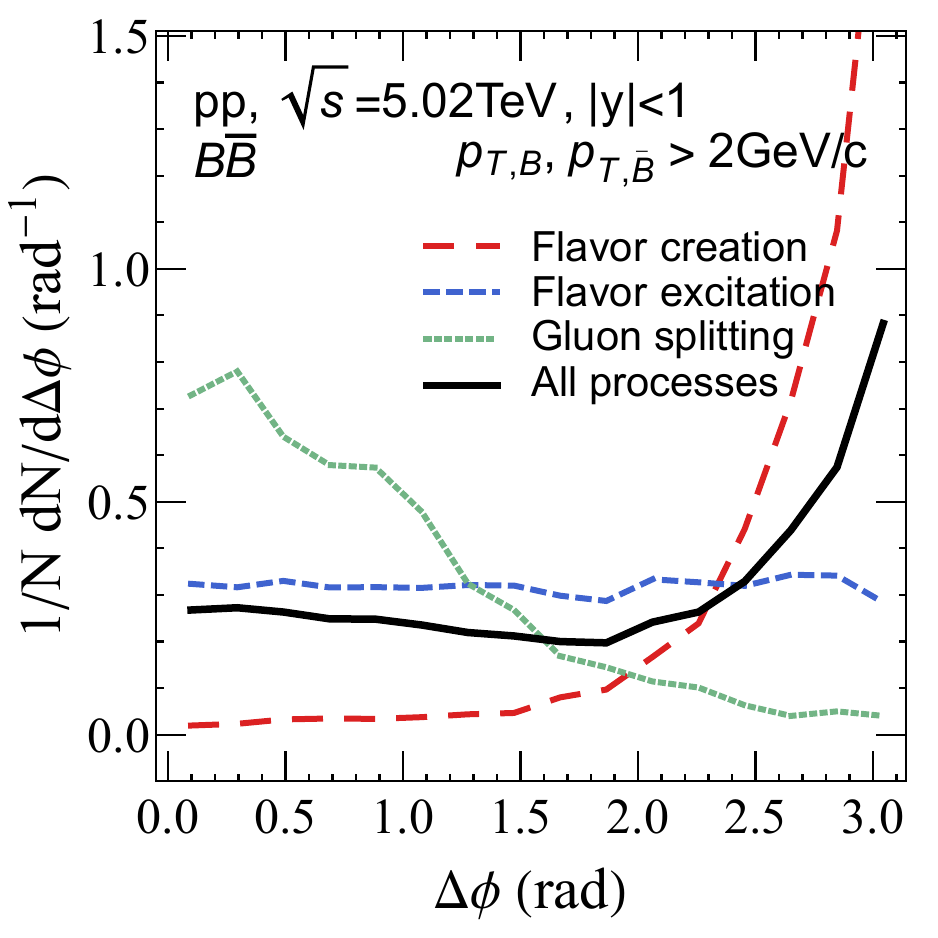}
\includegraphics[width=0.23\textwidth]{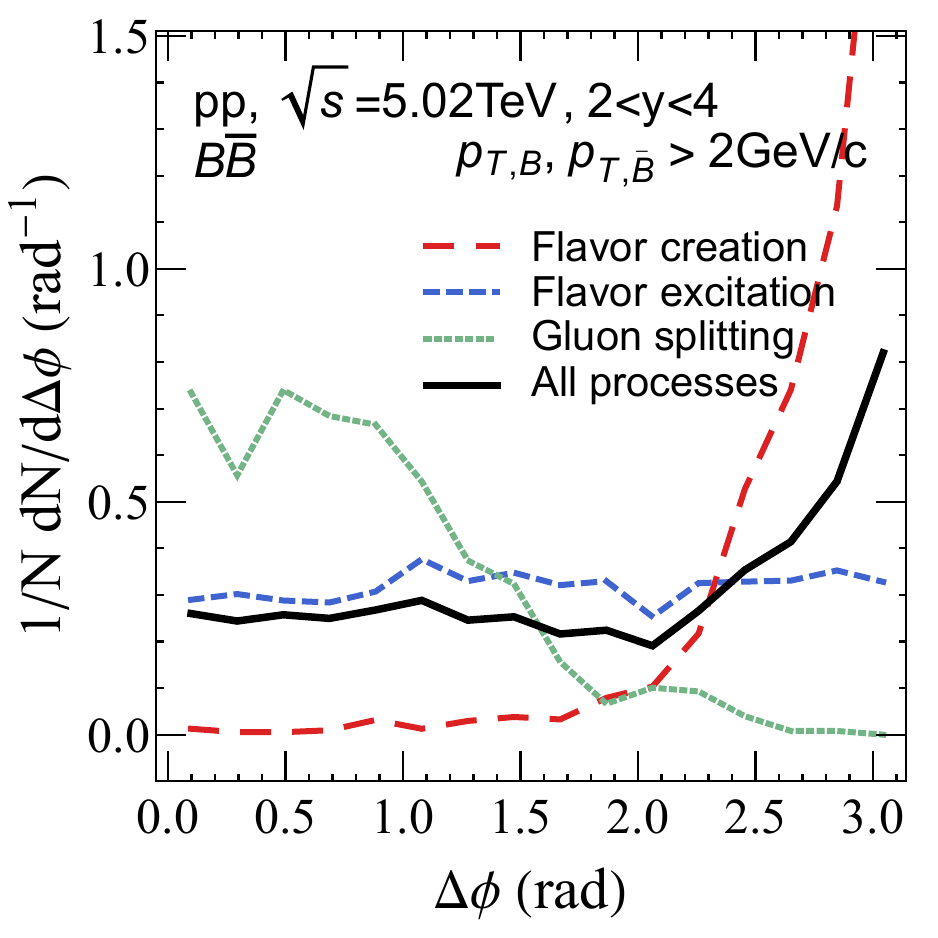}
\caption{Correlation of $b\bar b$ with $p_T>$ 2GeV from the same vertex at creation, before hadronization, and $B\bar B$ with $p_T>$ 2GeV in pp collisions at $\sqrt{s}=5.02~ \rm TeV$ and central rapidity (left) and forward rapidity (right).}
\label{fig.correlationsbb}
\end{figure}

The $p_T$ integrated azimuthal correlations between the $Q\bar Q$ pair for finally observed mesons for $p_{TQ},p_{T\bar Q} > $2 GeV is displayed in Fig.~\ref{fig.correlations} for charmed quarks (mesons) and in Fig.~\ref{fig.correlationsbb} for bottom quarks (mesons). We display in the top row the correlations of the heavy quarks at production, in the bottom row that for the finally observed open heavy flavor mesons. In the left row we see the results for mesons observed with $|y|<1$ and in the right row for forward emitted mesons  $2<y<4$. We separate the correlations for the different production processes. The dashed red line shows the results for quarks produced in a flavor creation process, the short dashed blue line those produced in a  flavor excitation processes and the dotted green line those from gluon splitting. The thick black line is the sum over all processes, as seen in experiment. We observe for all gluon creation mechanisms quite different correlations, which change, however, not much with centrality. As expected, the transverse momenta of $Q\bar Q$ pairs created by flavour creation are back to back and therefore the azimuthal opening angle peaks at $\pi$. On the contrary, those produced by gluon splitting have a small opening angle whereas those produced by flavor excitation show an almost flat distribution as a function of the azimuthal opening angle.
 %---------------------------------------------------------------------
%---------------------------------------------------------------------
%---------------------------------------------------------------------
%---------------------------------------------------------------------
%---------------------------------------------------------------------
\begin{figure}[!htb]
\includegraphics[width=0.23\textwidth]{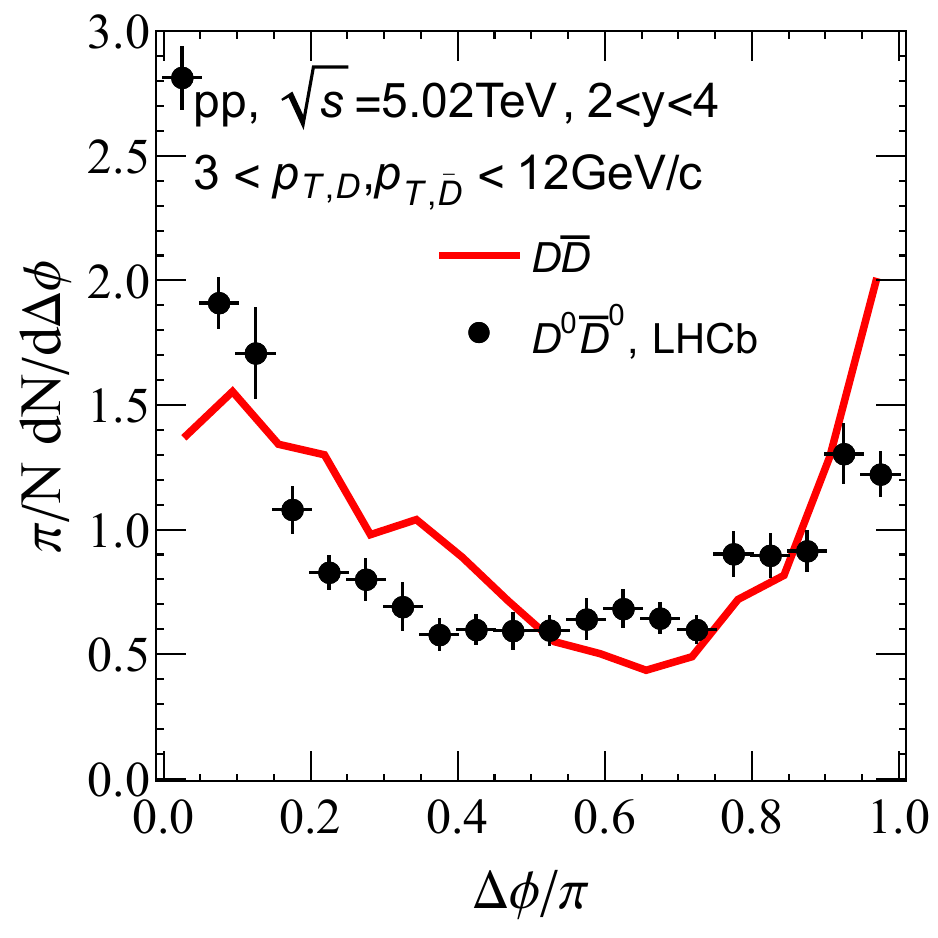}
\includegraphics[width=0.23\textwidth]{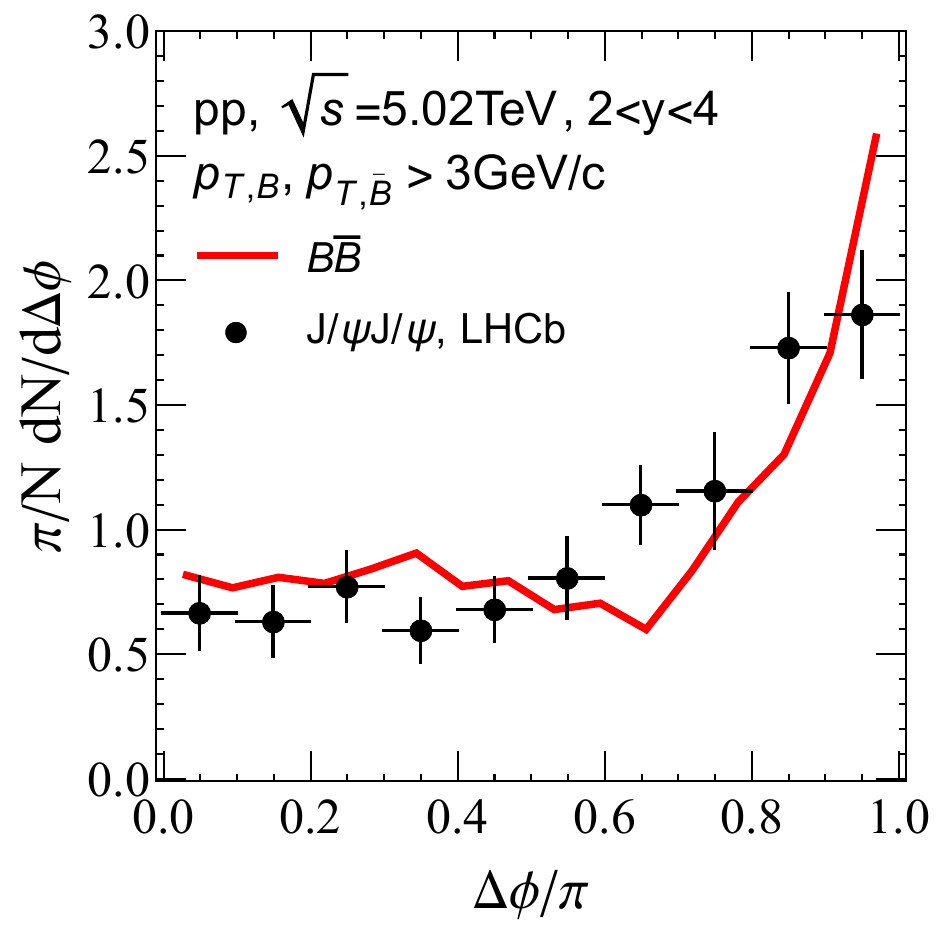}
\caption{Experimental correlations in comparison with EPOS4 results. Left: $D^0\bar D^0$ correlations in pp collisions at $\sqrt{s}=7~ \rm TeV$~in comparison with LHCb data  \cite{LHCb:2012aiv},  Right:  EPOS4HQ  $B\bar B$ correlations in comparison with the experimental correlations of  non-prompt $J/\psi$ from $B$ decays in pp collisions with $\sqrt{s}=7~ \rm TeV$~\cite{LHCb:2017bvf}.}
\label{fig.correlationsexp}
\end{figure}
%---------------------------------------------------------------------
%---------------------------------------------------------------------
\begin{figure}[!htb]
\includegraphics[width=0.23\textwidth]{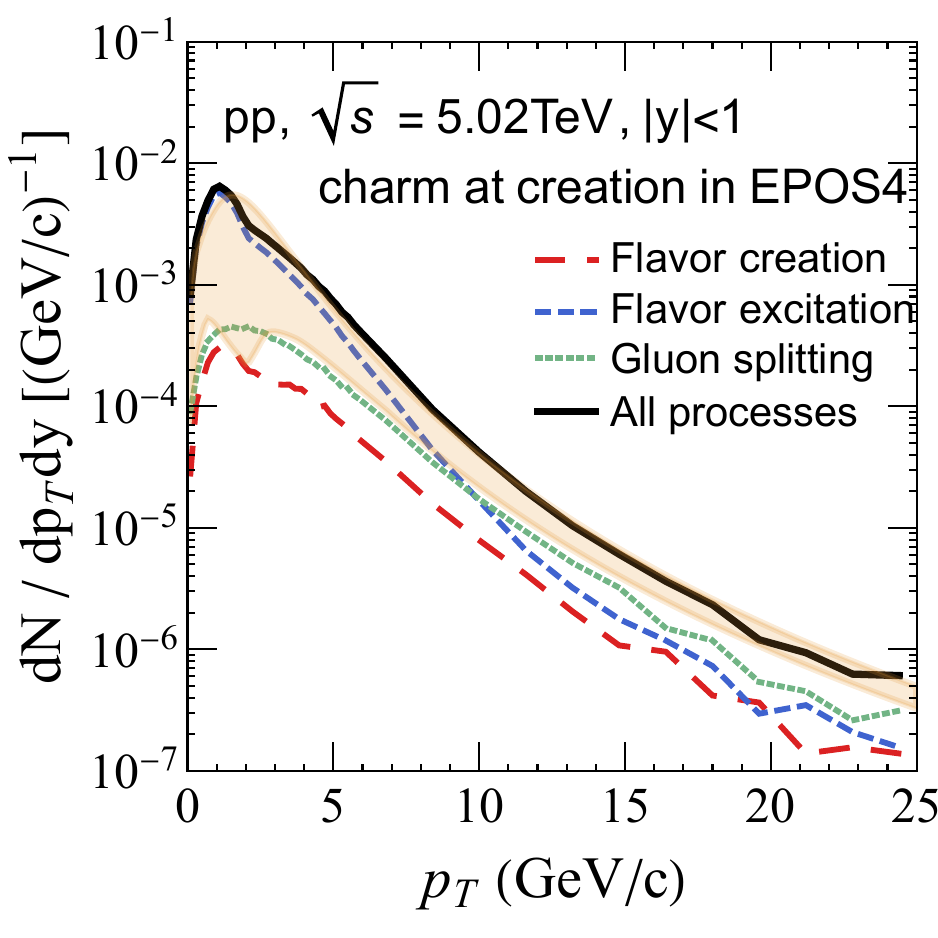}
\includegraphics[width=0.23\textwidth]{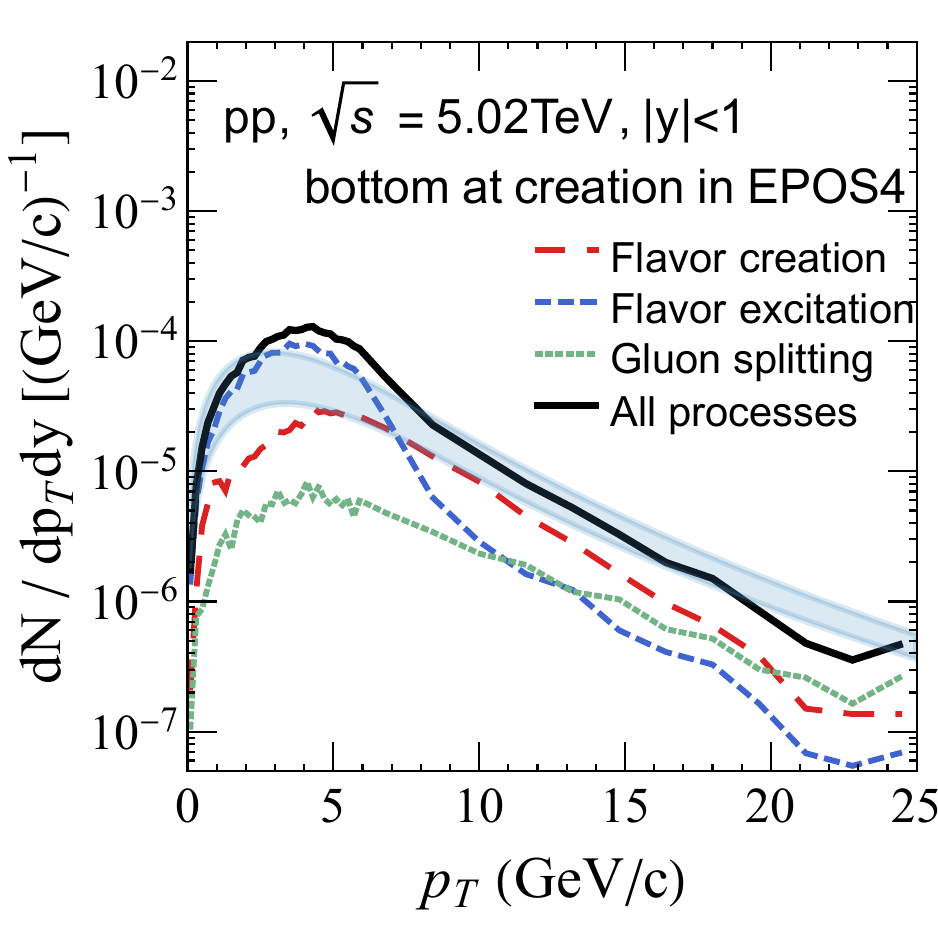}\\
\includegraphics[width=0.23\textwidth]{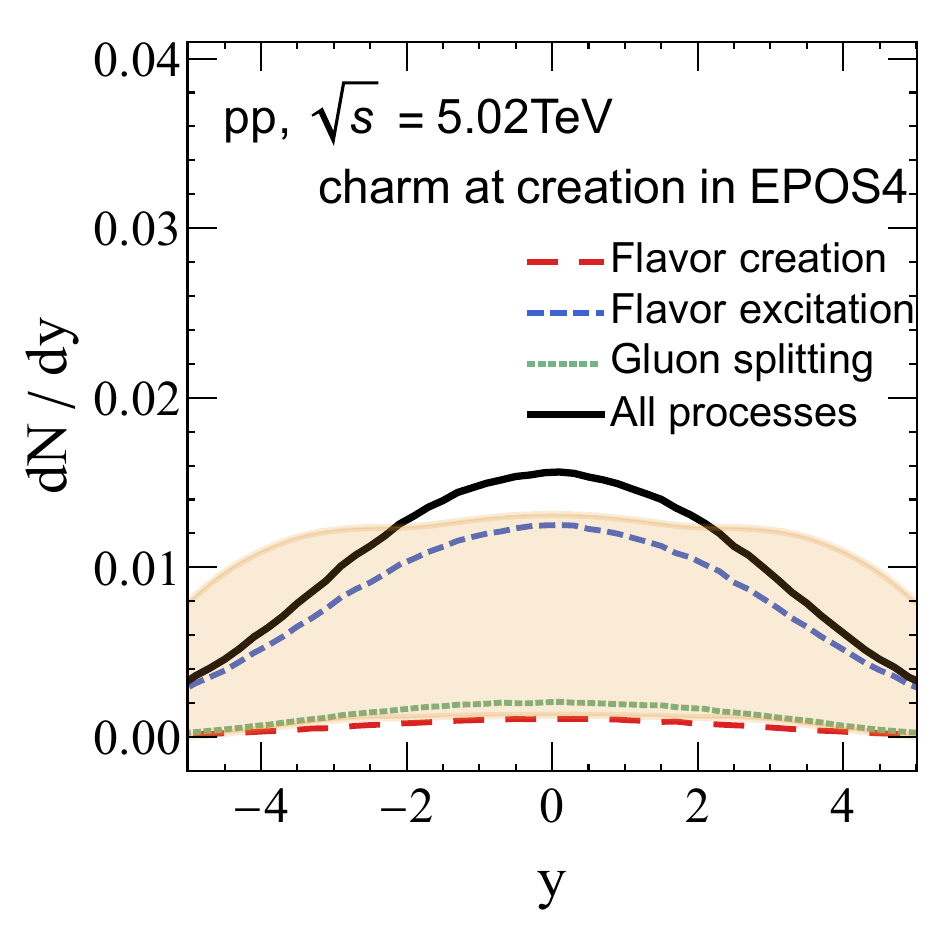}
\includegraphics[width=0.23\textwidth]{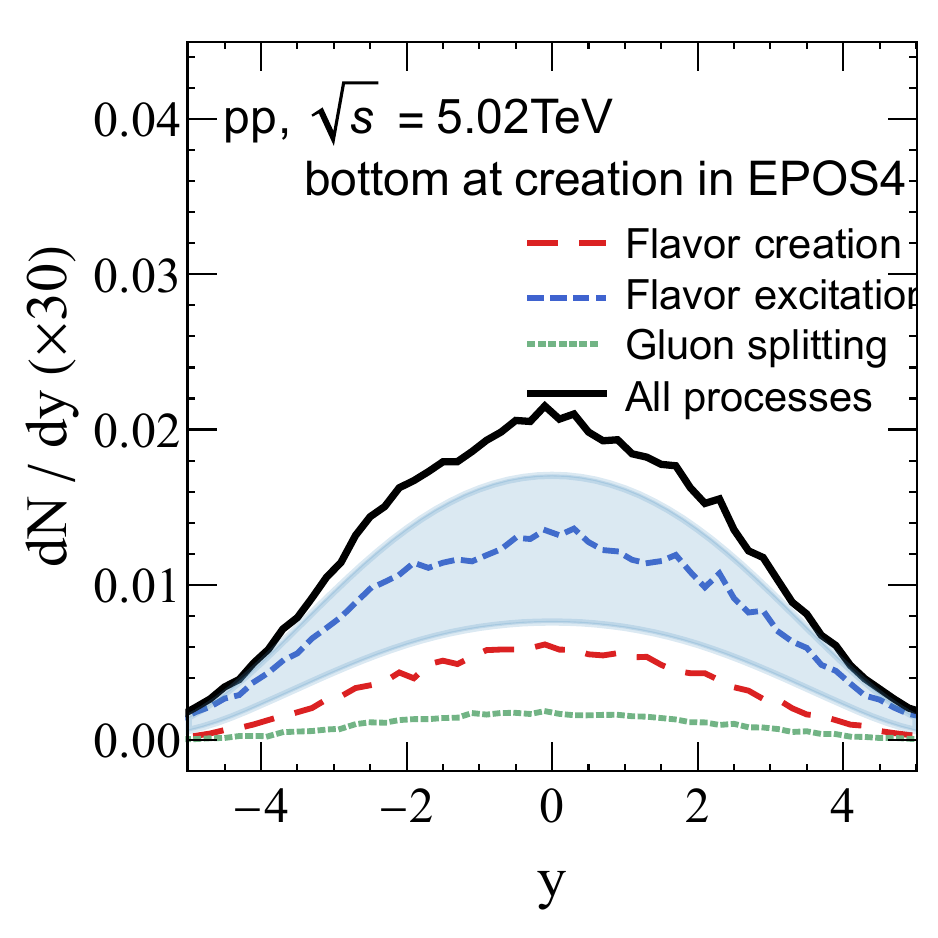}
\caption{$p_T$ spectrum and rapidity distribution of charm (left) and bottom quark (right) in pp collisions at $\sqrt{s}=5.02~ \rm TeV$. The band is the FONLL calculation~\cite{Cacciari:1998it}.}
\label{fig.candb.pt}
\end{figure}
%---------------------------------------------------------------------

If one sums over all processes the azimuthal correlations between $c\bar c$ pairs shows still a enhancement for $\Delta \phi= 0$ and  $\Delta \phi= \pi$. Since gluon splitting is more important than flavor creation the enhancement at $\Delta \phi= 0$ is larger. It is remarkable that the correlation function remains almost unchanged from the initial heavy quark production to the finally observed heavy meson in pp collisions. Hence, heavy-flavor correlation experiments offer the opportunity to study the importance of the different leading order and next to leading order heavy flavour creation processes and their kinematics. 

For the $b \bar b$ pairs the dominance of the back to back emission remains visible, even after summing over the all processes. For low $\Delta \phi$ the spectrum is almost flat due to the little  importance of the production by gluon splitting. Also the for bottom quarks the initial correlation is almost identical to that of the final $B$ mesons.

The different between the correlation functions of $D$ and $B$ mesons has also been observed experimentally. In Fig.~\ref{fig.correlationsexp} we compare our calculations with the available experimental results of LHCb. On the left hand side we compare the measured $D^0\bar D^0$ azimuthal correlations for $2<y<4$ and $3<p_T<12$ GeV with the $D\bar D$ results of EPOS4HQ. First of all, we see that due to the higher $p_T$ cut  ($p_T >$  3 GeV)  the $D\bar D$ correlations due to the three production processes become even more visible. The $B \bar B$ correlations, shown on the right hand side is different. We have to add here a warning.  The $b$ hadrons disintegrate into a $J/\psi$ via many different channels, for none of them the explicit kinematics is known.  Therefore we compare here the measured $J/\psi J/\psi$ correlation with the calculated $B\bar B$ correlation, neglecting the fact that the disintegration of the $B$ into a $J/\psi$ smears out the correlation.  We see clearly the enhancement due to back to back emission in the flavor creation process but, as discussed,  the enhancement of creation with a small azimuthal opening angle is much less visible than for $D\bar D$ creation. This is a consequence of the different relative contribution of the different heavy quark creation processes. Both, the $D\bar D$ as well as the $B\bar B$ azimuthal correlations are well reproduced in EPOS4HQ.

\subsection{Single particle spectra}
\begin{figure}[!htb]
\includegraphics[width=0.45\textwidth]{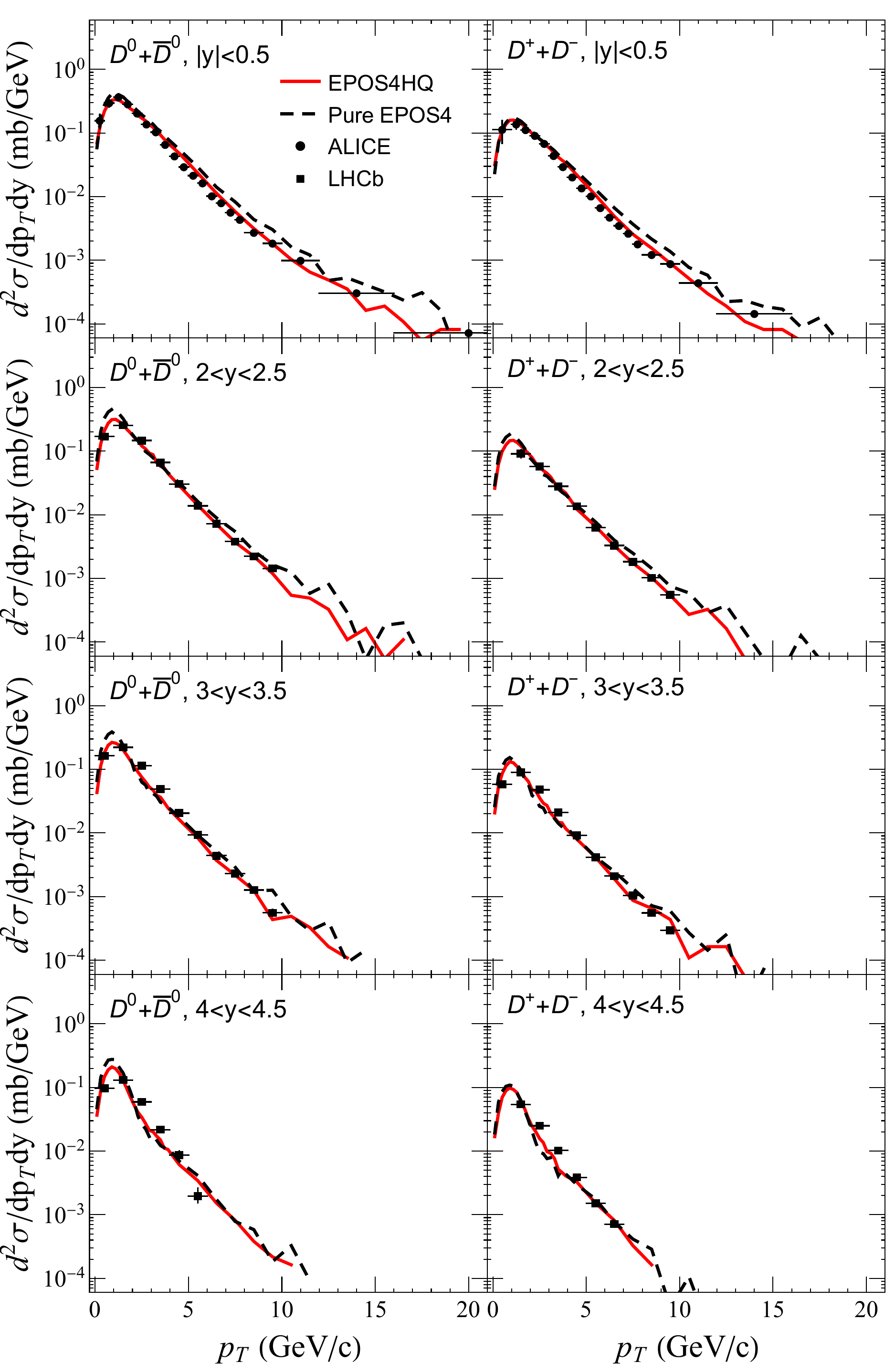}
\caption{$p_T$ spectra of $D^0+\bar D^0$ and $D^++D^-$ at central [-0.5,0.5] and forward rapidity bins [2,2.5], [3,3.5], and [4,4.5] in pp collisions at $\sqrt{s}=5.02~ \rm TeV$. The experimental data are from the ALICE~\cite{ALICE:2021mgk} and LHCb~\cite{LHCb:2016ikn}.}
\label{fig.Dpt_all}
\end{figure}

\begin{figure}[!htb]
\includegraphics[width=0.45\textwidth]{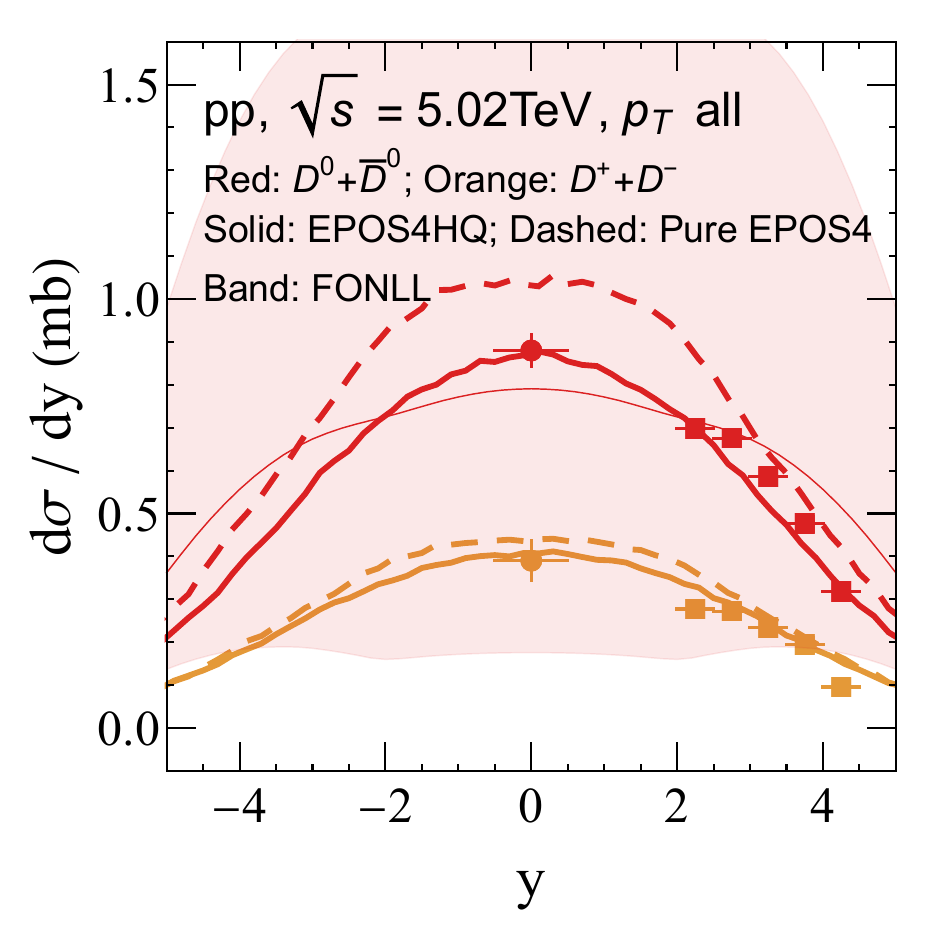}
\caption{Rapidity distribution of $D^0+\bar D^0$ and $D^++D^-$ in pp collisions at $\sqrt{s}=5.02~ \rm TeV$. The experimental data are from the ALICE~\cite{ALICE:2021mgk} and LHCb~\cite{LHCb:2016ikn}. The band is the $D^0+\bar D^0$ predicted by FONLL with the combined mass and scale uncertainties ~\cite{Cacciari:1998it}. The thin solid line is the central value of FONLL prediction.}
\label{fig.Drap_all}
\end{figure}

It is interesting to see how the different production mechanisms contribute to the EPOS4HQ single particle spectra, which have already been compared with the experimental results in~\cite{Zhao:2024ecc}. This is shown in Fig.~\ref{fig.candb.pt} where we compare the EPOS4HQ results also with FONLL calculations~\cite{Cacciari:1998it,Cacciari:2005rk} (shaded area). The $p_T$ spectra at midrapidity for $\sqrt{s}$ = 5.02 TeV are shown in the top row, the $p_T$ integrated rapidity distribution in the bottom row. Left we display the results for charm quarks and right for bottom quarks. We see  that  for $p_T > 10$ GeV  the charm quarks from the gluon splitting process are dominant, whereas for bottom quarks,  contributions from flavor creation are more important. Flavour excitation dominates both spectra at low $p_T$. The sum of all contributions is very similar to the FONLL prediction, for charm as well as for bottom quarks. The rapidity distributions are peaked at midrapidity and dominated due to the $p_T$ integration by the flavor excitation process.  In Fig.~\ref{fig.Dpt_all} we display the $D^0+ \bar D^0$, left, and the $D^++D^-$ spectra, right,  at midrapidity and forward rapidities. The different rows show different rapidity intervals, from central rapidities, measured by the ALICE collaboration~\cite{ALICE:2021mgk} until forward rapidities, measured by LHCb~\cite{LHCb:2016ikn}. These are the underlying spectra for the correlation calculations, presented in Fig.~\ref{fig.correlations}. We display as well the pure EPOS4 results. As explained in~\cite{Zhao:2024ecc}, all heavy mesons from EPOS4 are created by applying string decay whereas in EPOS4HQ a quark gluon plasma can be formed and heavy quarks, which emerge from this plasma can hadronize by coalescence. We see that the EPOS4 and EPOS4HQ calculations agree well with experiment. The differences are not visible in this logarithmic presentation.
The rapidity spectra of EPOS4 and EPOS4HQ of the $D$ mesons are shown in Fig.~\ref{fig.Drap_all} as dashed and solid lines, in comparison with the
FONLL calculations (shaded) and the experimental data. The deviation between theory and experiment come exclusively from the difference of the $p_T$ spectra at very low $p_T$, where pQCD based calculations reach their natural limitations. 

\begin{figure}[!htb]
\includegraphics[width=0.45\textwidth]{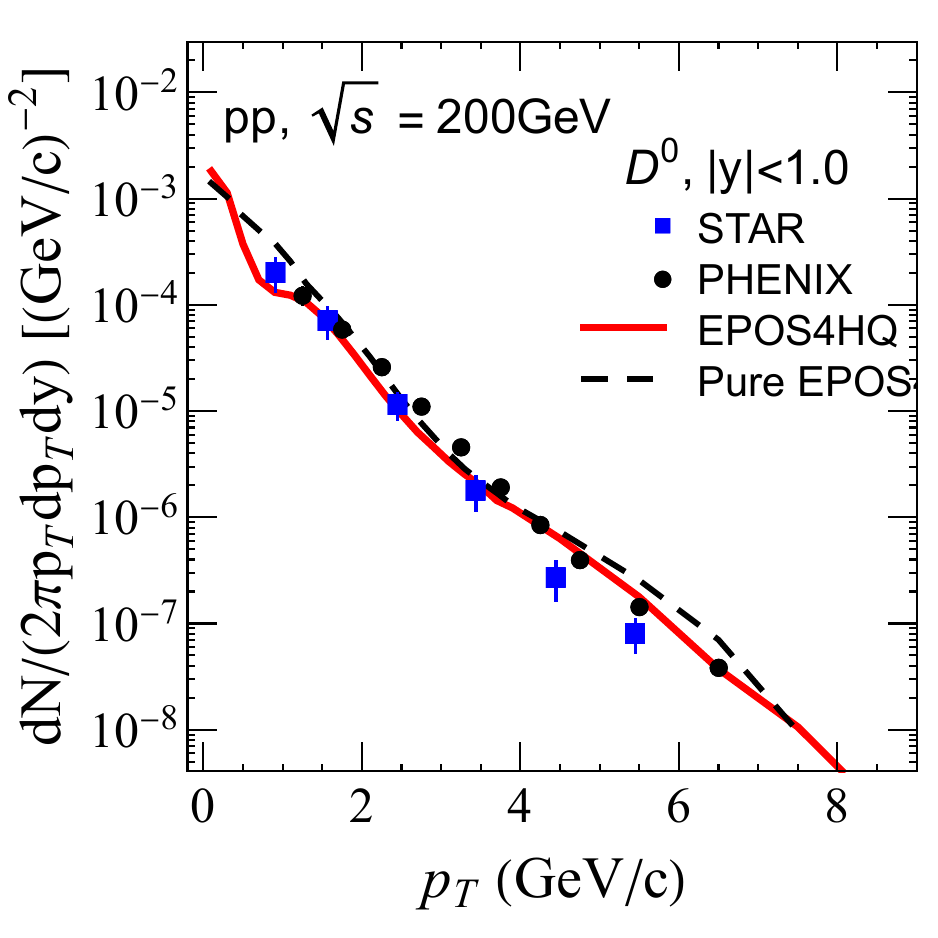}
\caption{$p_T$ spectra of $D^0$ in the central rapidity bin $|y|<1.0$ of pp collisions at $\sqrt{s}=200~ \rm GeV$. The experimental data are from the STAR~\cite{STAR:2014wif}}
\label{fig.Dpt_all_200}
\end{figure}

The $p_T$ spectra of $D^0$ at top RHIC energy are shown in Fig.~\ref{fig.Dpt_all_200} and compared with the results of PHENIX and STAR. We see a good agreement. The lack of data at low $p_T$ reduces the constraint of the low $p_T$ charm production, where the production is beyond the applicability of pQCD calculations. 
The spectra of $D$ mesons are different in EPOS4 and EPOS4HQ due to the different hadronization mechanisms. In EPOS4 heavy hadrons are produced by fragmentation only while in EPOS4HQ they can in addition be produced by coalescence. Hadronization by coalescence leads to an enhanced production of heavy baryons at low $p_T$ and hence to a reduction of the $D$ mesons.

%---------------------------------------------------------------------
%---------------------------------------------------------------------
%---------------------------------------------------------------------
%---------------------------------------------------------------------
%---------------------------------------------------------------------
%---------------------------------------------------------------------

%%%%%%%%%%%%%%%%%%%%%%%
\section{Quarkonium production}
\label{sec.III}
%%%%%%%%%%%%%%%%%%%%%%%

As said, the correlations between heavy quarks do not show up only in the correlation between open heavy flavor mesons but influence as well the production of hidden heavy flavor mesons because the probability that a $Q$ and a $\bar Q$ form a quarkonium depends on their kinematic variables. 

Before we investigate this in detail we introduce our model for quarkonium production, which is based on the quantal density matrix approach, expressed in terms of Wigner densities. The idea to use a density matrix approach to study bound two particle systems goes back to the seminal papers of Remler~\cite{Remler:1981du,Gyulassy:1982pe} and has recently been used to study quarkonium production in a thermal heat bath~\cite{Song:2023ywt} and in heavy-ion collisions \cite{Villar:2022sbv}. In its time independent form this method has been widely applied in the so called coalescence approach \cite{Fries:2008hs,Fries:2003kq}. The density matrix approach to study quarkonia in pp collisions has been introduced in \cite{Song:2017phm,Song:2023zma}. Here we present an advanced version of this approach, which we discuss now in detail. 
\subsection{The density matrix approach} 

The density matrix formalism is based on the quantal density operators
\begin{equation}
\rho^{\Phi_i}=|\Phi_i\rangle\langle\Phi_i| \quad \text{and} \quad \rho^{Q\bar Q}=|Q\bar Q\rangle\langle Q\bar Q| 
\end{equation}
where $\rho^{Q\bar Q}$ is the density matrix of the two body $Q\bar Q$ system  and  $|\Phi_i\rangle$ is wave function of the $i^{\rm th}$ quarkonium eigenstate. The production probability $P_i$ of a quarkonium state $\Phi_i$ in a $Q\bar Q$ system can be described by their density operators $\rho^{\Phi_i}$ and $\rho^{Q\bar Q}$ via
\begin{eqnarray}
P_i&\equiv& {\rm Tr}
(\rho^{\Phi_i}\rho^{Q\bar Q}).
\label{eq:prob}
\end{eqnarray}
A density matrix $\rho$  can be transformed into a Wigner density, which is defined as
\begin{eqnarray}
W({\bf x},{\bf q})&=&\int d^3ye^{-i{\bf q}\cdot {\bf y}/\hbar} \langle {\bf x}+{{\bf y}\over2 }|\rho|{\bf x}-{{\bf y}\over2 }\rangle\nonumber \\
&=& \int d^3ye^{-i{\bf q}\cdot {\bf y}/\hbar} \psi({\bf x}-{{\bf y}\over2 })\psi^*({\bf x}+{{\bf y}\over2 }). \label{eq.wignerdf}
\end{eqnarray} 
The Wigner function satisfies the normalization condition, 
\begin{eqnarray}
\int {d^3x\, d^3q \over (2\pi \hbar)^3}W({\bf x},{\bf q})=1. 
\end{eqnarray} 
and reduces to 
\begin{equation}
    W({\bf x},{\bf q})= \delta({\bf q}-{\bf q}_0).
\end{equation}
If the particle is in a momentum eigen state ${\bf q}_0$ and therefore described by a plane wave
$\psi({\bf x}) = \frac{1}{\sqrt{(2\pi \hbar)^3}} e^{i{\bf q}_0{\bf x}}$.
Introducing center-of-mass and relative coordinates
\begin{eqnarray}
&&{\bf R}=\frac{m_1{\bf r}_1+m_2{\bf r}_2}{m_1+m_2}, \quad {\bf r}={\bf r}_1-{\bf r}_2,\nonumber\\
&&{\bf P}={\bf p}_1+{\bf p}_2, \quad {\bf p}=\frac{m_2{\bf p}_1-m_1{\bf p}_2}{m_1+m_2}
\label{define1}
\end{eqnarray}
we can transform the density matrix $\rho^{\Phi_i}$ into a Wigner density,  which is defined as
\begin{eqnarray}
W({\bf R},{\bf P},{\bf r},{\bf p})&=&\int d^3yd^3Y\, e^{-i({\bf p}\cdot {\bf y} + {\bf P}\cdot {\bf Y})/\hbar} \nonumber \\
&\times &\Big\langle {\bf R}+{{\bf Y}\over2} , {\bf r}+{{\bf y}\over2} |\rho^{\Phi_i}|{\bf R}-{{\bf Y}\over2}, {\bf r}-{{\bf y}\over2} \Big \rangle \nonumber \\ &=& W^{\Phi_i}({\bf r},{\bf p})\, \delta({\bf P}-{\bf P}_0), 
 \label{eq.wignerdf}
\end{eqnarray} 
assuming that the center of mass motion of the bound state $i$ is described by a plane wave with momentum ${\bf P}_0$.
The Wigner density $W^{\Phi_i} $ of the quarkonium state $\Phi_i$  can be constructed via,
\begin{eqnarray}
W^{\Phi_i}({\bf r},{\bf p})=\int d^3ye^{-i{\bf p}\cdot {\bf y}} \psi_i({\bf r}+{{\bf y}\over2 })\psi_i^*({\bf r}-{{\bf y}\over2 }). 
\label{eq.wignerdf}
\end{eqnarray} 
Here, $\psi_i$ is the wave function of quarkonium state $i$, which can be obtained by solving the Schr\"odinger equation for the relative coordinates of the $Q\bar Q$ pair.  
The  momentum distribution of the center of mass motion of the quarkonium $i$, created  in a system, which is in the quantum state $\rho^{Q\bar Q}$,  is given by 
\begin{eqnarray}
{dP_i\over d^3 {P}}&=&g\int {d^3R\, d^3r\, d^3p \over (2\pi \hbar)^6} W^{\Phi_i}({\bf r},{\bf p})
W^{Q\bar Q}({\bf R},{\bf P},{\bf r},{\bf p})\nonumber \\
&\times &\delta({\bf P}-{\bf P}_0).
\label{eq.projection1}
\end{eqnarray}
 
In our semiclassical approach we cannot calculate the quantal two body Wigner density of a $Q\bar Q$ pair in the medium $W^{Q\bar Q}({\bf R},{\bf P},{\bf r},{\bf p})$ directly. Like in all transport approaches, we average the classical phase space density $W^{Q\bar Q}_{\rm class}({\bf R},{\bf P},{\bf r},{\bf p})=dN^{Q\bar Q}/(d^3Rd^3Pd^3pd^3r)$, which is given by EPOS4,  over many events. 

The projection on the discrete quantum numbers like spin and color are encoded in a degeneracy factor $g$. $W^{Q\bar Q}_{\rm class}$ contains all possibilities in spin $\otimes$ color space with the same probability. So, the degeneracy factor for a quarkonia with the total angular momentum $J$ is $g=(2J+1)/(9\times 4)$, assuming color and spin are randomly distributed.

The extension to a system, which contains $N/2$ heavy quarks and $N/2$ heavy antiquarks 
(and hence $N^2/4$ pairs of $Q \bar{Q}$), mostly $c\bar{c}$ pairs, is straightforward. Summing over all $Q\bar Q$ pairs $ik$ we obtain

\begin{eqnarray}
\frac{dP_{nl}}{d^{3}P} & = & g\int d^{3}R\,d^{3}r\,d^{3}p\,W_{nl}^{\Phi}({\bf r},{\bf p})\nonumber \\
 & \times & \sum_{N}\int\prod_{j=1}^{N}\frac{d^{3}r_{j}d^{3}p_{j}}{(2\pi\hbar)^{3}}W_{{\rm class}}^{Q\bar{Q}}({\bf r}_{1},{\bf p}_{1},...,{\bf r}_{N},{\bf p}_{N})\nonumber \\
 & \times & \sum_{{\rm pairs\,}i,k}\delta(\frac{{\bf p}_{i}-{\bf p}_{k}}{2}-{\bf p})\delta({\bf r}_{i}-{\bf r}_{k}-{\bf r})\nonumber \\
 & \times & \delta(\frac{{\bf r}_{i}+{\bf r}_{k}}{2}-{\bf R})\delta({\bf p}_{i}+{\bf p}_{k}-{\bf P})\delta({\bf P}-{\bf P}_{0}),\label{eq.projectionNN}
\end{eqnarray}
where ${\bf R},{\bf P},{\bf r},{\bf p}$ are the center of mass and relative coordinates of the considered pair.
We have introduced here the quantum numbers $n$ and $l$ of the state $i$. 
Both Eq.~\eqref{eq.projection1} and Eq.~\eqref{eq.projectionNN} are non-relativistic. For the relativistic version, which we apply in the calculations, we refer to the equation C9 in the Appendix C of Ref.~\cite{Villar:2022sbv}, where the details are given. In the relativistic version one has to Lorentz transform the momenta and positions of the heavy quarks, given by EPOS4 in the computational frame, into the center of mass system of the $Q\bar Q$ pair and determines there the probability $W_{nl}^\Phi({\bf r}_{cm}, {\bf p}_{cm})$ that this pair is in a state $\Phi_i$.~\footnote{strictly speaking, in Eq.~C8 of Ref.~\cite{Villar:2022sbv}, the quarkonium is assumed to have the same 4-velocity as the $c\bar{c}$ pair while the invariant mass was assumed to be modified, while here, momentum conservation is assumed, and energy conservation is violated. This difference should not modify the result in a noticeable way.}. In the computational frame the momentum distribution is then given by
~\cite{Villar:2022sbv}, 
\begin{eqnarray}
{dP_{nl}\over d^3 P}=\Big\langle g\sum_{Q\bar Q\ {\rm pairs}} W_{nl}^\Phi({\bf r}_{cm}, {\bf p}_{cm}) \delta^{(3)}({\bf P}-{\bf p}_1-{\bf p}_2)\Big\rangle,\nonumber\\
\label{eq.projectionmc2}
\end{eqnarray}
where the summation runs over all $N^2/4$ possible $Q\bar Q$ pairs and where the ${\bf r}_{cm}$ and ${\bf p}_{cm}$ are relative distance and momentum in the center-of-mass frame of each $Q\bar Q$ pair. $\langle \cdots \rangle$ indicates the averaging over all Monte Carlo events generated by EPOS4.

 %---------------------------------------------------------------------
\begin{table}
	\renewcommand\arraystretch{1.3}
	\setlength{\tabcolsep}{1.5mm}
	\begin{tabular}{c|c|c|c|c}
		\toprule[1pt]\toprule[1pt]
        \multicolumn{1}{c|}{States}&
		\multicolumn{1}{c|}{$M_{\rm Theo.}$(GeV)}& \multicolumn{1}{c|}{$M_{\rm Exp.}$(GeV)} &   \multicolumn{1}{c|}{$\langle r^2 \rangle(\rm fm^2)$} &   \multicolumn{1}{c}{$\sigma (\rm fm)$} \tabularnewline
		\midrule[1pt]
		$J/\psi$ & 3.071 & 3.097 & 0.182 & 0.348 \tabularnewline
        $\chi_c(1P)$ & 3.483 & 3.463 & 0.453 & 0.426 
        \tabularnewline
        $\psi(2S)$ & 3.652 & 3.686 & 0.714 & 0.452
        \tabularnewline
        \midrule[1pt]
        $\Upsilon(1S)$ & 9.390 & 9.460 & 0.042 & 0.167
        \tabularnewline
        $\chi_b(1P)$ & 9.870 & 9.876 & 0.153 & 0.247
        \tabularnewline
        $\chi_b(1D)$ & 10.109 & 10.163 & 0.284 & 0.285
        \tabularnewline
        $\Upsilon(2S)$ & 9.959 & 10.023 & 0.236 & 0.260
        \tabularnewline
        $\chi_b(2P)$ & 10.208 & 10.243 & 0.410 & 0.302
        \tabularnewline
        $\Upsilon(3S)$ & 10.288 & 10.355 & 0.520 & 0.307
        \tabularnewline
        \midrule[1pt]
        $B_c(1S)$ & 6.482 & 6.275 & 0.115 & 0.277
        \tabularnewline
        $B_c(1P)$ & 6.895 & - & 0.316 & 0.356
        \tabularnewline
        $B_c(1D)$ & 7.156 & - & 0.542 & 0.393
        \tabularnewline
        $B_c(2S)$ & 7.033 & 6.872 & 0.497 & 0.377
        \tabularnewline
		\bottomrule[1pt]
	\end{tabular}
	\caption{The masses, root-mean-square radius, and the Gaussian widths $\sigma$ of different charmonium and bottomonium states in vacuum. The experimental data for the masses are from Ref.~\cite{Workman:2022ynf}.}
	\label{table1}
\end{table}
%--------------------------------------------------------------------

%---------------------------------------------------------------------
\begin{figure}[!htb]
\includegraphics[width=0.23\textwidth]{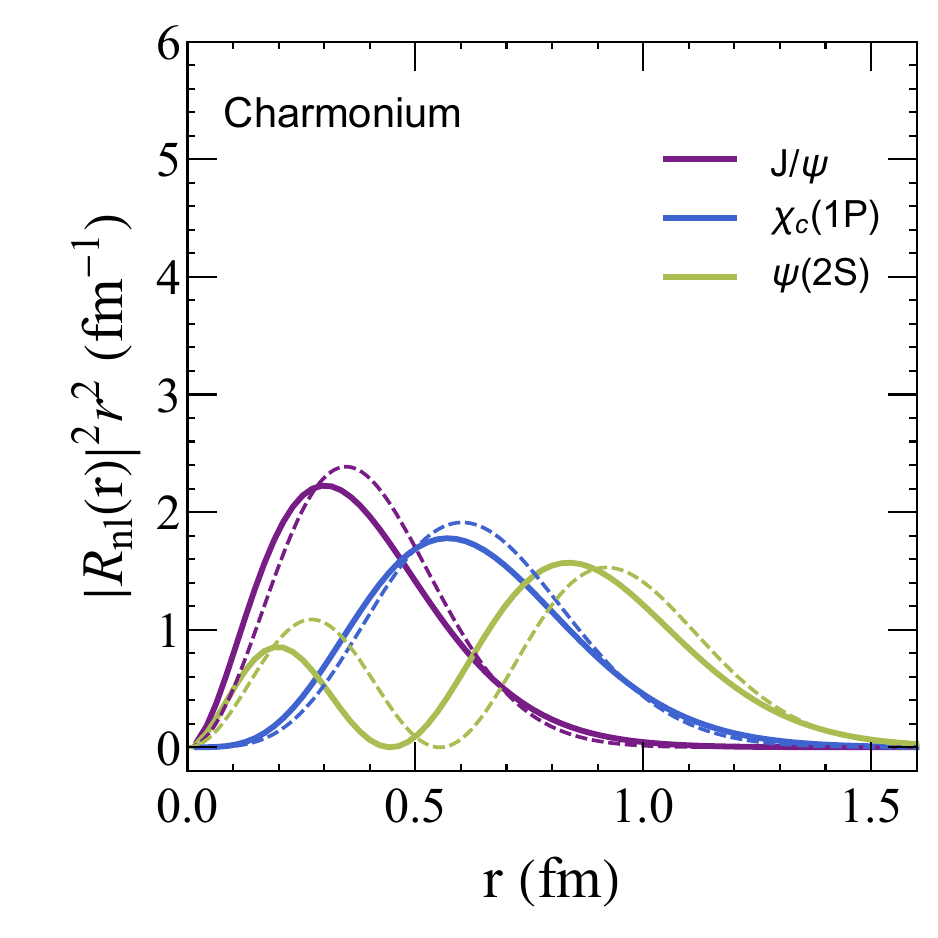}
\includegraphics[width=0.23\textwidth]{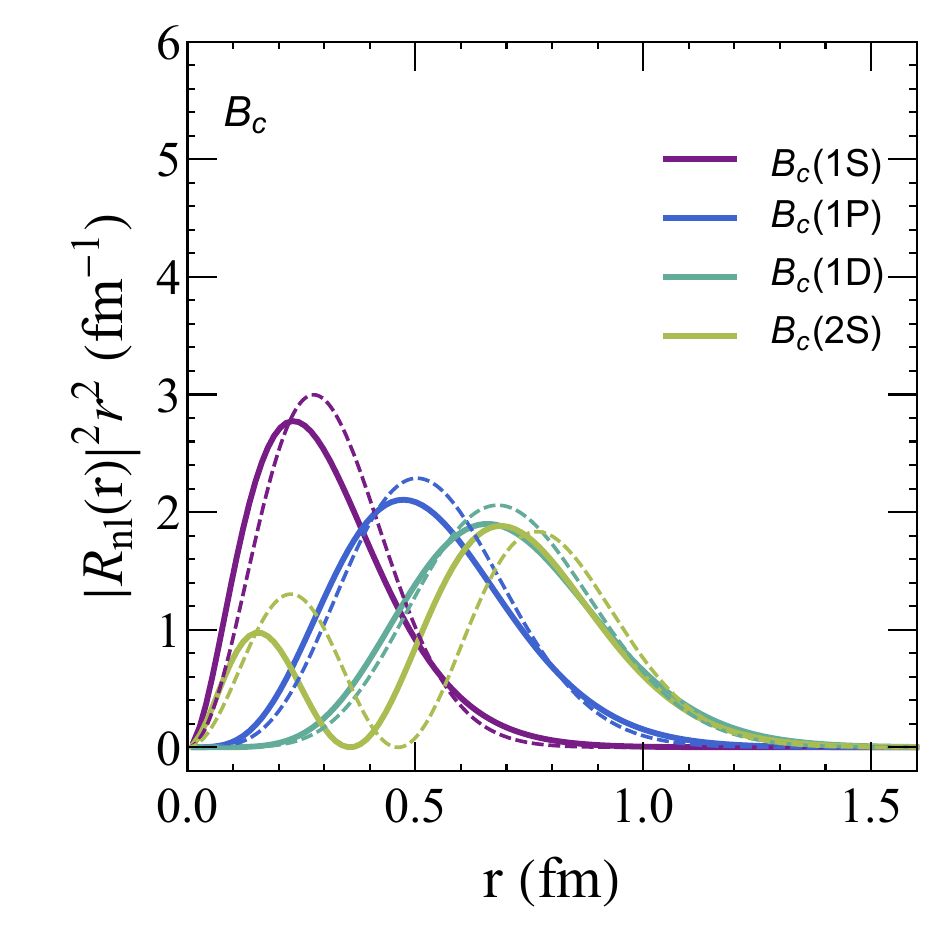}\\
\includegraphics[width=0.23\textwidth]{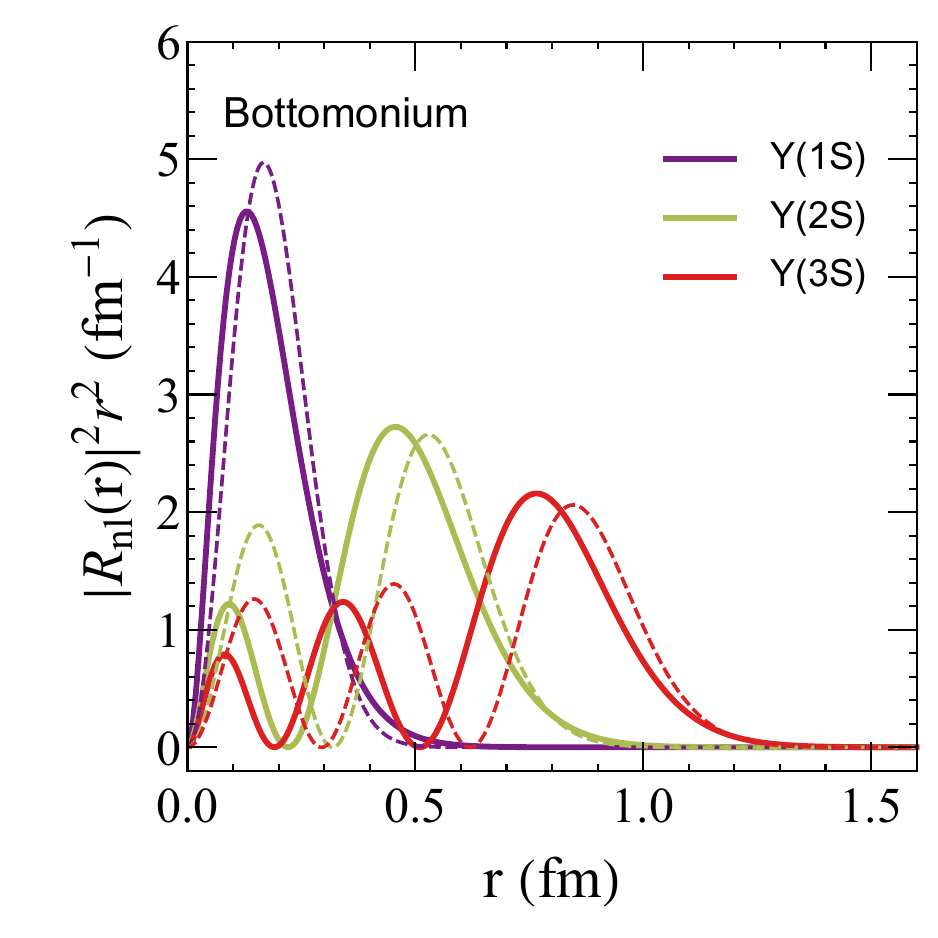}\includegraphics[width=0.23\textwidth]{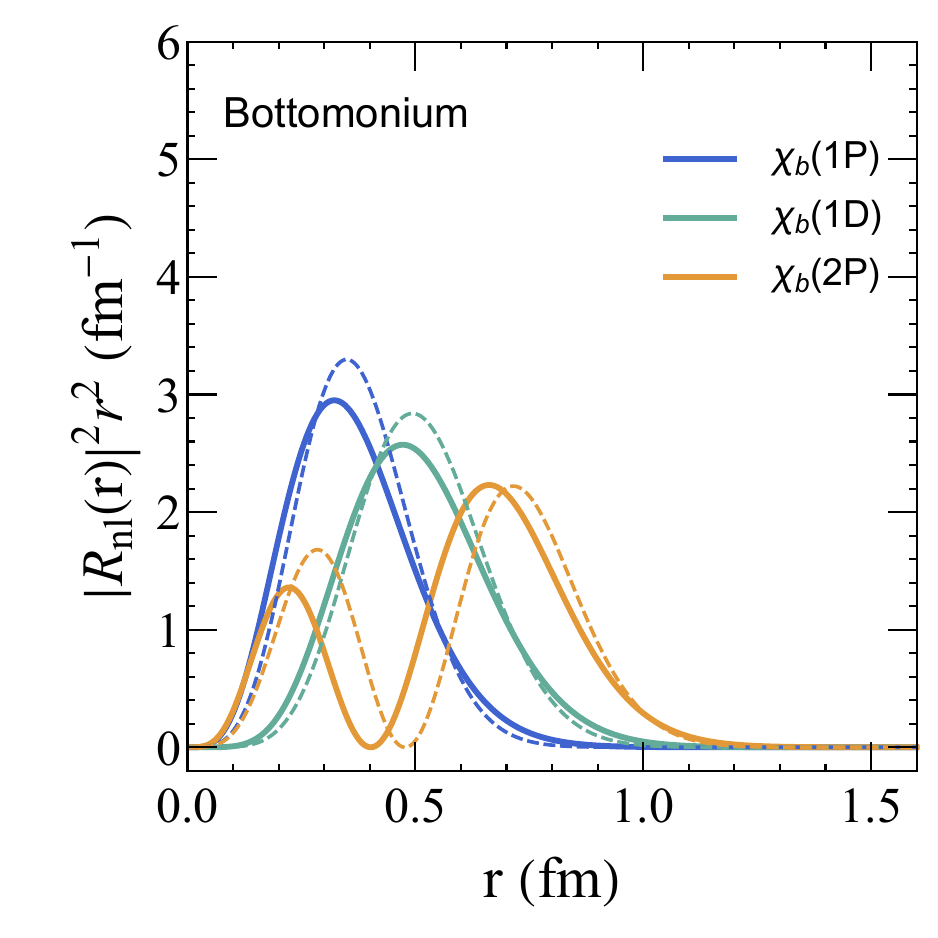}
\caption{Wavefunction of different charmonium (upper), $B_c$ (middle), and bottomonium (lower) states. Solid lines are from the Schrödinger equation, dashed lines are from the 3-D isotropic harmonic oscillator and given by Eq.~\eqref{eq.3dwf}.} 
\label{fig.wf}
\end{figure} 
%---------------------------------------------------------------------
We come now to the construction of  the quarkonium Wigner function $W_{nl}^\Phi$. The Wigner function is obtained from the quarkonium wave function via a Wigner transformation, and the wave function is the solution of the two-body Schr\"odinger equation. The interaction potential can be taken as Cornell potential in vacuum, $V(r)=-\alpha/|{\bf r}|+\sigma |{\bf r}|$ with $\alpha=0.513$ and $\sigma=0.17~ \rm GeV^2$, and the quark mass $m_c=1.5~ \rm GeV$, $m_b=5.2~\rm GeV$. Because the potential depends only on the relative distance of two quarks, we can separate the motion of the two-body system into a center-of-mass motion, which is just the motion of a free particle, and the relative motion. Furthermore, the Cornell potential is an isotropic potential. One can write the relative wave function $\psi({\bf r})$ into a radial part and an angular part, $\psi({\bf r})= R_{nl}(r)Y_{l,m}(\theta, \phi)$. Now the description of the two-body system is simplified to a one dimensional problem,
\begin{eqnarray}
&&\left[-{1\over 2\mu}\left({d^2\over dr^2}+{2\over 2}{d\over dr}\right)+{l(l+1)\over2\mu r^2}+V(r)\right]R_{nl}(r)\nonumber\\
&&=ER_{nl}(r),
\end{eqnarray}
where $\mu=m_im_j/(m_i+m_j)$ is the reduced mass. $(n,l)$ labels the main and orbital angular momentum number of quarkonium-states.
We solve the two-body Schr\"odinger equation for charmonium, bottomonium, and $B_c$. The wave functions are shown in Fig.~\ref{fig.wf}, and the masses and root-mean-square (rms) radii $\langle r^2\rangle$ are shown in Table~\ref{table1}. We can see that the masses are very close to the experimental values. 
However, the wave function is not analytical, and therefore the calculation of the Wigner density and of Eq.~\eqref{eq.wignerdf}  can only be performed numerically.
To avoid this we assume that the potential between the heavy quark pairs can be approximated by a 3-D isotropic harmonic oscillator (HO) potential, $V(r)=1/(2m_Q\sigma^4)r^2$. Then the wave functions are analytical. The wave  functions of a 3-D isotropic harmonic oscillator can be expressed as $\psi_{nlm}(r,\theta,\phi)=R_{nl}(r)Y_{l,m}(\theta,\phi)$, where $Y_{l,m}$ are the spherical harmonics. 
The radial part can be expressed as
\begin{eqnarray}
&&R_{nl}(r)\nonumber\\
&&=\left [{2 (n!) \over \sigma^3 \Gamma(n+l+3/2)} \right]^{1\over 2}\left({r\over \sigma} \right)^l e^{-{r^2\over 2\sigma^2}}L_n^{l+1/2}\left({r^2\over \sigma^2}\right),
\label{eq.3dwf}
\end{eqnarray}
where $L_n^{l+1/2}$ are the Laguerre polynomials. The width $\sigma $ of the harmonic oscillator potential is chosen that the numerically determined rms radius of the eigenfunction of the Cornell potential, tabulated in Table~\ref{table1} agrees with that of the harmonic oscillator potential.
The rms radii of the different harmonic oscillator wave functions are related to the width of these states by
$\langle r^2\rangle=3\sigma^2/2$ for $1S$, $\langle r^2\rangle=5\sigma^2/2$ for $1P$, $\langle r^2\rangle=7\sigma^2/2$ for $1D$ and $2S$,  $\langle r^2\rangle=9\sigma^2/2$ for $2P$, and $\langle r^2\rangle=11\sigma^2/2$ for $3S$ state. This procedure allows to connect the 3-D isotropic harmonic oscillator wave function with the real quarkonium wave function via the rms radii $\langle r^2\rangle$. The related widths are shown in Table~\ref{table1} and the wave functions are shown in Fig.~\ref{fig.wf} as dashed lines. We can see that the ground state and the low excited states can be well reproduced by the 3-D isotropic harmonic oscillator, while the difference for higher-order excited states, e.g. $2S$, $2P$, and $3S$ are larger.

%---------------------------------------------------------------------
With this analytical wave function, the Wigner function can be constructed via a Wigner transformation in spherical coordinates. The Wigner function for any given $(n,l)$ state, including the summation on all harmonics $m$, can be expressed as
\begin{eqnarray}
W^{\Phi}_{nl}({\bf r,p})&=&{1\over 2l+1}{(-1)^l\over 2\pi^3}\sum_{n'+N+l'=K}{(-1)^{n'+3l'/2}}\nonumber\\
&\times&\sqrt{\pi(2l+1)(2l'+1)n'!N!\over \Gamma(n'+l'+3/2)\Gamma(N+l'+3/2)}\nonumber\\
&\times&(n'l'Nl'0|nlnl0)L_N^{l'+1/2}(2{\bf r}^2)L_N^{l'+1/2}(2{\bf p}^2)\nonumber\\
&\times&(2\|{\bf r}\|\|{\bf p}\|)^{l'}P_{l'}(\cos\theta)e^{-({\bf r}^2+{\bf p}^2)},
\end{eqnarray}
where $K=2n+l$. The summation is from zero to $K$ and for all possible quantum numbers $n'$, $l'$, and $N$. The only restriction on these quantum numbers is that the sum is $K$. 
$\theta$ is the angle between $\bf r$ and $\bf p$, $L_n^l$ are the generalized Laguerre polynomials, $P_{l'}$ are the Legendre polynomials, and $(n'l'Nl'0|nlnl0)$ is the Talmi-Brody-Moshinsky (TBM) bracket. This expression has firstly been given in Ref.~\cite{Shlomo:1981ayz} to which we refer for the details.  

For the states up to 3S the Wigner densities have the concrete form
\begin{widetext}
\begin{eqnarray}
W^{\Phi}_{\rm 1S}({\bf r,p})&=&8e^{-\xi} , \\
W^{\Phi}_{\rm 1P}({\bf r,p})&=& {8\over 3}e^{-\xi }\Big(2\xi -3 \Big),\nonumber\\
W^{\Phi}_{\rm 1D}({\bf r,p})&=& {8\over 15}e^{-\xi}\Big(15+4\xi^2-20\xi +8[p^2r^2-({\bf p}\cdot{\bf r})^2] \Big),\nonumber\\
W^{\Phi}_{\rm 2S}({\bf r,p})&=& {8\over 3}e^{-\xi}\Big(3+2\xi^2-4\xi-8[p^2r^2-({\bf p}\cdot{\bf r})^2)] \Big),\nonumber\\
W^{\Phi}_{\rm 2P}({\bf r,p})&=& {8\over 15}e^{-\xi} \Big(-15+4\xi^3-22\xi^2+30\xi-8(2\xi-7)[p^2r^2-({\bf p}\cdot{\bf r})^2] \Big),\nonumber\\
W^{\Phi}_{\rm 3S}({\bf r,p})&=& {8\over 315}e^{-\xi} \Big(315+42\xi^4-336\xi^3+924\xi^2-840\xi \nonumber\\
&-&[2009+32p^2r^2+336r^4/\sigma^4-1400r^2/\sigma^2-896p^2\sigma^2+224p^4\sigma^4][p^2r^2-({\bf p}\cdot{\bf r})^2] \nonumber\\
&-&[686+608p^2r^2+112r^2/\sigma^2-896p^2\sigma^2+224p^4\sigma^4-672({\bf p}\cdot{\bf r})^2]({\bf p}\cdot{\bf r})^2\Big),\nonumber
\label{wigdens2} 
\end{eqnarray}
\end{widetext}
where $\xi={r^2\over \sigma^2}+p^2 \sigma^2$.
We can see that the Wigner densities of the excited states depend not only on  $r=|{\bf r}|$ and $p=|\bf p|$, but also on the angle between $\bf p$ and $\bf r$.

 %---------------------------------------------------------------------
\begin{table}
	\renewcommand\arraystretch{1.5}
	\setlength{\tabcolsep}{1.mm}
	\begin{tabular}{c|c|c|c}
		\toprule[1pt]\toprule[1pt]
        \multicolumn{1}{c|}{Fraction}&
	\multicolumn{1}{c|}{\rm Flavor creation}& \multicolumn{1}{c|}{\rm Flavor excitation} &    \multicolumn{1}{c}{\rm Gluon splitting} 
        \tabularnewline
		\midrule[1pt]
	$D^0$ & 6.12\% & 80.63\% & 13.25\%  
        \tabularnewline
        $J/\psi$ & 2.62\% & 76.02\% & 21.36\% 
        \tabularnewline
             \midrule[1pt]
	$B^0$ & 26.96\% & 65.72\% & 7.32\%  
        \tabularnewline
        $\Upsilon(1S)$ & 11.56\% & 73.20\% & 15.24\% 
        \tabularnewline
		\bottomrule[1pt]
	\end{tabular}
	\caption{The fractions of $D^0$ ($B^0$) and prompt $J/\psi$ ($\Upsilon(1S)$) from different processes.}
	\label{table2}
\end{table}
%--------------------------------------------------------------------
In EPOS4 the $Q$ and $\bar Q$  are produced at the same position in coordinate space with the momenta ${\bf p}_1$ and  ${\bf p}_2$.  The Wigner density $W_{nl}^\Phi({\bf r,p})$ can be locally negative but all differential rates should be positive if  the integration is performed over an elementary phase space cell $h^3$. To avoid negative probabilities and restore the compatibility with Heisenberg principle  we sample the relative distance ${\bf r}_{cm}$ between $Q$ and $\bar Q$ in their center-of-mass frame by a Gaussian distribution
around the EPOS4 production point of the pair,
\begin{eqnarray}
f({\bf r}_{cm})\sim r_{cm}^2\exp \left(-{r_{cm}^2 \over 2\sigma_{\rm Q\bar Q}^2} \right),
\end{eqnarray} 
where the relative distance is controlled by the effective width $\sigma_{\rm Q\bar Q}$, which is 
$\sigma_{c\bar c}=0.35~\rm fm$ for all charmonium states (as this parameter  enters only the $W^{c\bar{c}}$ distribution)  and $\sigma_{b\bar b}=0.2~\rm fm$ for all bottomonium states.

We can now calculate Eq.~\eqref{eq.projectionmc2} for all charmonium and bottomonium states as well for $B_c$ by averaging over many EPOS4HQ events. Because in this microscopic approach we follow all heavy quarks individually (and not only their distribution function) we can trace back the origin of all heavy quarks which are finally entrained in quarkonia states. Table~\ref{table2} presents the percentage distribution of the creation processes for the final $D^0, J/\psi, B^0$ and $Y(1S)$ mesons.  We observe that the gluon splitting contribution is, as expected in a Wigner projection approach, more important for the quarkonia than for the open heavy flavour mesons, whereas for flavor creation the opposite is the case, due to the unfavorable kinematics.  If we introduce a lower $p_T$ cut the influence of the flavor excitation process becomes smaller.

\subsection{Quarkonia observables}

\begin{figure}[!htb]
\includegraphics[width=0.23\textwidth]{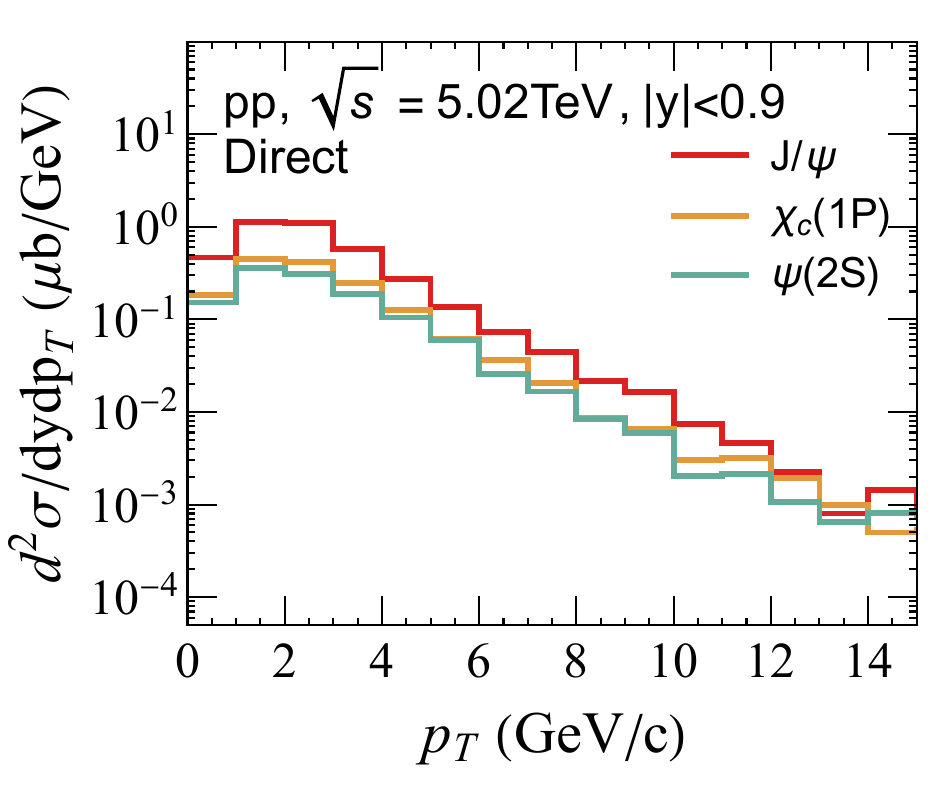}
\includegraphics[width=0.23\textwidth]{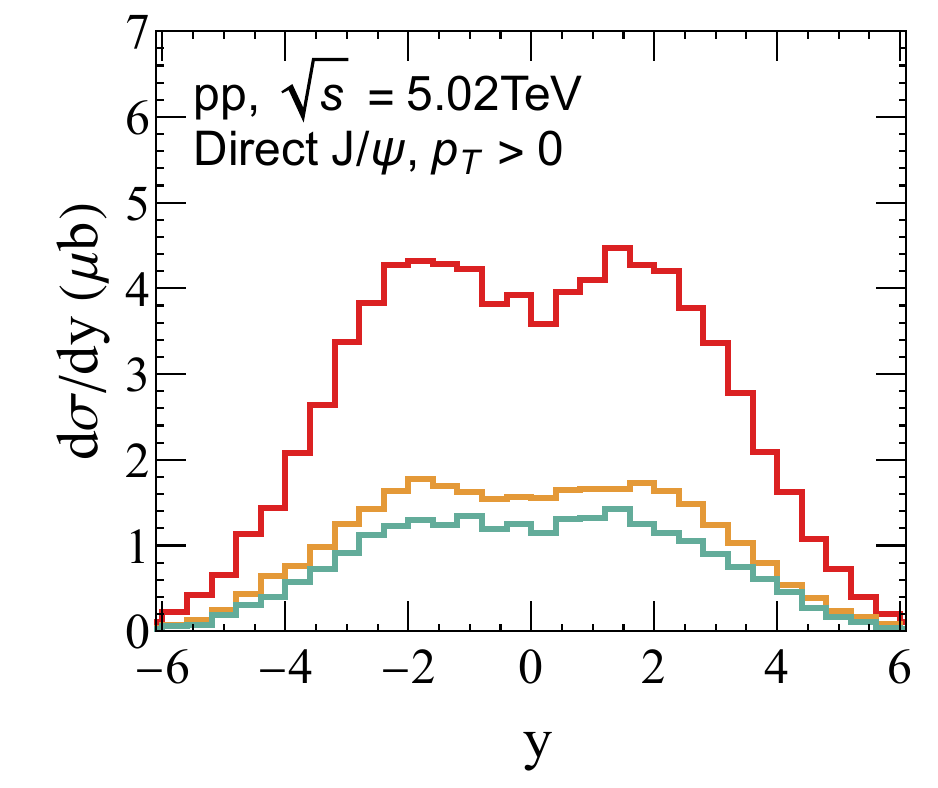}\\
\includegraphics[width=0.23\textwidth]{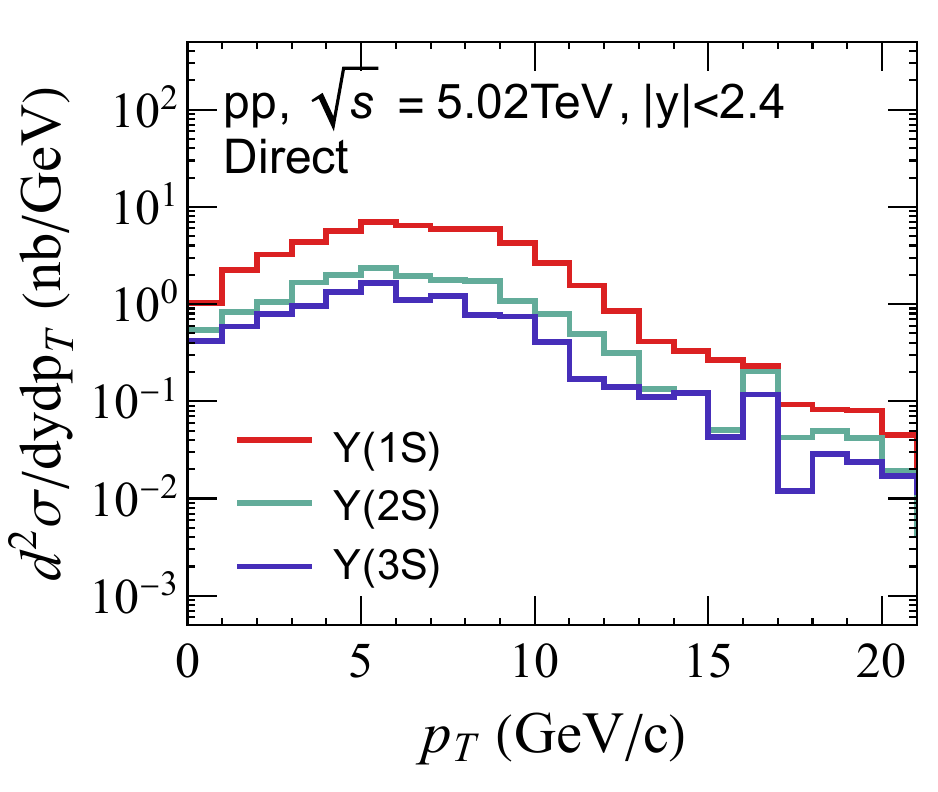}
\includegraphics[width=0.23\textwidth]{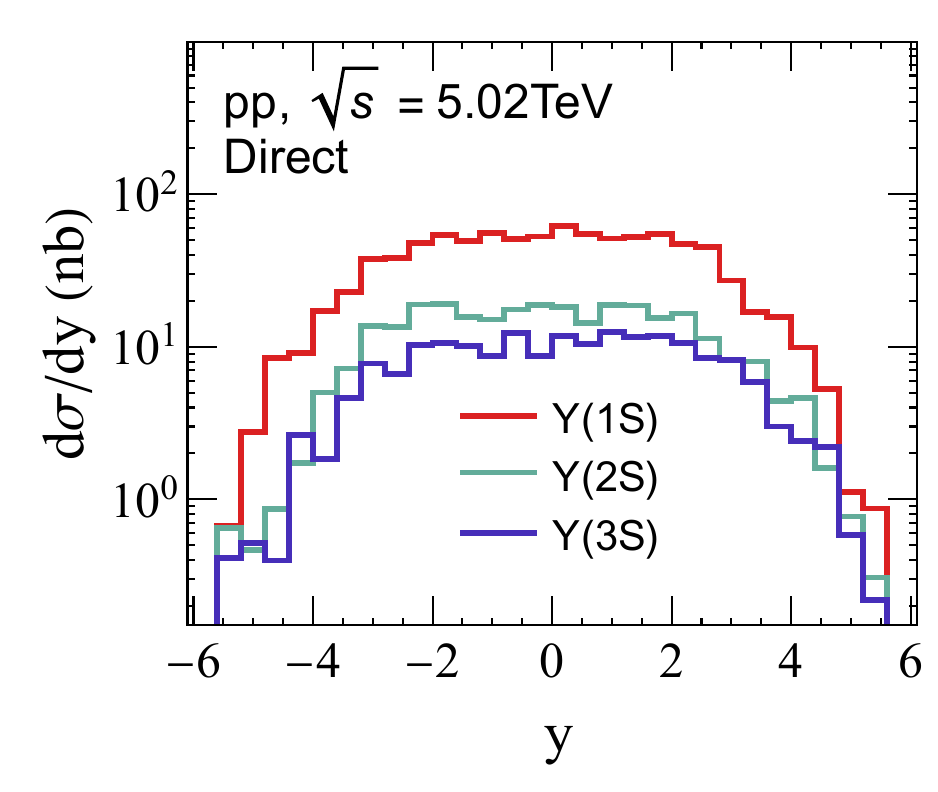}\\
\includegraphics[width=0.23\textwidth]{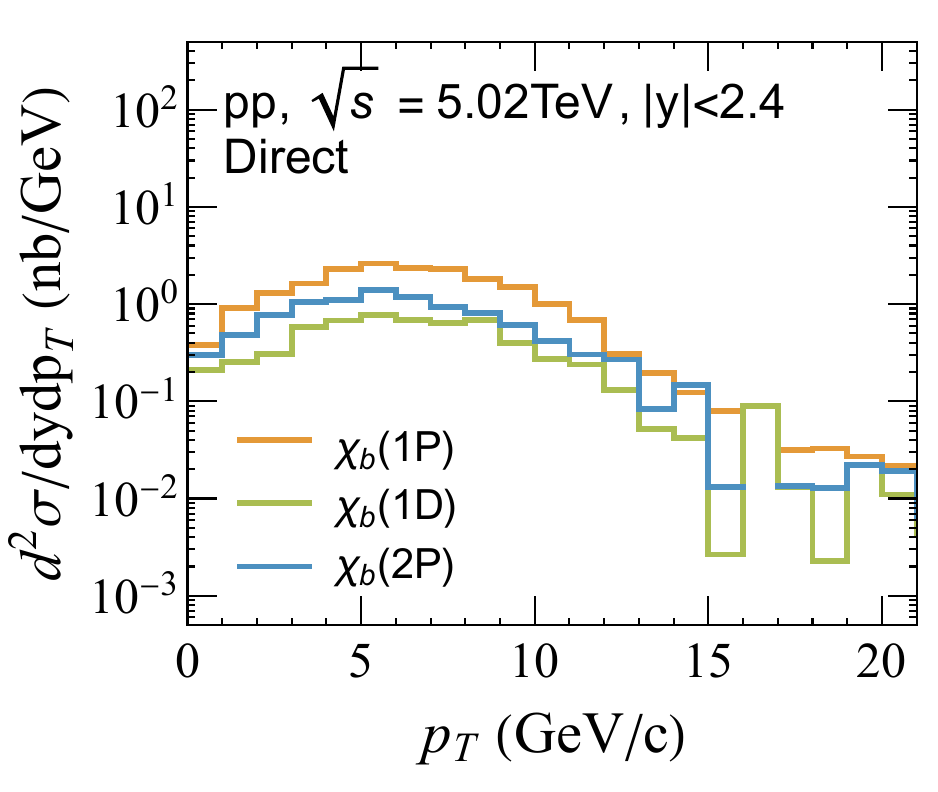}
\includegraphics[width=0.23\textwidth]{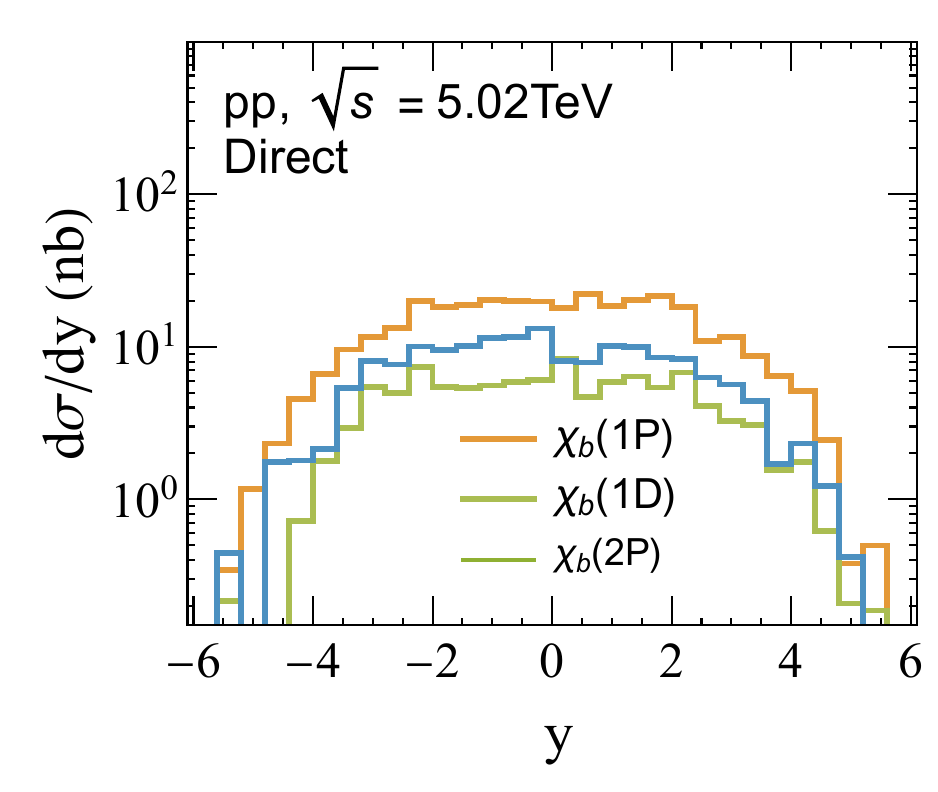}\\
\includegraphics[width=0.23\textwidth]{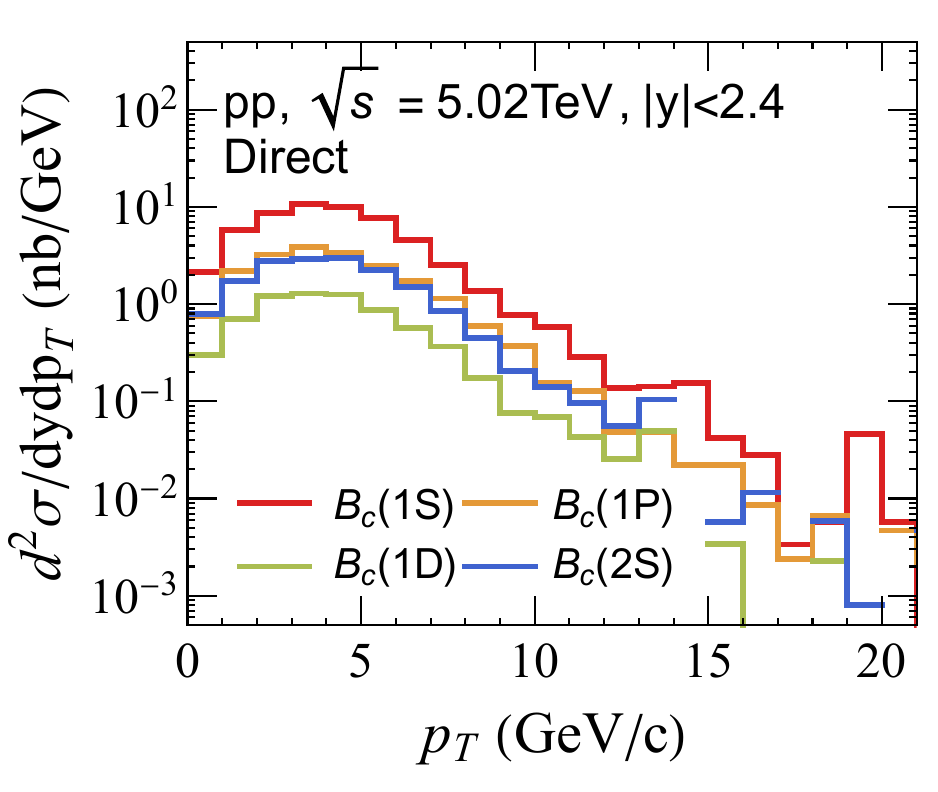}
\includegraphics[width=0.23\textwidth]{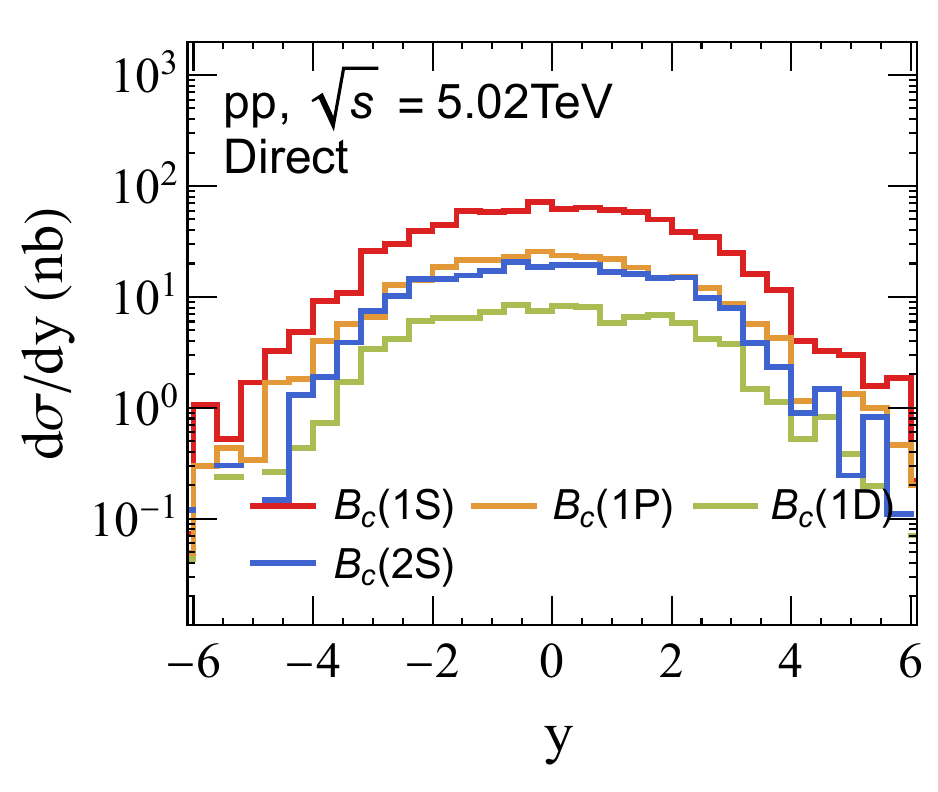}
\caption{$p_T$ spectra at midrapidity and rapidity distribution of different directly produced charmonium (upper), bottomonium states (middle for $S$-wave and $P$($D$)-wave), and $B_c$ states (lower) in pp collisions at $\sqrt{s}=5.02~ \rm TeV$. The color coding of the different states is noted in the figure.}
\label{fig.direct}
\end{figure}

In  Fig.~\ref{fig.direct}, we display the $p_T$ distribution at midrapidity (left) as well as the $p_T$ integrated rapidity distribution (right) of the different quarkonia. The top figures show the distributions for the charmonia, followed by that for bottomonia, for $\chi_b$ and  for $B_c$. The slopes of the $p_T$ spectra are rather different for charmonia, bottomonia and $B_c$, the rapidity distributions show a small double hump structure for charmonia and a rather flat distribution around midrapidity for bottomonia and $B_c$.
%---------------------------------------------------------------------
%-------------------------------------------------------------------
\begin{figure}[!htb]
\includegraphics[width=0.23\textwidth]{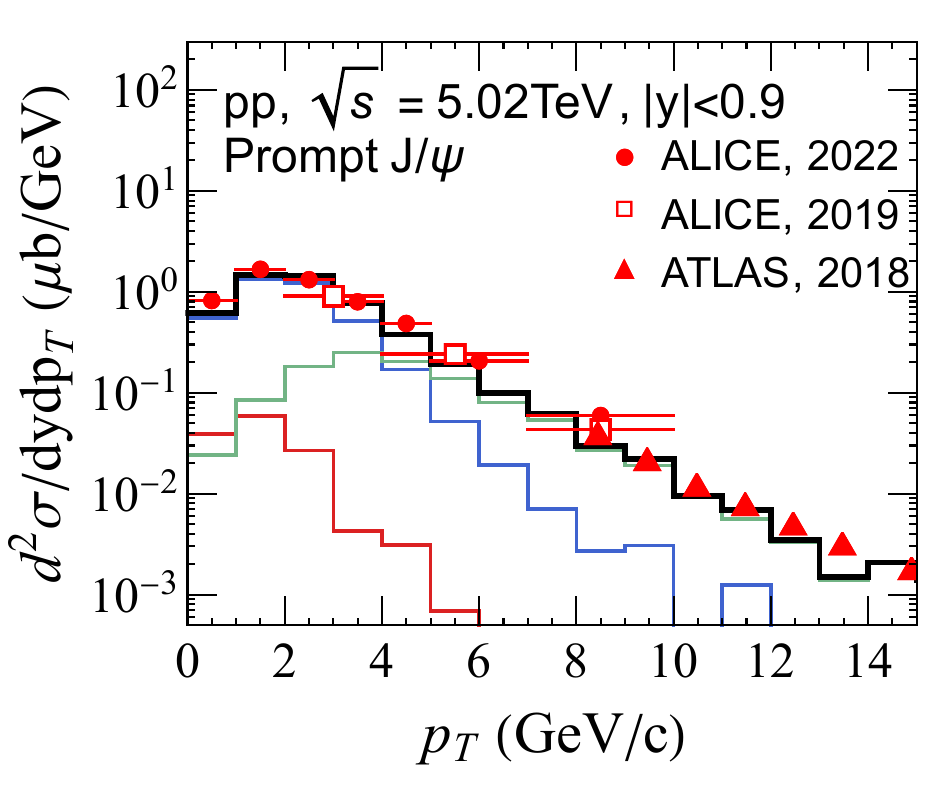}
\includegraphics[width=0.23\textwidth]{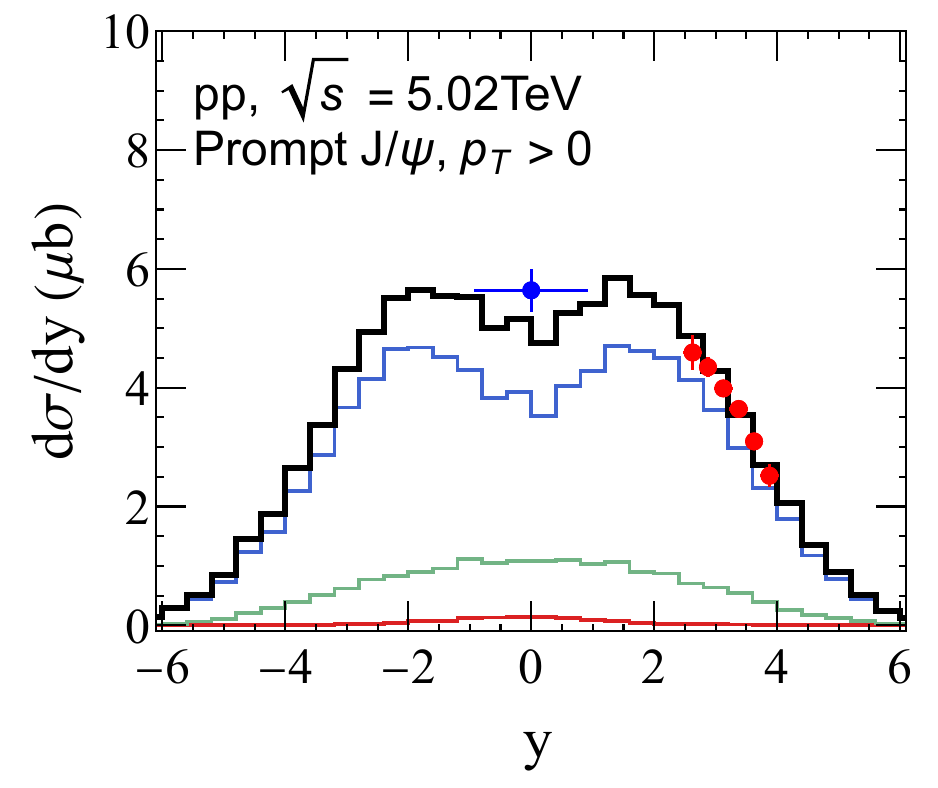}\\
\includegraphics[width=0.23\textwidth]{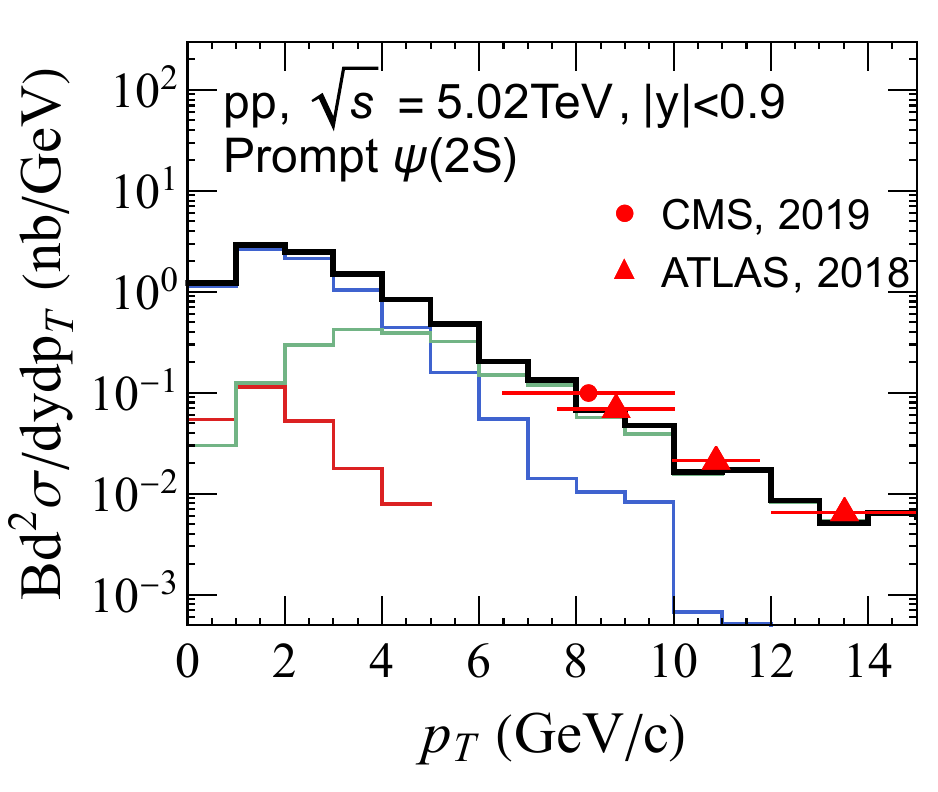}
\includegraphics[width=0.23\textwidth]{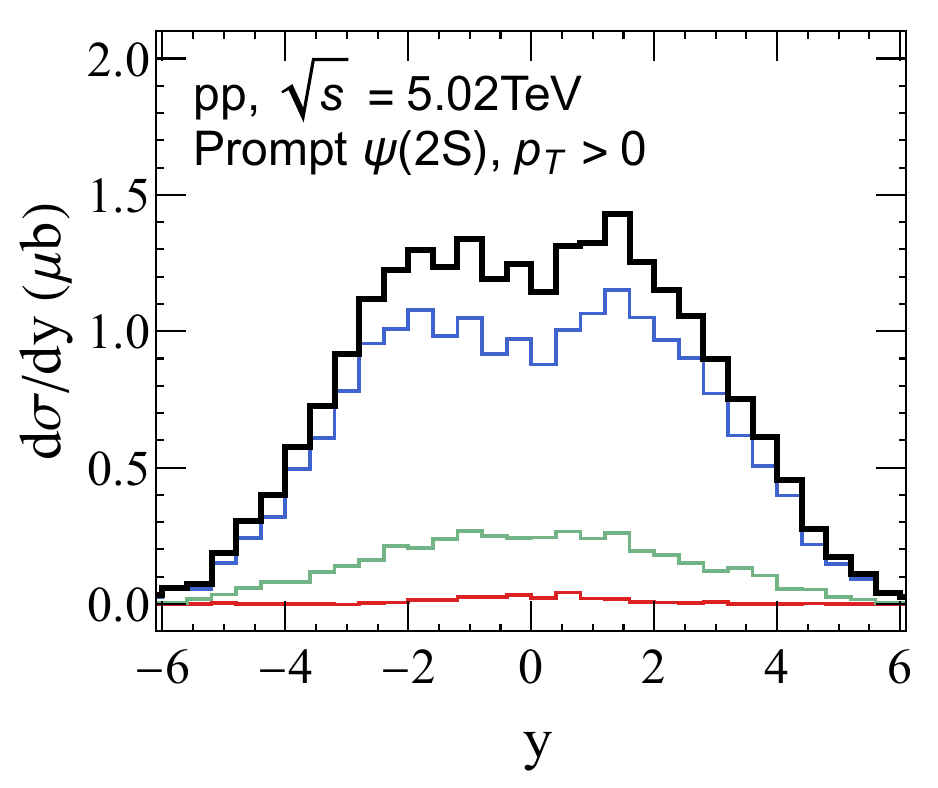}
\caption{$p_T$ spectra and rapidity distribution of prompt $J/\psi$ (upper) and $\psi(2S)$ (lower)  (thick black lines)  in pp collisions at $\sqrt{s}=5.02~ \rm TeV$. Red is the contribution from flavor creation, blue that from flavor excitation, and green from gluon splitting. The experimental data are from ALICE~\cite{ALICE:2021edd,ALICE:2019pid}, ATLAS~\cite{ATLAS:2017prf}, and CMS~\cite{CMS:2018gbb}. The branching ratio of $B(\psi(2S)\to \mu^+\mu^-)$ is taken as 0.8\%, as given by the PDG~\cite{Workman:2022ynf}.}
\label{fig.prompt.cc}
\end{figure}
%---------------------------------------------------------------------
%---------------------------------------------------------------------
\begin{figure}[!htb]
\includegraphics[width=0.23\textwidth]{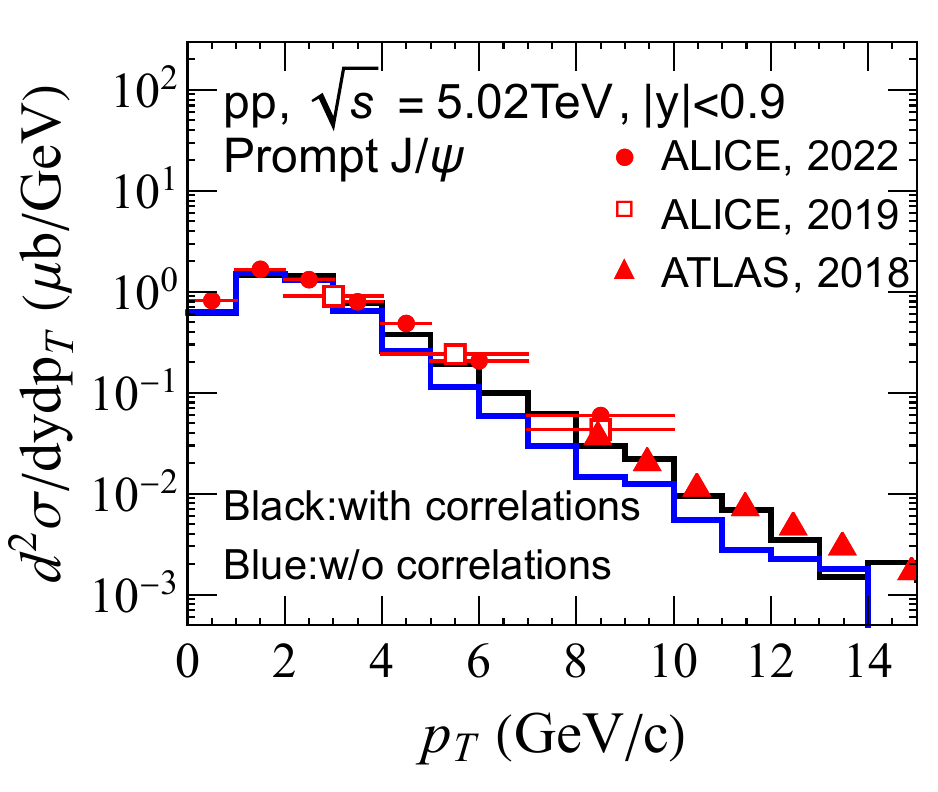}
\includegraphics[width=0.23\textwidth]{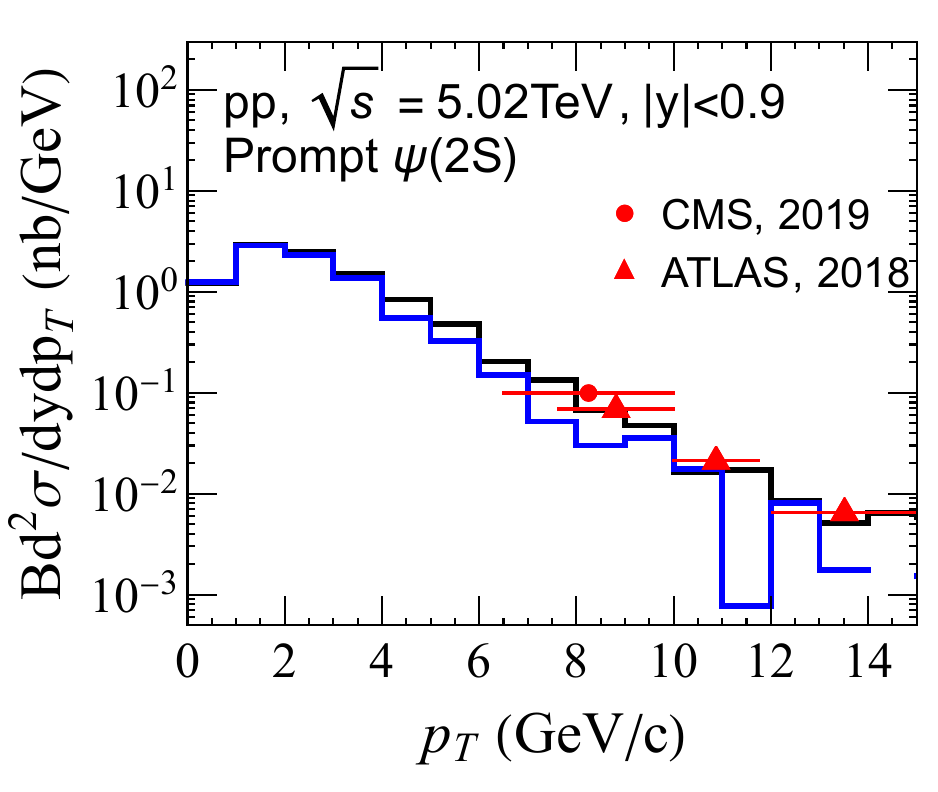}
\caption{$p_T$ spectra of prompt $J/\psi$ and $\psi(2S)$. The blue lines are obtained with the same framework and parameters but assuming the correlations between $c$ and $\bar c$ are random. }
\label{fig.wo.correlations}
\end{figure}
%---------------------------------------------------------------------
%-------------------------------------------------------------------
\begin{figure}[!htb]
\includegraphics[width=0.23\textwidth]{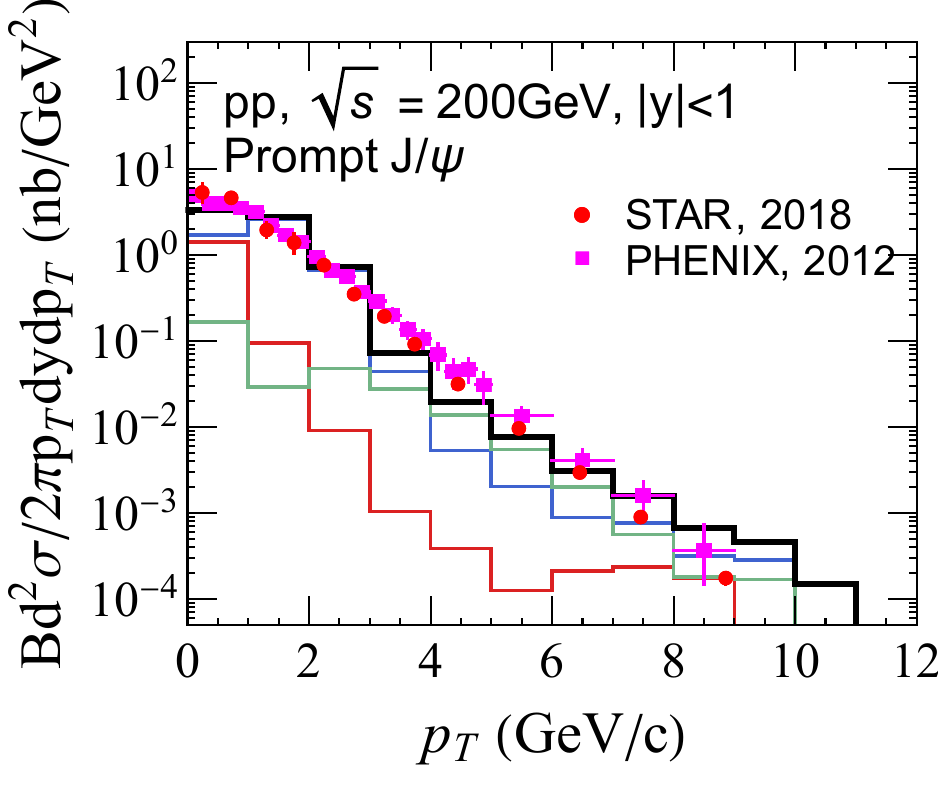}
\includegraphics[width=0.23\textwidth]{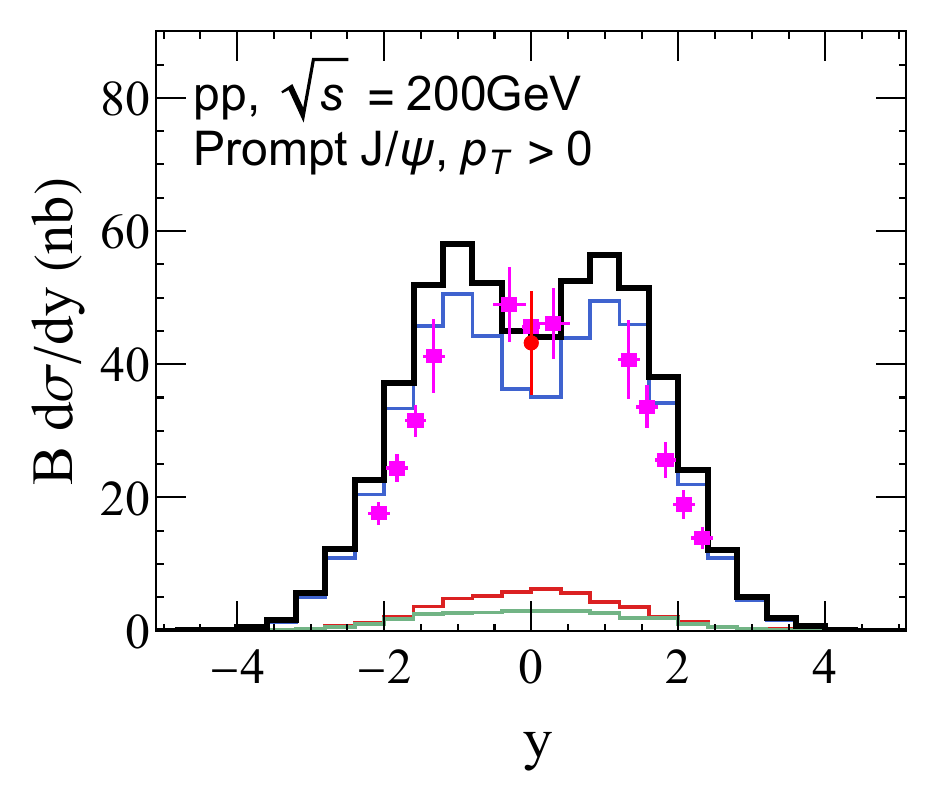}\\
\includegraphics[width=0.23\textwidth]{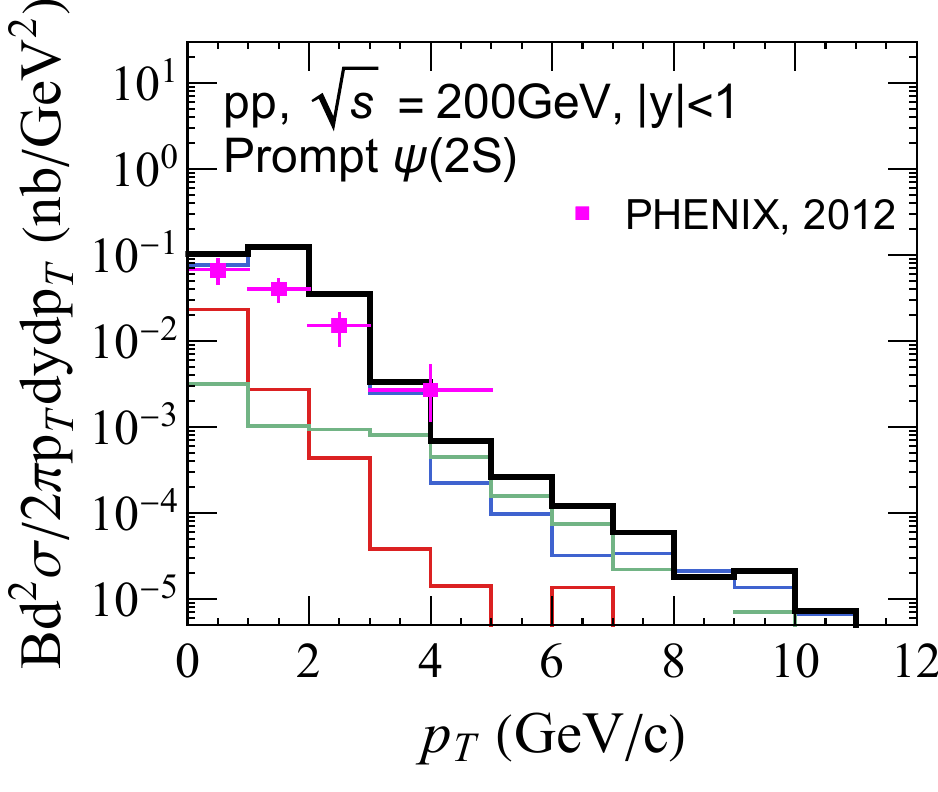}
\includegraphics[width=0.23\textwidth]{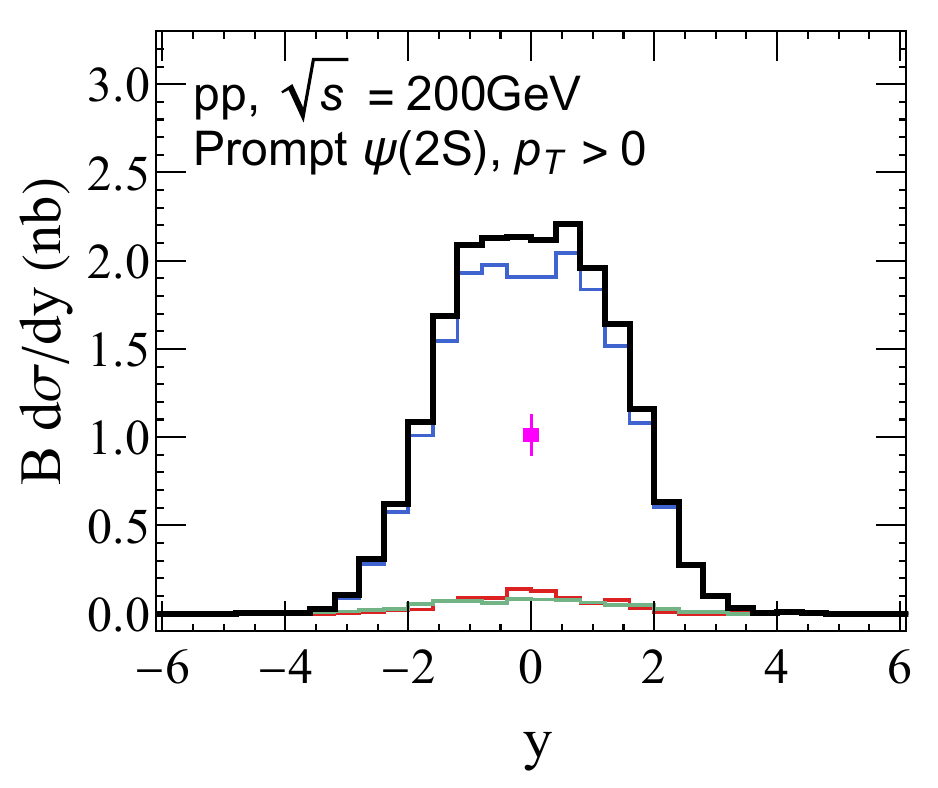}
\caption{$p_T$ spectra and rapidity distribution of prompt $J/\psi$ (upper) and $\psi(2S)$ (lower) ( thick black lines) in pp collisions at $\sqrt{s}=200~ \rm GeV$. Red is the contribution from flavor creation, blue that from flavor excitation, and green from gluon splitting.  The experimental data are from PHENIX~\cite{PHENIX:2011gyb} and STAR~\cite{STAR:2018smh}.}
\label{fig.prompt.cc200}
\end{figure}
%---------------------------------------------------------------------

%---------------------------------------------------------------------
\begin{figure}[!htb]
\includegraphics[width=0.23\textwidth]{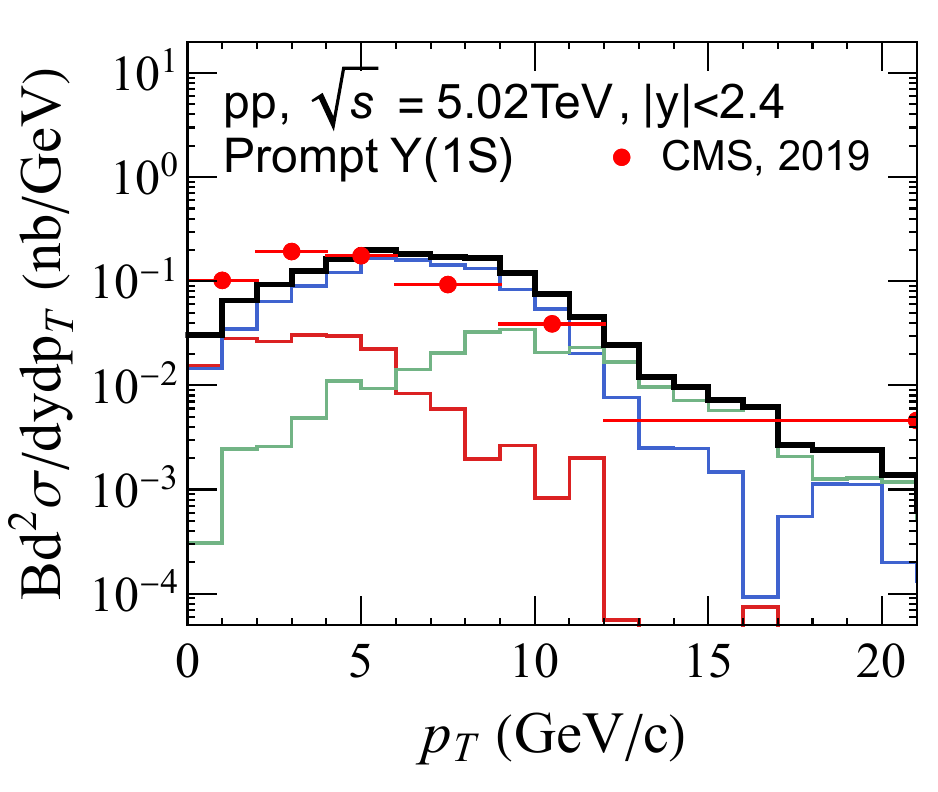}
\includegraphics[width=0.23\textwidth]{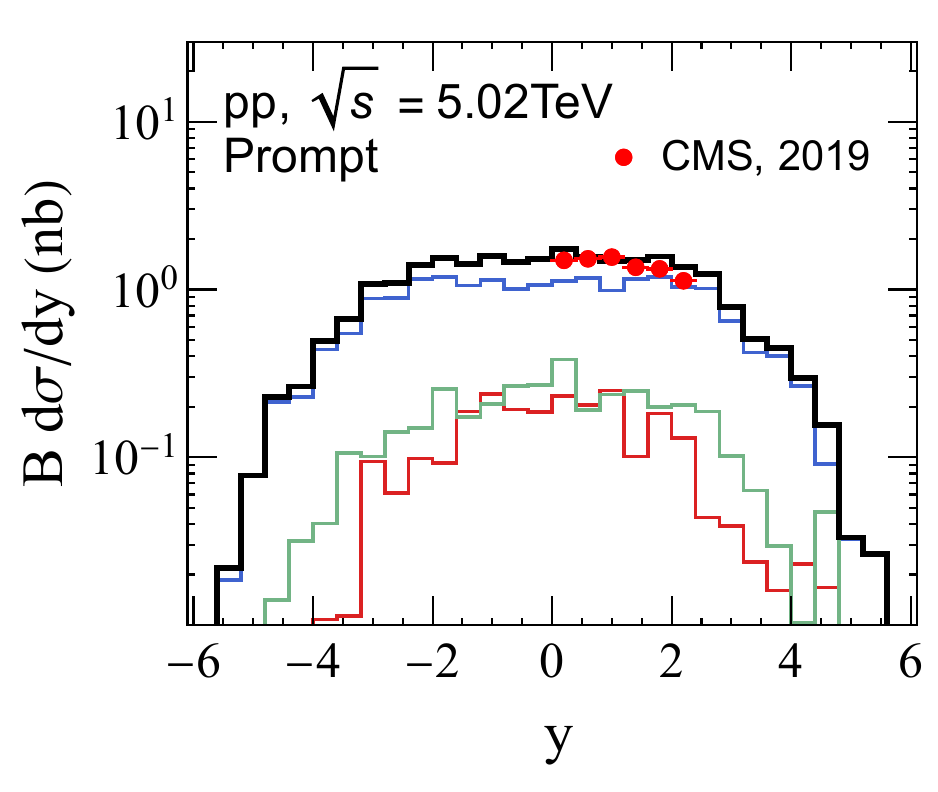}\\
\includegraphics[width=0.23\textwidth]{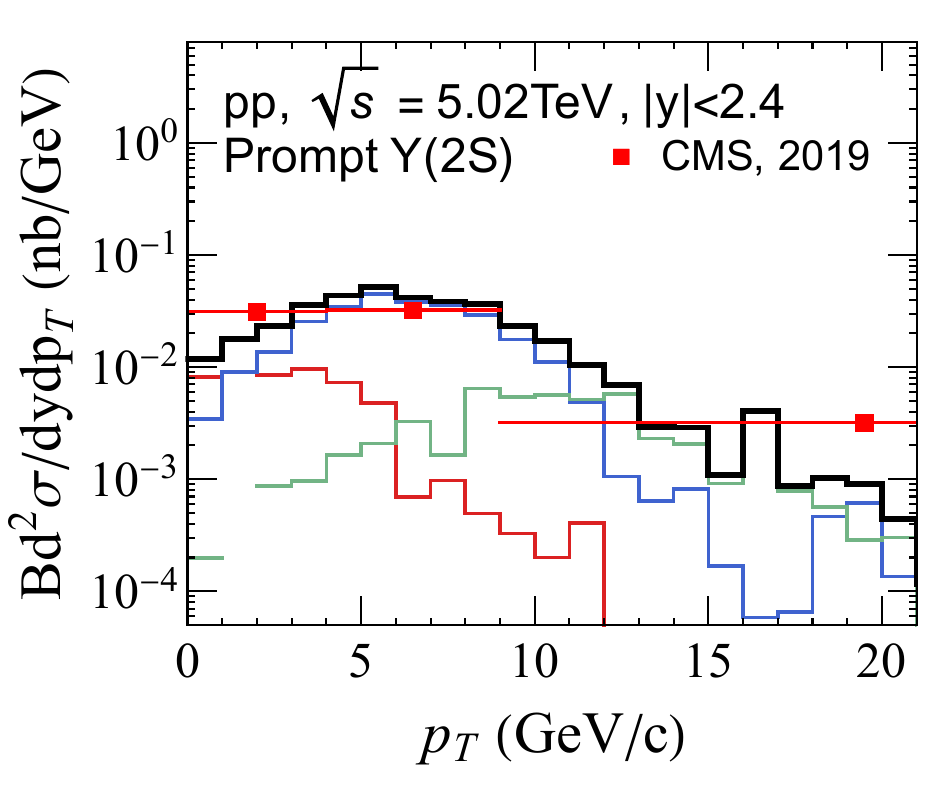}
\includegraphics[width=0.23\textwidth]{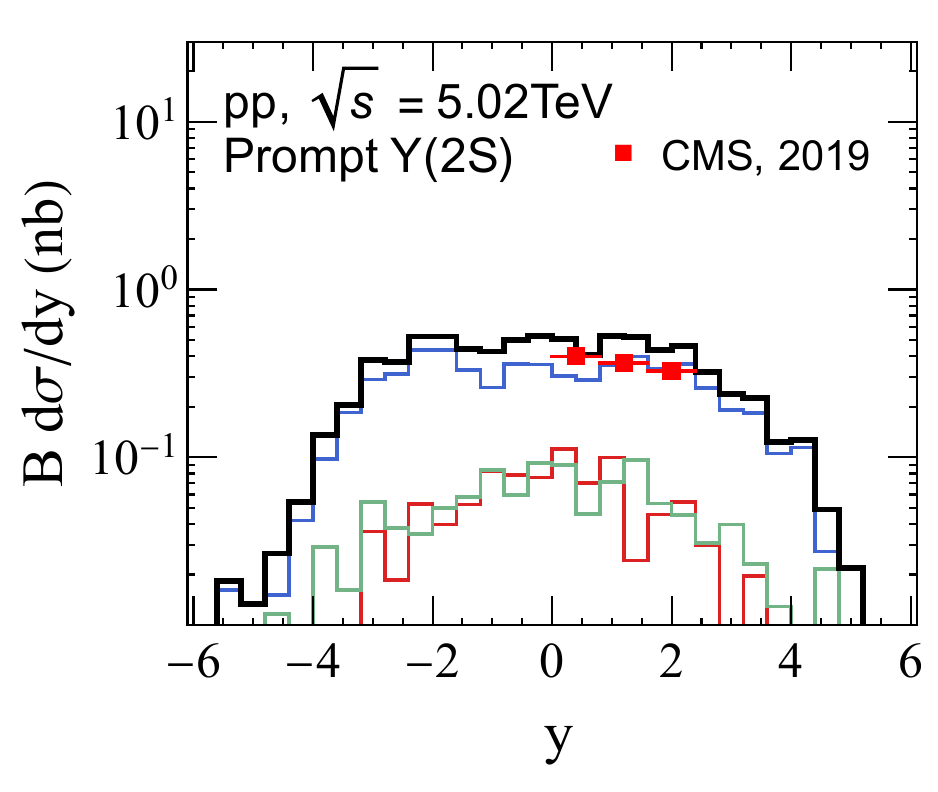}\\
\includegraphics[width=0.23\textwidth]{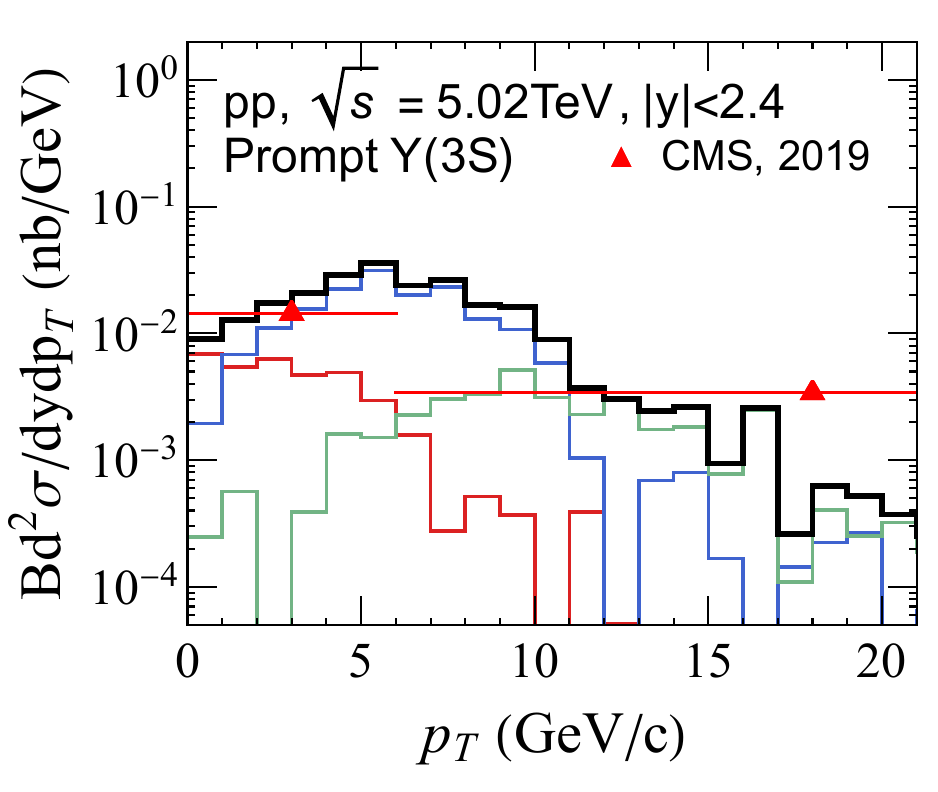}
\includegraphics[width=0.23\textwidth]{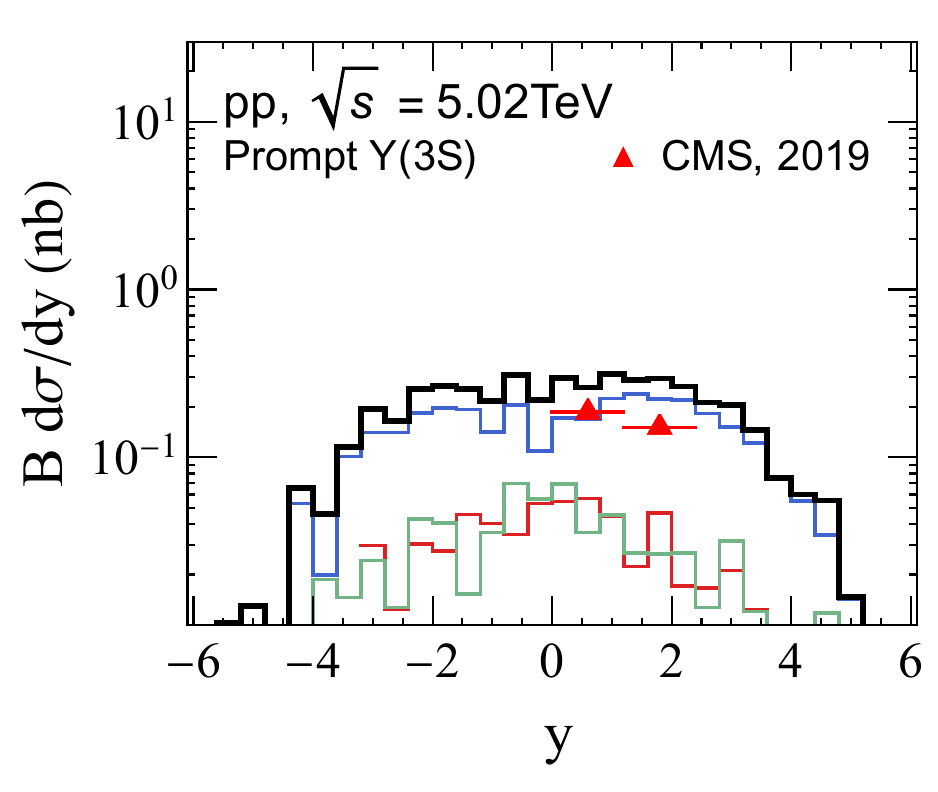}
\caption{$p_T$ spectra and rapidity $y$ dependence of prompt $\Upsilon(1S)$ (upper), $\Upsilon(2S)$ (middle), and $\Upsilon(3S)$ (lower) (thick black lines). Red is the contribution from flavor creation, blue that from flavor excitation, and green from gluon splitting. The experimental data are from CMS~\cite{CMS:2018zza}. The branching ratios of $B(\Upsilon(nS)\to \mu^+\mu^-)$ are taken as 2.48\% for $1S$, 1.93\% for $2S$, and 2.18\% for $3S$, which are given by PDG~\cite{Workman:2022ynf}. }
\label{fig.prompt.bb}
\end{figure}
%---------------------------------------------------------------------

The comparison of the EPOS4HQ calculations with the presently available charmonium data at $\sqrt{s}$ = 5.02 TeV is shown in Fig.~\ref{fig.prompt.cc}. The top figures display the
$p_T$ spectra for prompt $J/\psi$ at midrapidity and the $p_T$ integrated rapidity distribution, the bottom figures show the results for the prompt $\psi(2S)$, where experimentally only a transverse momentum spectrum is available. The prompt $J/\psi$ is the summation of the direct produced $J/\psi$ and the feed-down from the $\chi_c$ and $\psi(2S)$ to $J/\psi$ with the branching ratio 30\% and 61\%, respectively. 
We show the contributions from the  different $Q\bar Q$ production processes as well as the sum. The color coding is that of Fig.
\ref{fig.ytotal}. We see that quarkonia with $p_T >$ 4 GeV are almost exclusively produced by heavy quarks, which originate from the gluon splitting process in which the $Q\bar Q$ have a small azimuthal opening angle, favorable for the production of a charmonium. 
Only at very low $p_T$ flavour excitation is dominant and flavor creation contributes only marginally to the charmonium $p_T$ spectra.  Fig.~\ref{fig.wo.correlations} shows the influence of the $Q\bar Q$ correlations on the prompt $J/\psi$ transverse momentum distribution. Here 
we compare the spectrum with the correlations (black line) with that where we have randomized the azimuthal distributions between the $Q\bar Q$ pair. This randomization affects mostly the gluon splitting process, where the $Q\bar Q$ are strongly correlated and which is dominant at high $p_T$.
We see a visible modification of the slope, which is $p_T$ dependent because the different pQCD creation processes depend differently on $p_T$.

The spectra at the highest RHIC energy are shown in Fig.~\ref{fig.prompt.cc200}, in the top row for $J/\psi$ in the bottom row for $\psi(2s)$ . On the left hand side we display the $p_T$ spectra, on the right hand side the rapidity distribution. Our calculations are compared with PHENIX~\cite{PHENIX:2011gyb} and STAR~\cite{STAR:2018smh} data. A good agreement between the EPOS4HQ and $J/\psi$ data for almost all $p_T$ is found. 

The $p_T$ distribution of $\psi(2s)$ is slightly overpredicted. We display in this figure as well the spectra of the different production processes  of the $c$ quarks. All spectra are dominated by creation via the flavor excitation process. Flavour creation does not play a role at all whereas gluon splitting  contributes modestly at high $p_T$.

Fig.~\ref{fig.prompt.bb} shows the $p_T$ distribution at midrapidity (left) as well the $p_T$ integrated rapidity distribution (right) of the different prompt bottomonia in comparison with experimental data, when available. 
The prompt $\Upsilon(nS)$ comes from direct production and the feed-down from the excited states. In this study, we take the branching ratios 23\%, 20\%, 7\%, 7\%, and 1\%, respectively, for $\chi_b(1P)$, $\chi_b(1D)$, $\Upsilon(2S)$, $\chi_b(2P)$, and $\Upsilon(3S)$ to the ground state $\Upsilon(1S)$, 9.3\% and 10.6\% for $\chi_b(2P)$ and $\Upsilon(3S)$ to $\Upsilon(2S)$, respectively, as suggested by PDG~\cite{Workman:2022ynf}. 
From top to bottom we display the spectra for prompt $\Upsilon(1S)$, $\Upsilon(2S)$, and $\Upsilon(3S)$. Also here the bottomonia with a high transverse momentum 
stem from bottom quarks produced by gluon splitting, in contradistinction to the $B$-meson spectrum (Fig.~\ref{fig.candb.pt}). This is dominated by $b$ quarks coming from flavor creation, which contribute very little to the production of bottomonia. The EPOS4HQ results agree within the error bars with the experimental results for all bottomonia states.

\subsection{$B_c$ production}
We come now to the production of $B_c$ mesons. In heavy-ion collisions at low $p_T$ the creation of a significant yield of $B_c$ mesons is considered as a witness that a QGP is created~\cite{CMS:2022sxl} with which the heavy quarks interact and where the $B_c$ is  finally formed by hadronization.
In pp collisions, since the bottom and charm quark cannot be produced at the same vertex and a production due to hadronic scattering is very improbable, it can only appear in the rare pp collisions in which a $c\bar c$ and a $b \bar b$ pair is created in the same pp collision, They may be produced from the same pomeron \cite{Workman:2022ynf} or from two different pomerons. EPOS4HQ calculations show that the latter process is dominant. It is therefore interesting to study the $B_c$ yield in pp, where an eventually produced QGP does not modify the $B_c$ spectra substantially, not only to understand better the spatial structure of the interaction zone but also to provide a reference point for the comparison with heavy ion collisions.

\begin{figure}[!htb]
\includegraphics[width=0.23\textwidth]{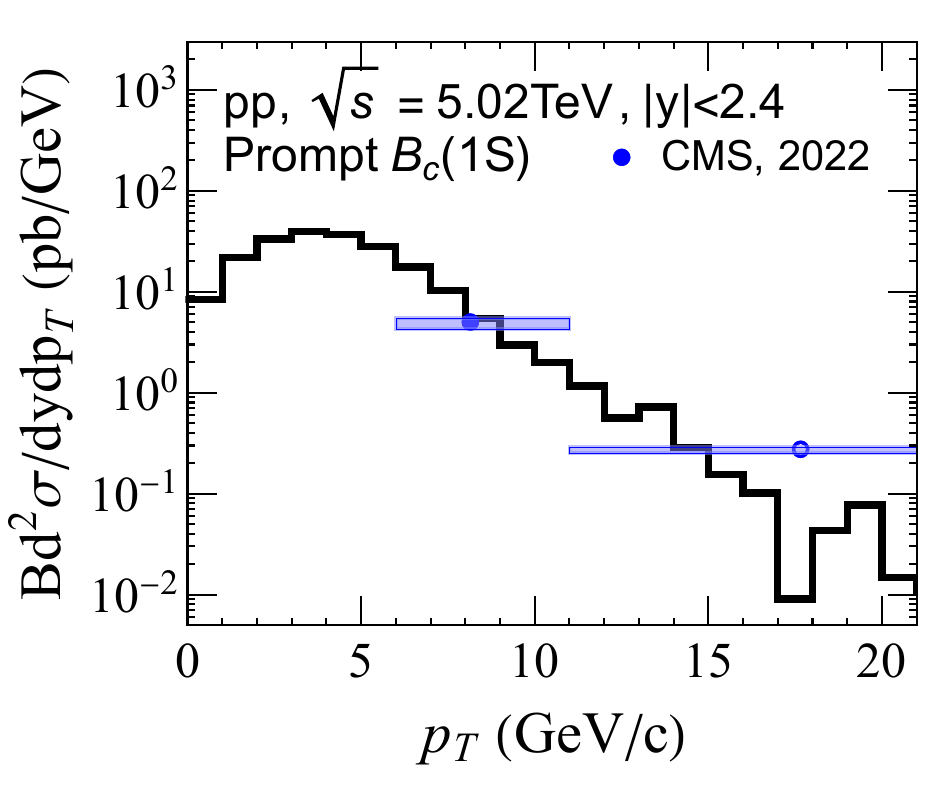}
\includegraphics[width=0.23\textwidth]{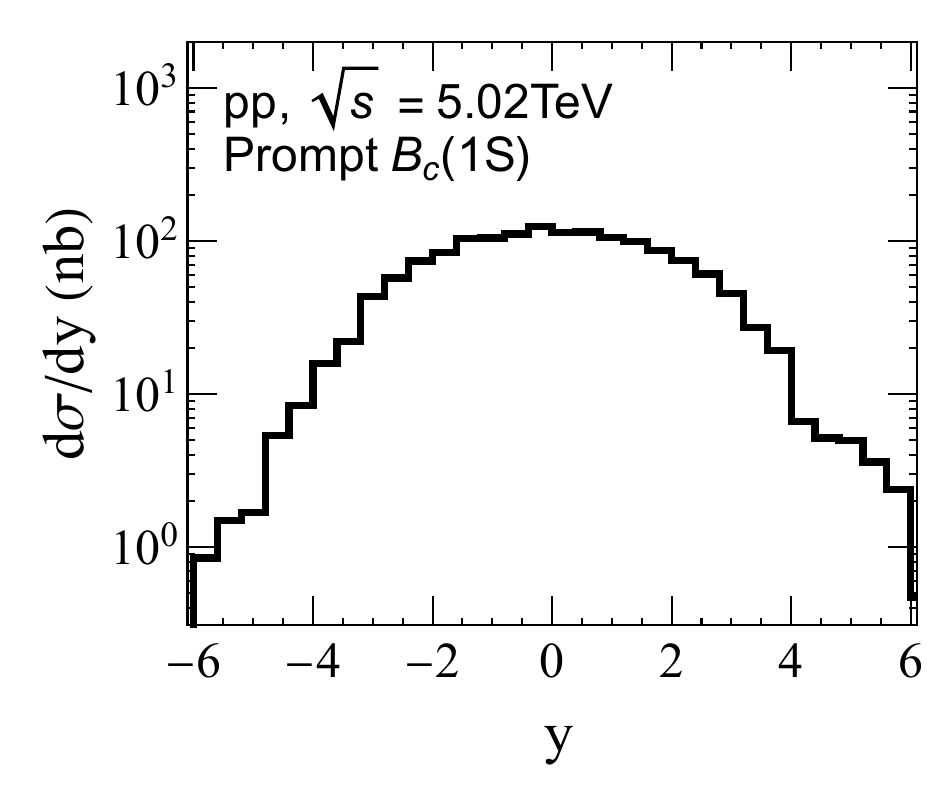}
\caption{$p_T$ spectrum and rapidity $y$ dependence of prompt $B_c(1S)$ state. The experimental data are from CMS~\cite{CMS:2022sxl}. The branching ratio is $B(B_c^+\to (J/\psi\to \mu^+\mu^-)\mu^+v_{\mu})$.}
\label{fig.bc}
\end{figure}

In Fig.~\ref{fig.bc} we display the EPOS4HQ results for the transverse momentum spectrum of the $B_c$ at midrapidity (left) and their rapidity distribution (right). The transverse momentum distribution is compared with CMS data.
The experimental data is displayed with the branching ratio $B(B_c\to (J/\psi\to \mu^+\mu^-)\mu^+v_{\mu})$, where $B(J/\psi\to \mu^+\mu^-)=5.96\%$ for $J/\psi$ is given by PDG~\cite{Workman:2022ynf}. However, the decay branching ratio $B(B_c\to J/\psi\mu^+v_{\mu})$ has not been observed in experiments. There are many theoretical predictions, which give a uncertainty range of from 1.2\% to 6.7\%~\cite{Ebert:2003cn,Hernandez:2006gt,Qiao:2012vt}. In our study, we find the spectrum can be well described by taking almost the central value $B(B_c\to J/\psi\mu^+v_{\mu})=3.5\%$. Due to the lack of information about the decay of the excited states, we take 100\% as the branching ratio for all excited states to the ground state $B_s$. 
As said, the  $\bar{b}$ and the $c$  quark cannot come from the same vertex. A detailed investigation shows that they do almost never come from the same pomeron, so $B_c$ is only produced when at least two pomerons are exchanged between the two incoming protons.  It is remarkable that with a source parameter $\sigma_{c\bar b}=\sqrt{(\sigma_{c\bar c}^2+\sigma_{b\bar b}^2)/2}=0.285~\rm fm$,
which controls the distance between $b$ and $c$ sources, we can also reproduce the $B_c$ using Eq.~\eqref{eq.projectionmc2}. In view of the fact that the heavy quarks in a $B_c$ come from different vertices and despite of the uncertainties of the branching ratio, this is an  indication that EPOS4HQ describes reasonably not only the momentum space observables but also the coordinate space distribution of the vertices. 
%---------------------------------------------------------------------
%---------------------------------------------------------------------
 %---------------------------------------------------------------------
\begin{table}
	\renewcommand\arraystretch{1.5}
	\setlength{\tabcolsep}{2.8mm}
	\begin{tabular}{c|c|c|c}
		\toprule[1pt]\toprule[1pt]
        \multicolumn{1}{c|}{Fraction}&
	\multicolumn{1}{c|}{\rm $\bar b$ from P1}& 
        \multicolumn{1}{c|}{\rm $\bar b$ from P2} &    \multicolumn{1}{c}{\rm $\bar b$ from P3} 
        \tabularnewline
		\midrule[1pt]
	\rm c from P1 & 2.48\% & 4.42\% & 0.32\%  
        \tabularnewline
        \rm c from  P2 & 17.66\% & 61.20\% & 5.68\% 
        \tabularnewline
	\rm c from  P3 & 2.56\% & 5.25\% & 0.43\%  
        \tabularnewline
		\bottomrule[1pt]
	\end{tabular}
	\caption{ The origin of the heavy mesons entrained in $B_c$ mesons.  We present the percentage distribution for the different heavy quark production mechanisms.  P1 is Flavor creation, P2 is Flavor excitation, and P3 is Gluon splitting. }
	\label{table3}
\end{table}
%--------------------------------------------------------------------
%--------------------------------------------------------------------
\begin{table}
	\renewcommand\arraystretch{1.5}
	\setlength{\tabcolsep}{2.8mm}
	\begin{tabular}{c|c|c|c}
		\toprule[1pt]\toprule[1pt]
        \multicolumn{1}{c|}{Fraction}&
	\multicolumn{1}{c|}{\rm $\bar b$ from P1}& 
        \multicolumn{1}{c|}{\rm $\bar b$ from P2} &    \multicolumn{1}{c}{\rm $\bar b$ from P3} 
        \tabularnewline
		\midrule[1pt]
	\rm c from P1 & 1.64\% & 5.12\% & 0.46\%  
        \tabularnewline
        \rm c from  P2 & 19.19\% & 59.91\% & 5.44\% 
        \tabularnewline
	\rm c from  P3 & 1.87\% & 5.84\% & 0.53\%  
        \tabularnewline
		\bottomrule[1pt]
	\end{tabular}
	\caption{ Expected percentage distribution of the origin of heavy quarks, entrained in $B_c$ mesons, if the production mechanisms of both heavy quarks are uncorrelated (see text).
  %\ref{table3} 
  }
	\label{table4}
\end{table}
%--------------------------------------------------------------------

The origin of heavy quarks, entrained in a $B_c$ is shown in Table~\ref{table3}. For the majority of $B_c$ both heavy quarks come from flavor excitation processes.  
Since the $b\bar b$ and $c\bar c$ stem from independent process, one would expect that each entry of Table~\ref{table3} can be written as ${\rm prob } (c)_i \times {\rm prob } (\bar{b})_j$, where ${\rm prob } ({c})_i$ and ${\rm prob }(\bar{b})_j$ are the cumulated distribution corresponding to Table~\ref{table3}, namely: ${\rm prob }(\bar{b})=\{22.70 \%, 70.87\%, 6.43\%\}$ and $  {\rm prob }(c)=\{7.22\%, 84.54\%, 8.24\%\}$. The corresponding results are displayed in Table~\ref{table4}. One sees that they are indeed close to those of Table~\ref{table3}, with small differences in the last column, what tends to confirm the independent production hypothesis.

%%%%%%%%%%%%%%%%%%%%%%%
\section{Summary}
\label{sec.IV}
%%%%%%%%%%%%%%%%%%%%%%%
In summary, we have investigated the influence of different pQCD processes for heavy flavor production on heavy flavor meson observables in pp collisions at RHIC and LHC energies. The different azimuthal correlations of these processes show up in the correlations between open heavy flavour mesons as well as in the production of quarkonia. 

We have verified that EPOS4HQ reproduces well the measured open heavy flavour meson spectra as well as the (different) correlations between $D\bar D$ and $B\bar B$  mesons. We have analyzed the influence of the different production processes on these correlations and have shown that the form of these correlations can be well identified with the different production processes, whose weight is different for $D$ and $B$ mesons.  It turns out that flavour excitation is the dominant process of heavy quark production. At high $p_T$ also flavor creation and gluon splitting contribute, differently for $b$ and $c$ quarks. This shows that the production of heavy mesons cannot be  described by simplified models which use parton distribution and fragmentation functions only.

The charmonium and bottomonium production in high energy $pp$ collisions is described via the Wigner density approach based on  the eigenfunctions of the Schr\"odinger equation of the different quarkonia states. Here we have included  all states up to $3S$ states. The similarity of
the eigenfunctions and the harmonic oscillator wave functions shows that the eigenfunctions are essentially characterized by the root mean square distance between the heavy quark and its antiquark, which is different for $c\bar c$ and $b\bar b$ pairs. With this one parameter all excited states can be calculated and the results show a good agreement with experimental data.

The correlations between heavy quarks play an important role for the quarkonia observables. The three production mechanisms contribute quite differently to the $p_T$ spectra  of quarkonia and only the sum of them creates agreement with experiment. Therefore, for the production of quarkonia the correlations between the quark and antiquark in the different production and the relative contribution of the different processes are essential.  

Finally we have shown that the same density matrix approach, without any additional parameter, allows also to reproduce quantitatively the experimental $p_T$-spectra of $B_c$ mesons, which are formed by heavy quarks coming from different pomerons, although the uncertainty of the branching ratio is large. This confirms that the spatial quark distribution in the interaction zone of a pp collision is described by the prescription adopted in the EPOS4HQ approach. Future progress on the spatial distribution of partons in hadrons will be helpful to impose more constraints on these parameters.

\vspace{1cm}
\noindent {\bf Acknowledgement}: We thank E. Bratkovskaya and T. Song for fruitful discussions. This work is funded by the European Union’s Horizon 2020 research and innovation program under grant agreement No. 824093 (STRONG-2020). JX is also the Helmholtz Research Academy Hessen for FAIR (HFHF).

%=====================
\bibliographystyle{apsrev4-1.bst}
\bibliography{Ref}

\begin{thebibliography}{71}%
\makeatletter
\providecommand \@ifxundefined [1]{%
 \@ifx{#1\undefined}
}%
\providecommand \@ifnum [1]{%
 \ifnum #1\expandafter \@firstoftwo
 \else \expandafter \@secondoftwo
 \fi
}%
\providecommand \@ifx [1]{%
 \ifx #1\expandafter \@firstoftwo
 \else \expandafter \@secondoftwo
 \fi
}%
\providecommand \natexlab [1]{#1}%
\providecommand \enquote  [1]{``#1''}%
\providecommand \bibnamefont  [1]{#1}%
\providecommand \bibfnamefont [1]{#1}%
\providecommand \citenamefont [1]{#1}%
\providecommand \href@noop [0]{\@secondoftwo}%
\providecommand \href [0]{\begingroup \@sanitize@url \@href}%
\providecommand \@href[1]{\@@startlink{#1}\@@href}%
\providecommand \@@href[1]{\endgroup#1\@@endlink}%
\providecommand \@sanitize@url [0]{\catcode `\\12\catcode `\$12\catcode
  `\&12\catcode `\#12\catcode `\^12\catcode `\_12\catcode `\%12\relax}%
\providecommand \@@startlink[1]{}%
\providecommand \@@endlink[0]{}%
\providecommand \url  [0]{\begingroup\@sanitize@url \@url }%
\providecommand \@url [1]{\endgroup\@href {#1}{\urlprefix }}%
\providecommand \urlprefix  [0]{URL }%
\providecommand \Eprint [0]{\href }%
\providecommand \doibase [0]{http://dx.doi.org/}%
\providecommand \selectlanguage [0]{\@gobble}%
\providecommand \bibinfo  [0]{\@secondoftwo}%
\providecommand \bibfield  [0]{\@secondoftwo}%
\providecommand \translation [1]{[#1]}%
\providecommand \BibitemOpen [0]{}%
\providecommand \bibitemStop [0]{}%
\providecommand \bibitemNoStop [0]{.\EOS\space}%
\providecommand \EOS [0]{\spacefactor3000\relax}%
\providecommand \BibitemShut  [1]{\csname bibitem#1\endcsname}%
\let\auto@bib@innerbib\@empty
%</preamble>
\bibitem [{\citenamefont {van Hees}\ \emph {et~al.}(2006)\citenamefont {van
  Hees}, \citenamefont {Greco},\ and\ \citenamefont {Rapp}}]{vanHees:2005wb}%
  \BibitemOpen
  \bibfield  {author} {\bibinfo {author} {\bibfnamefont {H.}~\bibnamefont {van
  Hees}}, \bibinfo {author} {\bibfnamefont {V.}~\bibnamefont {Greco}}, \ and\
  \bibinfo {author} {\bibfnamefont {R.}~\bibnamefont {Rapp}},\ }\href {\doibase
  10.1103/PhysRevC.73.034913} {\bibfield  {journal} {\bibinfo  {journal} {Phys.
  Rev. C}\ }\textbf {\bibinfo {volume} {73}},\ \bibinfo {pages} {034913}
  (\bibinfo {year} {2006})},\ \Eprint {http://arxiv.org/abs/nucl-th/0508055}
  {arXiv:nucl-th/0508055} \BibitemShut {NoStop}%
\bibitem [{\citenamefont {He}\ \emph {et~al.}(2012)\citenamefont {He},
  \citenamefont {Fries},\ and\ \citenamefont {Rapp}}]{He:2011qa}%
  \BibitemOpen
  \bibfield  {author} {\bibinfo {author} {\bibfnamefont {M.}~\bibnamefont
  {He}}, \bibinfo {author} {\bibfnamefont {R.~J.}\ \bibnamefont {Fries}}, \
  and\ \bibinfo {author} {\bibfnamefont {R.}~\bibnamefont {Rapp}},\ }\href
  {\doibase 10.1103/PhysRevC.86.014903} {\bibfield  {journal} {\bibinfo
  {journal} {Phys. Rev. C}\ }\textbf {\bibinfo {volume} {86}},\ \bibinfo
  {pages} {014903} (\bibinfo {year} {2012})},\ \Eprint
  {http://arxiv.org/abs/1106.6006} {arXiv:1106.6006 [nucl-th]} \BibitemShut
  {NoStop}%
\bibitem [{\citenamefont {Minissale}\ \emph {et~al.}(2021)\citenamefont
  {Minissale}, \citenamefont {Plumari},\ and\ \citenamefont
  {Greco}}]{Minissale:2020bif}%
  \BibitemOpen
  \bibfield  {author} {\bibinfo {author} {\bibfnamefont {V.}~\bibnamefont
  {Minissale}}, \bibinfo {author} {\bibfnamefont {S.}~\bibnamefont {Plumari}},
  \ and\ \bibinfo {author} {\bibfnamefont {V.}~\bibnamefont {Greco}},\ }\href
  {\doibase 10.1016/j.physletb.2021.136622} {\bibfield  {journal} {\bibinfo
  {journal} {Phys. Lett. B}\ }\textbf {\bibinfo {volume} {821}},\ \bibinfo
  {pages} {136622} (\bibinfo {year} {2021})},\ \Eprint
  {http://arxiv.org/abs/2012.12001} {arXiv:2012.12001 [hep-ph]} \BibitemShut
  {NoStop}%
\bibitem [{\citenamefont {Cao}\ \emph {et~al.}(2015)\citenamefont {Cao},
  \citenamefont {Qin},\ and\ \citenamefont {Bass}}]{Cao:2015hia}%
  \BibitemOpen
  \bibfield  {author} {\bibinfo {author} {\bibfnamefont {S.}~\bibnamefont
  {Cao}}, \bibinfo {author} {\bibfnamefont {G.-Y.}\ \bibnamefont {Qin}}, \ and\
  \bibinfo {author} {\bibfnamefont {S.~A.}\ \bibnamefont {Bass}},\ }\href
  {\doibase 10.1103/PhysRevC.92.024907} {\bibfield  {journal} {\bibinfo
  {journal} {Phys. Rev. C}\ }\textbf {\bibinfo {volume} {92}},\ \bibinfo
  {pages} {024907} (\bibinfo {year} {2015})},\ \Eprint
  {http://arxiv.org/abs/1505.01413} {arXiv:1505.01413 [nucl-th]} \BibitemShut
  {NoStop}%
\bibitem [{\citenamefont {Cao}\ \emph {et~al.}(2016)\citenamefont {Cao},
  \citenamefont {Luo}, \citenamefont {Qin},\ and\ \citenamefont
  {Wang}}]{Cao:2016gvr}%
  \BibitemOpen
  \bibfield  {author} {\bibinfo {author} {\bibfnamefont {S.}~\bibnamefont
  {Cao}}, \bibinfo {author} {\bibfnamefont {T.}~\bibnamefont {Luo}}, \bibinfo
  {author} {\bibfnamefont {G.-Y.}\ \bibnamefont {Qin}}, \ and\ \bibinfo
  {author} {\bibfnamefont {X.-N.}\ \bibnamefont {Wang}},\ }\href {\doibase
  10.1103/PhysRevC.94.014909} {\bibfield  {journal} {\bibinfo  {journal} {Phys.
  Rev. C}\ }\textbf {\bibinfo {volume} {94}},\ \bibinfo {pages} {014909}
  (\bibinfo {year} {2016})},\ \Eprint {http://arxiv.org/abs/1605.06447}
  {arXiv:1605.06447 [nucl-th]} \BibitemShut {NoStop}%
\bibitem [{\citenamefont {Cao}\ \emph {et~al.}(2019)\citenamefont {Cao} \emph
  {et~al.}}]{Cao:2018ews}%
  \BibitemOpen
  \bibfield  {author} {\bibinfo {author} {\bibfnamefont {S.}~\bibnamefont
  {Cao}} \emph {et~al.},\ }\href {\doibase 10.1103/PhysRevC.99.054907}
  {\bibfield  {journal} {\bibinfo  {journal} {Phys. Rev. C}\ }\textbf {\bibinfo
  {volume} {99}},\ \bibinfo {pages} {054907} (\bibinfo {year} {2019})},\
  \Eprint {http://arxiv.org/abs/1809.07894} {arXiv:1809.07894 [nucl-th]}
  \BibitemShut {NoStop}%
\bibitem [{\citenamefont {Cao}\ \emph {et~al.}(2020)\citenamefont {Cao},
  \citenamefont {Sun}, \citenamefont {Li}, \citenamefont {Liu}, \citenamefont
  {Xing}, \citenamefont {Qin},\ and\ \citenamefont {Ko}}]{Cao:2019iqs}%
  \BibitemOpen
  \bibfield  {author} {\bibinfo {author} {\bibfnamefont {S.}~\bibnamefont
  {Cao}}, \bibinfo {author} {\bibfnamefont {K.-J.}\ \bibnamefont {Sun}},
  \bibinfo {author} {\bibfnamefont {S.-Q.}\ \bibnamefont {Li}}, \bibinfo
  {author} {\bibfnamefont {S.~Y.~F.}\ \bibnamefont {Liu}}, \bibinfo {author}
  {\bibfnamefont {W.-J.}\ \bibnamefont {Xing}}, \bibinfo {author}
  {\bibfnamefont {G.-Y.}\ \bibnamefont {Qin}}, \ and\ \bibinfo {author}
  {\bibfnamefont {C.~M.}\ \bibnamefont {Ko}},\ }\href {\doibase
  10.1016/j.physletb.2020.135561} {\bibfield  {journal} {\bibinfo  {journal}
  {Phys. Lett. B}\ }\textbf {\bibinfo {volume} {807}},\ \bibinfo {pages}
  {135561} (\bibinfo {year} {2020})},\ \Eprint
  {http://arxiv.org/abs/1911.00456} {arXiv:1911.00456 [nucl-th]} \BibitemShut
  {NoStop}%
\bibitem [{\citenamefont {Gossiaux}\ \emph {et~al.}(2009)\citenamefont
  {Gossiaux}, \citenamefont {Bierkandt},\ and\ \citenamefont
  {Aichelin}}]{Gossiaux:2009mk}%
  \BibitemOpen
  \bibfield  {author} {\bibinfo {author} {\bibfnamefont {P.~B.}\ \bibnamefont
  {Gossiaux}}, \bibinfo {author} {\bibfnamefont {R.}~\bibnamefont {Bierkandt}},
  \ and\ \bibinfo {author} {\bibfnamefont {J.}~\bibnamefont {Aichelin}},\
  }\href {\doibase 10.1103/PhysRevC.79.044906} {\bibfield  {journal} {\bibinfo
  {journal} {Phys. Rev. C}\ }\textbf {\bibinfo {volume} {79}},\ \bibinfo
  {pages} {044906} (\bibinfo {year} {2009})},\ \Eprint
  {http://arxiv.org/abs/0901.0946} {arXiv:0901.0946 [hep-ph]} \BibitemShut
  {NoStop}%
\bibitem [{\citenamefont {Song}\ \emph {et~al.}(2015)\citenamefont {Song},
  \citenamefont {Berrehrah}, \citenamefont {Cabrera}, \citenamefont
  {Torres-Rincon}, \citenamefont {Tolos}, \citenamefont {Cassing},\ and\
  \citenamefont {Bratkovskaya}}]{Song:2015sfa}%
  \BibitemOpen
  \bibfield  {author} {\bibinfo {author} {\bibfnamefont {T.}~\bibnamefont
  {Song}}, \bibinfo {author} {\bibfnamefont {H.}~\bibnamefont {Berrehrah}},
  \bibinfo {author} {\bibfnamefont {D.}~\bibnamefont {Cabrera}}, \bibinfo
  {author} {\bibfnamefont {J.~M.}\ \bibnamefont {Torres-Rincon}}, \bibinfo
  {author} {\bibfnamefont {L.}~\bibnamefont {Tolos}}, \bibinfo {author}
  {\bibfnamefont {W.}~\bibnamefont {Cassing}}, \ and\ \bibinfo {author}
  {\bibfnamefont {E.}~\bibnamefont {Bratkovskaya}},\ }\href {\doibase
  10.1103/PhysRevC.92.014910} {\bibfield  {journal} {\bibinfo  {journal} {Phys.
  Rev. C}\ }\textbf {\bibinfo {volume} {92}},\ \bibinfo {pages} {014910}
  (\bibinfo {year} {2015})},\ \Eprint {http://arxiv.org/abs/1503.03039}
  {arXiv:1503.03039 [nucl-th]} \BibitemShut {NoStop}%
\bibitem [{\citenamefont {Song}\ \emph {et~al.}(2016)\citenamefont {Song},
  \citenamefont {Berrehrah}, \citenamefont {Cabrera}, \citenamefont {Cassing},\
  and\ \citenamefont {Bratkovskaya}}]{Song:2015ykw}%
  \BibitemOpen
  \bibfield  {author} {\bibinfo {author} {\bibfnamefont {T.}~\bibnamefont
  {Song}}, \bibinfo {author} {\bibfnamefont {H.}~\bibnamefont {Berrehrah}},
  \bibinfo {author} {\bibfnamefont {D.}~\bibnamefont {Cabrera}}, \bibinfo
  {author} {\bibfnamefont {W.}~\bibnamefont {Cassing}}, \ and\ \bibinfo
  {author} {\bibfnamefont {E.}~\bibnamefont {Bratkovskaya}},\ }\href {\doibase
  10.1103/PhysRevC.93.034906} {\bibfield  {journal} {\bibinfo  {journal} {Phys.
  Rev. C}\ }\textbf {\bibinfo {volume} {93}},\ \bibinfo {pages} {034906}
  (\bibinfo {year} {2016})},\ \Eprint {http://arxiv.org/abs/1512.00891}
  {arXiv:1512.00891 [nucl-th]} \BibitemShut {NoStop}%
\bibitem [{\citenamefont {He}\ and\ \citenamefont {Rapp}(2020)}]{He:2019vgs}%
  \BibitemOpen
  \bibfield  {author} {\bibinfo {author} {\bibfnamefont {M.}~\bibnamefont
  {He}}\ and\ \bibinfo {author} {\bibfnamefont {R.}~\bibnamefont {Rapp}},\
  }\href {\doibase 10.1103/PhysRevLett.124.042301} {\bibfield  {journal}
  {\bibinfo  {journal} {Phys. Rev. Lett.}\ }\textbf {\bibinfo {volume} {124}},\
  \bibinfo {pages} {042301} (\bibinfo {year} {2020})},\ \Eprint
  {http://arxiv.org/abs/1905.09216} {arXiv:1905.09216 [nucl-th]} \BibitemShut
  {NoStop}%
\bibitem [{\citenamefont {Li}\ \emph {et~al.}(2021)\citenamefont {Li},
  \citenamefont {Liu},\ and\ \citenamefont {Vitev}}]{Li:2020zbk}%
  \BibitemOpen
  \bibfield  {author} {\bibinfo {author} {\bibfnamefont {H.~T.}\ \bibnamefont
  {Li}}, \bibinfo {author} {\bibfnamefont {Z.~L.}\ \bibnamefont {Liu}}, \ and\
  \bibinfo {author} {\bibfnamefont {I.}~\bibnamefont {Vitev}},\ }\href
  {\doibase 10.1016/j.physletb.2021.136261} {\bibfield  {journal} {\bibinfo
  {journal} {Phys. Lett. B}\ }\textbf {\bibinfo {volume} {816}},\ \bibinfo
  {pages} {136261} (\bibinfo {year} {2021})},\ \Eprint
  {http://arxiv.org/abs/2007.10994} {arXiv:2007.10994 [hep-ph]} \BibitemShut
  {NoStop}%
\bibitem [{\citenamefont {Beraudo}\ \emph {et~al.}(2022)\citenamefont
  {Beraudo}, \citenamefont {De~Pace}, \citenamefont {Monteno}, \citenamefont
  {Nardi},\ and\ \citenamefont {Prino}}]{Beraudo:2022dpz}%
  \BibitemOpen
  \bibfield  {author} {\bibinfo {author} {\bibfnamefont {A.}~\bibnamefont
  {Beraudo}}, \bibinfo {author} {\bibfnamefont {A.}~\bibnamefont {De~Pace}},
  \bibinfo {author} {\bibfnamefont {M.}~\bibnamefont {Monteno}}, \bibinfo
  {author} {\bibfnamefont {M.}~\bibnamefont {Nardi}}, \ and\ \bibinfo {author}
  {\bibfnamefont {F.}~\bibnamefont {Prino}},\ }\href {\doibase
  10.1140/epjc/s10052-022-10482-y} {\bibfield  {journal} {\bibinfo  {journal}
  {Eur. Phys. J. C}\ }\textbf {\bibinfo {volume} {82}},\ \bibinfo {pages} {607}
  (\bibinfo {year} {2022})},\ \Eprint {http://arxiv.org/abs/2202.08732}
  {arXiv:2202.08732 [hep-ph]} \BibitemShut {NoStop}%
\bibitem [{\citenamefont {Zhao}\ \emph {et~al.}(2024)\citenamefont {Zhao},
  \citenamefont {Aichelin}, \citenamefont {Gossiaux}, \citenamefont
  {Ozvenchuk},\ and\ \citenamefont {Werner}}]{Zhao:2024ecc}%
  \BibitemOpen
  \bibfield  {author} {\bibinfo {author} {\bibfnamefont {J.}~\bibnamefont
  {Zhao}}, \bibinfo {author} {\bibfnamefont {J.}~\bibnamefont {Aichelin}},
  \bibinfo {author} {\bibfnamefont {P.~B.}\ \bibnamefont {Gossiaux}}, \bibinfo
  {author} {\bibfnamefont {V.}~\bibnamefont {Ozvenchuk}}, \ and\ \bibinfo
  {author} {\bibfnamefont {K.}~\bibnamefont {Werner}},\ }\href@noop {} {\
  (\bibinfo {year} {2024})},\ \Eprint {http://arxiv.org/abs/2401.17096}
  {arXiv:2401.17096 [hep-ph]} \BibitemShut {NoStop}%
\bibitem [{\citenamefont {Adamczyk}\ \emph {et~al.}(2017)\citenamefont
  {Adamczyk} \emph {et~al.}}]{STAR:2017kkh}%
  \BibitemOpen
  \bibfield  {author} {\bibinfo {author} {\bibfnamefont {L.}~\bibnamefont
  {Adamczyk}} \emph {et~al.} (\bibinfo {collaboration} {STAR}),\ }\href
  {\doibase 10.1103/PhysRevLett.118.212301} {\bibfield  {journal} {\bibinfo
  {journal} {Phys. Rev. Lett.}\ }\textbf {\bibinfo {volume} {118}},\ \bibinfo
  {pages} {212301} (\bibinfo {year} {2017})},\ \Eprint
  {http://arxiv.org/abs/1701.06060} {arXiv:1701.06060 [nucl-ex]} \BibitemShut
  {NoStop}%
\bibitem [{\citenamefont {Adam}\ \emph
  {et~al.}(2019{\natexlab{a}})\citenamefont {Adam} \emph
  {et~al.}}]{STAR:2018zdy}%
  \BibitemOpen
  \bibfield  {author} {\bibinfo {author} {\bibfnamefont {J.}~\bibnamefont
  {Adam}} \emph {et~al.} (\bibinfo {collaboration} {STAR}),\ }\href {\doibase
  10.1103/PhysRevC.99.034908} {\bibfield  {journal} {\bibinfo  {journal} {Phys.
  Rev. C}\ }\textbf {\bibinfo {volume} {99}},\ \bibinfo {pages} {034908}
  (\bibinfo {year} {2019}{\natexlab{a}})},\ \Eprint
  {http://arxiv.org/abs/1812.10224} {arXiv:1812.10224 [nucl-ex]} \BibitemShut
  {NoStop}%
\bibitem [{\citenamefont {Acharya}\ \emph
  {et~al.}(2022{\natexlab{a}})\citenamefont {Acharya} \emph
  {et~al.}}]{ALICE:2021rxa}%
  \BibitemOpen
  \bibfield  {author} {\bibinfo {author} {\bibfnamefont {S.}~\bibnamefont
  {Acharya}} \emph {et~al.} (\bibinfo {collaboration} {ALICE}),\ }\href
  {\doibase 10.1007/JHEP01(2022)174} {\bibfield  {journal} {\bibinfo  {journal}
  {JHEP}\ }\textbf {\bibinfo {volume} {01}},\ \bibinfo {pages} {174} (\bibinfo
  {year} {2022}{\natexlab{a}})},\ \Eprint {http://arxiv.org/abs/2110.09420}
  {arXiv:2110.09420 [nucl-ex]} \BibitemShut {NoStop}%
\bibitem [{\citenamefont {Acharya}\ \emph
  {et~al.}(2021{\natexlab{a}})\citenamefont {Acharya} \emph
  {et~al.}}]{ALICE:2020iug}%
  \BibitemOpen
  \bibfield  {author} {\bibinfo {author} {\bibfnamefont {S.}~\bibnamefont
  {Acharya}} \emph {et~al.} (\bibinfo {collaboration} {ALICE}),\ }\href
  {\doibase 10.1016/j.physletb.2020.136054} {\bibfield  {journal} {\bibinfo
  {journal} {Phys. Lett. B}\ }\textbf {\bibinfo {volume} {813}},\ \bibinfo
  {pages} {136054} (\bibinfo {year} {2021}{\natexlab{a}})},\ \Eprint
  {http://arxiv.org/abs/2005.11131} {arXiv:2005.11131 [nucl-ex]} \BibitemShut
  {NoStop}%
\bibitem [{\citenamefont {Sirunyan}\ \emph {et~al.}(2021)\citenamefont
  {Sirunyan} \emph {et~al.}}]{CMS:2020bnz}%
  \BibitemOpen
  \bibfield  {author} {\bibinfo {author} {\bibfnamefont {A.~M.}\ \bibnamefont
  {Sirunyan}} \emph {et~al.} (\bibinfo {collaboration} {CMS}),\ }\href
  {\doibase 10.1016/j.physletb.2021.136253} {\bibfield  {journal} {\bibinfo
  {journal} {Phys. Lett. B}\ }\textbf {\bibinfo {volume} {816}},\ \bibinfo
  {pages} {136253} (\bibinfo {year} {2021})},\ \Eprint
  {http://arxiv.org/abs/2009.12628} {arXiv:2009.12628 [hep-ex]} \BibitemShut
  {NoStop}%
\bibitem [{\citenamefont {Sirunyan}\ \emph {et~al.}(2018)\citenamefont
  {Sirunyan} \emph {et~al.}}]{CMS:2017qjw}%
  \BibitemOpen
  \bibfield  {author} {\bibinfo {author} {\bibfnamefont {A.~M.}\ \bibnamefont
  {Sirunyan}} \emph {et~al.} (\bibinfo {collaboration} {CMS}),\ }\href
  {\doibase 10.1016/j.physletb.2018.05.074} {\bibfield  {journal} {\bibinfo
  {journal} {Phys. Lett. B}\ }\textbf {\bibinfo {volume} {782}},\ \bibinfo
  {pages} {474} (\bibinfo {year} {2018})},\ \Eprint
  {http://arxiv.org/abs/1708.04962} {arXiv:1708.04962 [nucl-ex]} \BibitemShut
  {NoStop}%
\bibitem [{\citenamefont {Aaij}\ \emph {et~al.}(2012)\citenamefont {Aaij} \emph
  {et~al.}}]{LHCb:2012aiv}%
  \BibitemOpen
  \bibfield  {author} {\bibinfo {author} {\bibfnamefont {R.}~\bibnamefont
  {Aaij}} \emph {et~al.} (\bibinfo {collaboration} {LHCb}),\ }\href {\doibase
  10.1007/JHEP06(2012)141} {\bibfield  {journal} {\bibinfo  {journal} {JHEP}\
  }\textbf {\bibinfo {volume} {06}},\ \bibinfo {pages} {141} (\bibinfo {year}
  {2012})},\ \bibinfo {note} {[Addendum: JHEP 03, 108 (2014)]},\ \Eprint
  {http://arxiv.org/abs/1205.0975} {arXiv:1205.0975 [hep-ex]} \BibitemShut
  {NoStop}%
\bibitem [{\citenamefont {Adam}\ \emph
  {et~al.}(2019{\natexlab{b}})\citenamefont {Adam} \emph
  {et~al.}}]{STAR:2019fge}%
  \BibitemOpen
  \bibfield  {author} {\bibinfo {author} {\bibfnamefont {J.}~\bibnamefont
  {Adam}} \emph {et~al.} (\bibinfo {collaboration} {STAR}),\ }\href {\doibase
  10.1016/j.physletb.2019.134917} {\bibfield  {journal} {\bibinfo  {journal}
  {Phys. Lett. B}\ }\textbf {\bibinfo {volume} {797}},\ \bibinfo {pages}
  {134917} (\bibinfo {year} {2019}{\natexlab{b}})},\ \Eprint
  {http://arxiv.org/abs/1905.13669} {arXiv:1905.13669 [nucl-ex]} \BibitemShut
  {NoStop}%
\bibitem [{\citenamefont {Abelev}\ \emph {et~al.}(2014)\citenamefont {Abelev}
  \emph {et~al.}}]{ALICE:2013osk}%
  \BibitemOpen
  \bibfield  {author} {\bibinfo {author} {\bibfnamefont {B.~B.}\ \bibnamefont
  {Abelev}} \emph {et~al.} (\bibinfo {collaboration} {ALICE}),\ }\href
  {\doibase 10.1016/j.physletb.2014.05.064} {\bibfield  {journal} {\bibinfo
  {journal} {Phys. Lett. B}\ }\textbf {\bibinfo {volume} {734}},\ \bibinfo
  {pages} {314} (\bibinfo {year} {2014})},\ \Eprint
  {http://arxiv.org/abs/1311.0214} {arXiv:1311.0214 [nucl-ex]} \BibitemShut
  {NoStop}%
\bibitem [{\citenamefont {Acharya}\ \emph {et~al.}(2020)\citenamefont {Acharya}
  \emph {et~al.}}]{ALICE:2020pvw}%
  \BibitemOpen
  \bibfield  {author} {\bibinfo {author} {\bibfnamefont {S.}~\bibnamefont
  {Acharya}} \emph {et~al.} (\bibinfo {collaboration} {ALICE}),\ }\href
  {\doibase 10.1007/JHEP10(2020)141} {\bibfield  {journal} {\bibinfo  {journal}
  {JHEP}\ }\textbf {\bibinfo {volume} {10}},\ \bibinfo {pages} {141} (\bibinfo
  {year} {2020})},\ \Eprint {http://arxiv.org/abs/2005.14518} {arXiv:2005.14518
  [nucl-ex]} \BibitemShut {NoStop}%
\bibitem [{\citenamefont {Adamczyk}\ \emph {et~al.}(2013)\citenamefont
  {Adamczyk} \emph {et~al.}}]{STAR:2012jzy}%
  \BibitemOpen
  \bibfield  {author} {\bibinfo {author} {\bibfnamefont {L.}~\bibnamefont
  {Adamczyk}} \emph {et~al.} (\bibinfo {collaboration} {STAR}),\ }\href
  {\doibase 10.1103/PhysRevLett.111.052301} {\bibfield  {journal} {\bibinfo
  {journal} {Phys. Rev. Lett.}\ }\textbf {\bibinfo {volume} {111}},\ \bibinfo
  {pages} {052301} (\bibinfo {year} {2013})},\ \Eprint
  {http://arxiv.org/abs/1212.3304} {arXiv:1212.3304 [nucl-ex]} \BibitemShut
  {NoStop}%
\bibitem [{\citenamefont {Cacciari}\ \emph {et~al.}(1998)\citenamefont
  {Cacciari}, \citenamefont {Greco},\ and\ \citenamefont
  {Nason}}]{Cacciari:1998it}%
  \BibitemOpen
  \bibfield  {author} {\bibinfo {author} {\bibfnamefont {M.}~\bibnamefont
  {Cacciari}}, \bibinfo {author} {\bibfnamefont {M.}~\bibnamefont {Greco}}, \
  and\ \bibinfo {author} {\bibfnamefont {P.}~\bibnamefont {Nason}},\ }\href
  {\doibase 10.1088/1126-6708/1998/05/007} {\bibfield  {journal} {\bibinfo
  {journal} {JHEP}\ }\textbf {\bibinfo {volume} {05}},\ \bibinfo {pages} {007}
  (\bibinfo {year} {1998})},\ \Eprint {http://arxiv.org/abs/hep-ph/9803400}
  {arXiv:hep-ph/9803400} \BibitemShut {NoStop}%
\bibitem [{\citenamefont {Cacciari}\ \emph {et~al.}(2005)\citenamefont
  {Cacciari}, \citenamefont {Nason},\ and\ \citenamefont
  {Vogt}}]{Cacciari:2005rk}%
  \BibitemOpen
  \bibfield  {author} {\bibinfo {author} {\bibfnamefont {M.}~\bibnamefont
  {Cacciari}}, \bibinfo {author} {\bibfnamefont {P.}~\bibnamefont {Nason}}, \
  and\ \bibinfo {author} {\bibfnamefont {R.}~\bibnamefont {Vogt}},\ }\href
  {\doibase 10.1103/PhysRevLett.95.122001} {\bibfield  {journal} {\bibinfo
  {journal} {Phys. Rev. Lett.}\ }\textbf {\bibinfo {volume} {95}},\ \bibinfo
  {pages} {122001} (\bibinfo {year} {2005})},\ \Eprint
  {http://arxiv.org/abs/hep-ph/0502203} {arXiv:hep-ph/0502203} \BibitemShut
  {NoStop}%
\bibitem [{\citenamefont {Norrbin}\ and\ \citenamefont
  {Sjostrand}(2000)}]{Norrbin:2000zc}%
  \BibitemOpen
  \bibfield  {author} {\bibinfo {author} {\bibfnamefont {E.}~\bibnamefont
  {Norrbin}}\ and\ \bibinfo {author} {\bibfnamefont {T.}~\bibnamefont
  {Sjostrand}},\ }\href {\doibase 10.1007/s100520000460} {\bibfield  {journal}
  {\bibinfo  {journal} {Eur. Phys. J. C}\ }\textbf {\bibinfo {volume} {17}},\
  \bibinfo {pages} {137} (\bibinfo {year} {2000})},\ \Eprint
  {http://arxiv.org/abs/hep-ph/0005110} {arXiv:hep-ph/0005110} \BibitemShut
  {NoStop}%
\bibitem [{\citenamefont {Vogt}(2018)}]{Vogt:2018oje}%
  \BibitemOpen
  \bibfield  {author} {\bibinfo {author} {\bibfnamefont {R.}~\bibnamefont
  {Vogt}},\ }\href {\doibase 10.1103/PhysRevC.98.034907} {\bibfield  {journal}
  {\bibinfo  {journal} {Phys. Rev. C}\ }\textbf {\bibinfo {volume} {98}},\
  \bibinfo {pages} {034907} (\bibinfo {year} {2018})},\ \Eprint
  {http://arxiv.org/abs/1806.01904} {arXiv:1806.01904 [hep-ph]} \BibitemShut
  {NoStop}%
\bibitem [{\citenamefont {Souza}\ and\ \citenamefont
  {Brook}(2016)}]{Souza:2015dgh}%
  \BibitemOpen
  \bibfield  {author} {\bibinfo {author} {\bibfnamefont {D.}~\bibnamefont
  {Souza}}\ and\ \bibinfo {author} {\bibfnamefont {N.~H.}\ \bibnamefont
  {Brook}},\ }\href {\doibase 10.1088/0954-3899/43/1/015001} {\bibfield
  {journal} {\bibinfo  {journal} {J. Phys. G}\ }\textbf {\bibinfo {volume}
  {43}},\ \bibinfo {pages} {015001} (\bibinfo {year} {2016})},\ \Eprint
  {http://arxiv.org/abs/1511.06226} {arXiv:1511.06226 [hep-ph]} \BibitemShut
  {NoStop}%
\bibitem [{\citenamefont {Maciula}\ and\ \citenamefont
  {Szczurek}(2014)}]{Maciula:2014oya}%
  \BibitemOpen
  \bibfield  {author} {\bibinfo {author} {\bibfnamefont {R.}~\bibnamefont
  {Maciula}}\ and\ \bibinfo {author} {\bibfnamefont {A.}~\bibnamefont
  {Szczurek}},\ }\href {\doibase 10.1051/epjconf/20148101007} {\bibfield
  {journal} {\bibinfo  {journal} {EPJ Web Conf.}\ }\textbf {\bibinfo {volume}
  {81}},\ \bibinfo {pages} {01007} (\bibinfo {year} {2014})}\BibitemShut
  {NoStop}%
\bibitem [{\citenamefont {Fritzsch}(1977)}]{Fritzsch:1977ay}%
  \BibitemOpen
  \bibfield  {author} {\bibinfo {author} {\bibfnamefont {H.}~\bibnamefont
  {Fritzsch}},\ }\href {\doibase 10.1016/0370-2693(77)90108-3} {\bibfield
  {journal} {\bibinfo  {journal} {Phys. Lett. B}\ }\textbf {\bibinfo {volume}
  {67}},\ \bibinfo {pages} {217} (\bibinfo {year} {1977})}\BibitemShut
  {NoStop}%
\bibitem [{\citenamefont {Amundson}\ \emph {et~al.}(1996)\citenamefont
  {Amundson}, \citenamefont {Eboli}, \citenamefont {Gregores},\ and\
  \citenamefont {Halzen}}]{Amundson:1995em}%
  \BibitemOpen
  \bibfield  {author} {\bibinfo {author} {\bibfnamefont {J.~F.}\ \bibnamefont
  {Amundson}}, \bibinfo {author} {\bibfnamefont {O.~J.~P.}\ \bibnamefont
  {Eboli}}, \bibinfo {author} {\bibfnamefont {E.~M.}\ \bibnamefont {Gregores}},
  \ and\ \bibinfo {author} {\bibfnamefont {F.}~\bibnamefont {Halzen}},\ }\href
  {\doibase 10.1016/0370-2693(96)00035-4} {\bibfield  {journal} {\bibinfo
  {journal} {Phys. Lett. B}\ }\textbf {\bibinfo {volume} {372}},\ \bibinfo
  {pages} {127} (\bibinfo {year} {1996})},\ \Eprint
  {http://arxiv.org/abs/hep-ph/9512248} {arXiv:hep-ph/9512248} \BibitemShut
  {NoStop}%
\bibitem [{\citenamefont {Cheung}\ and\ \citenamefont
  {Vogt}(2018)}]{Cheung:2018tvq}%
  \BibitemOpen
  \bibfield  {author} {\bibinfo {author} {\bibfnamefont {V.}~\bibnamefont
  {Cheung}}\ and\ \bibinfo {author} {\bibfnamefont {R.}~\bibnamefont {Vogt}},\
  }\href {\doibase 10.1103/PhysRevD.98.114029} {\bibfield  {journal} {\bibinfo
  {journal} {Phys. Rev. D}\ }\textbf {\bibinfo {volume} {98}},\ \bibinfo
  {pages} {114029} (\bibinfo {year} {2018})},\ \Eprint
  {http://arxiv.org/abs/1808.02909} {arXiv:1808.02909 [hep-ph]} \BibitemShut
  {NoStop}%
\bibitem [{\citenamefont {Chang}(1980)}]{Chang:1979nn}%
  \BibitemOpen
  \bibfield  {author} {\bibinfo {author} {\bibfnamefont {C.-H.}\ \bibnamefont
  {Chang}},\ }\href {\doibase 10.1016/0550-3213(80)90175-3} {\bibfield
  {journal} {\bibinfo  {journal} {Nucl. Phys. B}\ }\textbf {\bibinfo {volume}
  {172}},\ \bibinfo {pages} {425} (\bibinfo {year} {1980})}\BibitemShut
  {NoStop}%
\bibitem [{\citenamefont {Baier}\ and\ \citenamefont
  {Ruckl}(1981)}]{Baier:1981uk}%
  \BibitemOpen
  \bibfield  {author} {\bibinfo {author} {\bibfnamefont {R.}~\bibnamefont
  {Baier}}\ and\ \bibinfo {author} {\bibfnamefont {R.}~\bibnamefont {Ruckl}},\
  }\href {\doibase 10.1016/0370-2693(81)90636-5} {\bibfield  {journal}
  {\bibinfo  {journal} {Phys. Lett. B}\ }\textbf {\bibinfo {volume} {102}},\
  \bibinfo {pages} {364} (\bibinfo {year} {1981})}\BibitemShut {NoStop}%
\bibitem [{\citenamefont {Bodwin}\ \emph {et~al.}(1992)\citenamefont {Bodwin},
  \citenamefont {Braaten}, \citenamefont {Yuan},\ and\ \citenamefont
  {Lepage}}]{Bodwin:1992qr}%
  \BibitemOpen
  \bibfield  {author} {\bibinfo {author} {\bibfnamefont {G.~T.}\ \bibnamefont
  {Bodwin}}, \bibinfo {author} {\bibfnamefont {E.}~\bibnamefont {Braaten}},
  \bibinfo {author} {\bibfnamefont {T.~C.}\ \bibnamefont {Yuan}}, \ and\
  \bibinfo {author} {\bibfnamefont {G.~P.}\ \bibnamefont {Lepage}},\ }\href
  {\doibase 10.1103/PhysRevD.46.R3703} {\bibfield  {journal} {\bibinfo
  {journal} {Phys. Rev. D}\ }\textbf {\bibinfo {volume} {46}},\ \bibinfo
  {pages} {R3703} (\bibinfo {year} {1992})},\ \Eprint
  {http://arxiv.org/abs/hep-ph/9208254} {arXiv:hep-ph/9208254} \BibitemShut
  {NoStop}%
\bibitem [{\citenamefont {Bodwin}\ \emph {et~al.}(1995)\citenamefont {Bodwin},
  \citenamefont {Braaten},\ and\ \citenamefont {Lepage}}]{Bodwin:1994jh}%
  \BibitemOpen
  \bibfield  {author} {\bibinfo {author} {\bibfnamefont {G.~T.}\ \bibnamefont
  {Bodwin}}, \bibinfo {author} {\bibfnamefont {E.}~\bibnamefont {Braaten}}, \
  and\ \bibinfo {author} {\bibfnamefont {G.~P.}\ \bibnamefont {Lepage}},\
  }\href {\doibase 10.1103/PhysRevD.55.5853} {\bibfield  {journal} {\bibinfo
  {journal} {Phys. Rev. D}\ }\textbf {\bibinfo {volume} {51}},\ \bibinfo
  {pages} {1125} (\bibinfo {year} {1995})},\ \bibinfo {note} {[Erratum:
  Phys.Rev.D 55, 5853 (1997)]},\ \Eprint {http://arxiv.org/abs/hep-ph/9407339}
  {arXiv:hep-ph/9407339} \BibitemShut {NoStop}%
\bibitem [{\citenamefont {Butenschoen}\ and\ \citenamefont
  {Kniehl}(2012)}]{Butenschoen:2012px}%
  \BibitemOpen
  \bibfield  {author} {\bibinfo {author} {\bibfnamefont {M.}~\bibnamefont
  {Butenschoen}}\ and\ \bibinfo {author} {\bibfnamefont {B.~A.}\ \bibnamefont
  {Kniehl}},\ }\href {\doibase 10.1103/PhysRevLett.108.172002} {\bibfield
  {journal} {\bibinfo  {journal} {Phys. Rev. Lett.}\ }\textbf {\bibinfo
  {volume} {108}},\ \bibinfo {pages} {172002} (\bibinfo {year} {2012})},\
  \Eprint {http://arxiv.org/abs/1201.1872} {arXiv:1201.1872 [hep-ph]}
  \BibitemShut {NoStop}%
\bibitem [{\citenamefont {Gong}\ \emph {et~al.}(2013)\citenamefont {Gong},
  \citenamefont {Wan}, \citenamefont {Wang},\ and\ \citenamefont
  {Zhang}}]{Gong:2012ug}%
  \BibitemOpen
  \bibfield  {author} {\bibinfo {author} {\bibfnamefont {B.}~\bibnamefont
  {Gong}}, \bibinfo {author} {\bibfnamefont {L.-P.}\ \bibnamefont {Wan}},
  \bibinfo {author} {\bibfnamefont {J.-X.}\ \bibnamefont {Wang}}, \ and\
  \bibinfo {author} {\bibfnamefont {H.-F.}\ \bibnamefont {Zhang}},\ }\href
  {\doibase 10.1103/PhysRevLett.110.042002} {\bibfield  {journal} {\bibinfo
  {journal} {Phys. Rev. Lett.}\ }\textbf {\bibinfo {volume} {110}},\ \bibinfo
  {pages} {042002} (\bibinfo {year} {2013})},\ \Eprint
  {http://arxiv.org/abs/1205.6682} {arXiv:1205.6682 [hep-ph]} \BibitemShut
  {NoStop}%
\bibitem [{\citenamefont {Chao}\ \emph {et~al.}(2012)\citenamefont {Chao},
  \citenamefont {Ma}, \citenamefont {Shao}, \citenamefont {Wang},\ and\
  \citenamefont {Zhang}}]{Chao:2012iv}%
  \BibitemOpen
  \bibfield  {author} {\bibinfo {author} {\bibfnamefont {K.-T.}\ \bibnamefont
  {Chao}}, \bibinfo {author} {\bibfnamefont {Y.-Q.}\ \bibnamefont {Ma}},
  \bibinfo {author} {\bibfnamefont {H.-S.}\ \bibnamefont {Shao}}, \bibinfo
  {author} {\bibfnamefont {K.}~\bibnamefont {Wang}}, \ and\ \bibinfo {author}
  {\bibfnamefont {Y.-J.}\ \bibnamefont {Zhang}},\ }\href {\doibase
  10.1103/PhysRevLett.108.242004} {\bibfield  {journal} {\bibinfo  {journal}
  {Phys. Rev. Lett.}\ }\textbf {\bibinfo {volume} {108}},\ \bibinfo {pages}
  {242004} (\bibinfo {year} {2012})},\ \Eprint {http://arxiv.org/abs/1201.2675}
  {arXiv:1201.2675 [hep-ph]} \BibitemShut {NoStop}%
\bibitem [{\citenamefont {Song}\ \emph {et~al.}(2017)\citenamefont {Song},
  \citenamefont {Aichelin},\ and\ \citenamefont {Bratkovskaya}}]{Song:2017phm}%
  \BibitemOpen
  \bibfield  {author} {\bibinfo {author} {\bibfnamefont {T.}~\bibnamefont
  {Song}}, \bibinfo {author} {\bibfnamefont {J.}~\bibnamefont {Aichelin}}, \
  and\ \bibinfo {author} {\bibfnamefont {E.}~\bibnamefont {Bratkovskaya}},\
  }\href {\doibase 10.1103/PhysRevC.96.014907} {\bibfield  {journal} {\bibinfo
  {journal} {Phys. Rev. C}\ }\textbf {\bibinfo {volume} {96}},\ \bibinfo
  {pages} {014907} (\bibinfo {year} {2017})},\ \Eprint
  {http://arxiv.org/abs/1705.00046} {arXiv:1705.00046 [nucl-th]} \BibitemShut
  {NoStop}%
\bibitem [{\citenamefont {Villar}\ \emph {et~al.}(2023)\citenamefont {Villar},
  \citenamefont {Zhao}, \citenamefont {Aichelin},\ and\ \citenamefont
  {Gossiaux}}]{Villar:2022sbv}%
  \BibitemOpen
  \bibfield  {author} {\bibinfo {author} {\bibfnamefont {D.~Y.~A.}\
  \bibnamefont {Villar}}, \bibinfo {author} {\bibfnamefont {J.}~\bibnamefont
  {Zhao}}, \bibinfo {author} {\bibfnamefont {J.}~\bibnamefont {Aichelin}}, \
  and\ \bibinfo {author} {\bibfnamefont {P.~B.}\ \bibnamefont {Gossiaux}},\
  }\href {\doibase 10.1103/PhysRevC.107.054913} {\bibfield  {journal} {\bibinfo
   {journal} {Phys. Rev. C}\ }\textbf {\bibinfo {volume} {107}},\ \bibinfo
  {pages} {054913} (\bibinfo {year} {2023})},\ \Eprint
  {http://arxiv.org/abs/2206.01308} {arXiv:2206.01308 [nucl-th]} \BibitemShut
  {NoStop}%
\bibitem [{\citenamefont {Song}\ \emph
  {et~al.}(2023{\natexlab{a}})\citenamefont {Song}, \citenamefont {Aichelin},
  \citenamefont {Zhao}, \citenamefont {Gossiaux},\ and\ \citenamefont
  {Bratkovskaya}}]{Song:2023zma}%
  \BibitemOpen
  \bibfield  {author} {\bibinfo {author} {\bibfnamefont {T.}~\bibnamefont
  {Song}}, \bibinfo {author} {\bibfnamefont {J.}~\bibnamefont {Aichelin}},
  \bibinfo {author} {\bibfnamefont {J.}~\bibnamefont {Zhao}}, \bibinfo {author}
  {\bibfnamefont {P.~B.}\ \bibnamefont {Gossiaux}}, \ and\ \bibinfo {author}
  {\bibfnamefont {E.}~\bibnamefont {Bratkovskaya}},\ }\href {\doibase
  10.1103/PhysRevC.108.054908} {\bibfield  {journal} {\bibinfo  {journal}
  {Phys. Rev. C}\ }\textbf {\bibinfo {volume} {108}},\ \bibinfo {pages}
  {054908} (\bibinfo {year} {2023}{\natexlab{a}})},\ \Eprint
  {http://arxiv.org/abs/2305.10750} {arXiv:2305.10750 [nucl-th]} \BibitemShut
  {NoStop}%
\bibitem [{\citenamefont {Werner}(2023)}]{Werner:2023zvo}%
  \BibitemOpen
  \bibfield  {author} {\bibinfo {author} {\bibfnamefont {K.}~\bibnamefont
  {Werner}},\ }\href {\doibase 10.1103/PhysRevC.108.064903} {\bibfield
  {journal} {\bibinfo  {journal} {Phys. Rev. C}\ }\textbf {\bibinfo {volume}
  {108}},\ \bibinfo {pages} {064903} (\bibinfo {year} {2023})},\ \Eprint
  {http://arxiv.org/abs/2301.12517} {arXiv:2301.12517 [hep-ph]} \BibitemShut
  {NoStop}%
\bibitem [{\citenamefont {Werner}\ and\ \citenamefont
  {Guiot}(2023)}]{Werner:2023fne}%
  \BibitemOpen
  \bibfield  {author} {\bibinfo {author} {\bibfnamefont {K.}~\bibnamefont
  {Werner}}\ and\ \bibinfo {author} {\bibfnamefont {B.}~\bibnamefont {Guiot}},\
  }\href {\doibase 10.1103/PhysRevC.108.034904} {\bibfield  {journal} {\bibinfo
   {journal} {Phys. Rev. C}\ }\textbf {\bibinfo {volume} {108}},\ \bibinfo
  {pages} {034904} (\bibinfo {year} {2023})},\ \Eprint
  {http://arxiv.org/abs/2306.02396} {arXiv:2306.02396 [hep-ph]} \BibitemShut
  {NoStop}%
\bibitem [{\citenamefont
  {Werner}(2024{\natexlab{a}})}]{werner:2023-epos4-smatrix}%
  \BibitemOpen
  \bibfield  {author} {\bibinfo {author} {\bibfnamefont {K.}~\bibnamefont
  {Werner}},\ }\href {\doibase 10.1103/PhysRevC.109.034918} {\bibfield
  {journal} {\bibinfo  {journal} {Phys. Rev. C}\ }\textbf {\bibinfo {volume}
  {109}},\ \bibinfo {pages} {034918} (\bibinfo {year} {2024}{\natexlab{a}})},\
  \Eprint {http://arxiv.org/abs/2310.09380} {arXiv:2310.09380 [hep-ph]}
  \BibitemShut {NoStop}%
\bibitem [{\citenamefont
  {Werner}(2024{\natexlab{b}})}]{werner:2023-epos4-micro}%
  \BibitemOpen
  \bibfield  {author} {\bibinfo {author} {\bibfnamefont {K.}~\bibnamefont
  {Werner}},\ }\href {\doibase 10.1103/PhysRevC.109.014910} {\bibfield
  {journal} {\bibinfo  {journal} {Phys. Rev. C}\ }\textbf {\bibinfo {volume}
  {109}},\ \bibinfo {pages} {014910} (\bibinfo {year} {2024}{\natexlab{b}})},\
  \Eprint {http://arxiv.org/abs/2306.10277} {arXiv:2306.10277 [hep-ph]}
  \BibitemShut {NoStop}%
\bibitem [{\citenamefont {Zhao}\ \emph {et~al.}(2023)\citenamefont {Zhao},
  \citenamefont {Aichelin}, \citenamefont {Gossiaux},\ and\ \citenamefont
  {Werner}}]{Zhao:2023ucp}%
  \BibitemOpen
  \bibfield  {author} {\bibinfo {author} {\bibfnamefont {J.}~\bibnamefont
  {Zhao}}, \bibinfo {author} {\bibfnamefont {J.}~\bibnamefont {Aichelin}},
  \bibinfo {author} {\bibfnamefont {P.~B.}\ \bibnamefont {Gossiaux}}, \ and\
  \bibinfo {author} {\bibfnamefont {K.}~\bibnamefont {Werner}},\ }\href@noop {}
  {\  (\bibinfo {year} {2023})},\ \Eprint {http://arxiv.org/abs/2310.08684}
  {arXiv:2310.08684 [hep-ph]} \BibitemShut {NoStop}%
\bibitem [{\citenamefont {Aaij}\ \emph
  {et~al.}(2017{\natexlab{a}})\citenamefont {Aaij} \emph
  {et~al.}}]{LHCb:2017bvf}%
  \BibitemOpen
  \bibfield  {author} {\bibinfo {author} {\bibfnamefont {R.}~\bibnamefont
  {Aaij}} \emph {et~al.} (\bibinfo {collaboration} {LHCb}),\ }\href {\doibase
  10.1007/JHEP11(2017)030} {\bibfield  {journal} {\bibinfo  {journal} {JHEP}\
  }\textbf {\bibinfo {volume} {11}},\ \bibinfo {pages} {030} (\bibinfo {year}
  {2017}{\natexlab{a}})},\ \Eprint {http://arxiv.org/abs/1708.05994}
  {arXiv:1708.05994 [hep-ex]} \BibitemShut {NoStop}%
\bibitem [{\citenamefont {Acharya}\ \emph
  {et~al.}(2021{\natexlab{b}})\citenamefont {Acharya} \emph
  {et~al.}}]{ALICE:2021mgk}%
  \BibitemOpen
  \bibfield  {author} {\bibinfo {author} {\bibfnamefont {S.}~\bibnamefont
  {Acharya}} \emph {et~al.} (\bibinfo {collaboration} {ALICE}),\ }\href
  {\doibase 10.1007/JHEP05(2021)220} {\bibfield  {journal} {\bibinfo  {journal}
  {JHEP}\ }\textbf {\bibinfo {volume} {05}},\ \bibinfo {pages} {220} (\bibinfo
  {year} {2021}{\natexlab{b}})},\ \Eprint {http://arxiv.org/abs/2102.13601}
  {arXiv:2102.13601 [nucl-ex]} \BibitemShut {NoStop}%
\bibitem [{\citenamefont {Aaij}\ \emph
  {et~al.}(2017{\natexlab{b}})\citenamefont {Aaij} \emph
  {et~al.}}]{LHCb:2016ikn}%
  \BibitemOpen
  \bibfield  {author} {\bibinfo {author} {\bibfnamefont {R.}~\bibnamefont
  {Aaij}} \emph {et~al.} (\bibinfo {collaboration} {LHCb}),\ }\href {\doibase
  10.1007/JHEP06(2017)147} {\bibfield  {journal} {\bibinfo  {journal} {JHEP}\
  }\textbf {\bibinfo {volume} {06}},\ \bibinfo {pages} {147} (\bibinfo {year}
  {2017}{\natexlab{b}})},\ \Eprint {http://arxiv.org/abs/1610.02230}
  {arXiv:1610.02230 [hep-ex]} \BibitemShut {NoStop}%
\bibitem [{\citenamefont {Adamczyk}\ \emph {et~al.}(2014)\citenamefont
  {Adamczyk} \emph {et~al.}}]{STAR:2014wif}%
  \BibitemOpen
  \bibfield  {author} {\bibinfo {author} {\bibfnamefont {L.}~\bibnamefont
  {Adamczyk}} \emph {et~al.} (\bibinfo {collaboration} {STAR}),\ }\href
  {\doibase 10.1103/PhysRevLett.113.142301} {\bibfield  {journal} {\bibinfo
  {journal} {Phys. Rev. Lett.}\ }\textbf {\bibinfo {volume} {113}},\ \bibinfo
  {pages} {142301} (\bibinfo {year} {2014})},\ \bibinfo {note} {[Erratum:
  Phys.Rev.Lett. 121, 229901 (2018)]},\ \Eprint
  {http://arxiv.org/abs/1404.6185} {arXiv:1404.6185 [nucl-ex]} \BibitemShut
  {NoStop}%
\bibitem [{\citenamefont {Remler}(1981)}]{Remler:1981du}%
  \BibitemOpen
  \bibfield  {author} {\bibinfo {author} {\bibfnamefont {E.~A.}\ \bibnamefont
  {Remler}},\ }\href {\doibase 10.1016/0003-4916(81)90100-7} {\bibfield
  {journal} {\bibinfo  {journal} {Annals Phys.}\ }\textbf {\bibinfo {volume}
  {136}},\ \bibinfo {pages} {293} (\bibinfo {year} {1981})}\BibitemShut
  {NoStop}%
\bibitem [{\citenamefont {Gyulassy}\ \emph {et~al.}(1983)\citenamefont
  {Gyulassy}, \citenamefont {Frankel},\ and\ \citenamefont
  {Remler}}]{Gyulassy:1982pe}%
  \BibitemOpen
  \bibfield  {author} {\bibinfo {author} {\bibfnamefont {M.}~\bibnamefont
  {Gyulassy}}, \bibinfo {author} {\bibfnamefont {K.}~\bibnamefont {Frankel}}, \
  and\ \bibinfo {author} {\bibfnamefont {E.~a.}\ \bibnamefont {Remler}},\
  }\href {\doibase 10.1016/0375-9474(83)90222-1} {\bibfield  {journal}
  {\bibinfo  {journal} {Nucl. Phys. A}\ }\textbf {\bibinfo {volume} {402}},\
  \bibinfo {pages} {596} (\bibinfo {year} {1983})}\BibitemShut {NoStop}%
\bibitem [{\citenamefont {Song}\ \emph
  {et~al.}(2023{\natexlab{b}})\citenamefont {Song}, \citenamefont {Aichelin},\
  and\ \citenamefont {Bratkovskaya}}]{Song:2023ywt}%
  \BibitemOpen
  \bibfield  {author} {\bibinfo {author} {\bibfnamefont {T.}~\bibnamefont
  {Song}}, \bibinfo {author} {\bibfnamefont {J.}~\bibnamefont {Aichelin}}, \
  and\ \bibinfo {author} {\bibfnamefont {E.}~\bibnamefont {Bratkovskaya}},\
  }\href {\doibase 10.1103/PhysRevC.107.054906} {\bibfield  {journal} {\bibinfo
   {journal} {Phys. Rev. C}\ }\textbf {\bibinfo {volume} {107}},\ \bibinfo
  {pages} {054906} (\bibinfo {year} {2023}{\natexlab{b}})},\ \Eprint
  {http://arxiv.org/abs/2302.14001} {arXiv:2302.14001 [hep-ph]} \BibitemShut
  {NoStop}%
\bibitem [{\citenamefont {Fries}\ \emph {et~al.}(2008)\citenamefont {Fries},
  \citenamefont {Greco},\ and\ \citenamefont {Sorensen}}]{Fries:2008hs}%
  \BibitemOpen
  \bibfield  {author} {\bibinfo {author} {\bibfnamefont {R.~J.}\ \bibnamefont
  {Fries}}, \bibinfo {author} {\bibfnamefont {V.}~\bibnamefont {Greco}}, \ and\
  \bibinfo {author} {\bibfnamefont {P.}~\bibnamefont {Sorensen}},\ }\href
  {\doibase 10.1146/annurev.nucl.58.110707.171134} {\bibfield  {journal}
  {\bibinfo  {journal} {Ann. Rev. Nucl. Part. Sci.}\ }\textbf {\bibinfo
  {volume} {58}},\ \bibinfo {pages} {177} (\bibinfo {year} {2008})},\ \Eprint
  {http://arxiv.org/abs/0807.4939} {arXiv:0807.4939 [nucl-th]} \BibitemShut
  {NoStop}%
\bibitem [{\citenamefont {Fries}\ \emph {et~al.}(2003)\citenamefont {Fries},
  \citenamefont {Muller}, \citenamefont {Nonaka},\ and\ \citenamefont
  {Bass}}]{Fries:2003kq}%
  \BibitemOpen
  \bibfield  {author} {\bibinfo {author} {\bibfnamefont {R.~J.}\ \bibnamefont
  {Fries}}, \bibinfo {author} {\bibfnamefont {B.}~\bibnamefont {Muller}},
  \bibinfo {author} {\bibfnamefont {C.}~\bibnamefont {Nonaka}}, \ and\ \bibinfo
  {author} {\bibfnamefont {S.~A.}\ \bibnamefont {Bass}},\ }\href {\doibase
  10.1103/PhysRevC.68.044902} {\bibfield  {journal} {\bibinfo  {journal} {Phys.
  Rev. C}\ }\textbf {\bibinfo {volume} {68}},\ \bibinfo {pages} {044902}
  (\bibinfo {year} {2003})},\ \Eprint {http://arxiv.org/abs/nucl-th/0306027}
  {arXiv:nucl-th/0306027} \BibitemShut {NoStop}%
\bibitem [{\citenamefont {Workman}\ and\ \citenamefont
  {Others}(2022)}]{Workman:2022ynf}%
  \BibitemOpen
  \bibfield  {author} {\bibinfo {author} {\bibfnamefont {R.~L.}\ \bibnamefont
  {Workman}}\ and\ \bibinfo {author} {\bibnamefont {Others}} (\bibinfo
  {collaboration} {Particle Data Group}),\ }\href {\doibase
  10.1093/ptep/ptac097} {\bibfield  {journal} {\bibinfo  {journal} {PTEP}\
  }\textbf {\bibinfo {volume} {2022}},\ \bibinfo {pages} {083C01} (\bibinfo
  {year} {2022})}\BibitemShut {NoStop}%
\bibitem [{\citenamefont {Shlomo}\ and\ \citenamefont
  {Prakash}(1981)}]{Shlomo:1981ayz}%
  \BibitemOpen
  \bibfield  {author} {\bibinfo {author} {\bibfnamefont {S.}~\bibnamefont
  {Shlomo}}\ and\ \bibinfo {author} {\bibfnamefont {M.}~\bibnamefont
  {Prakash}},\ }\href {\doibase 10.1016/0375-9474(81)90631-X} {\bibfield
  {journal} {\bibinfo  {journal} {Nucl. Phys. A}\ }\textbf {\bibinfo {volume}
  {357}},\ \bibinfo {pages} {157} (\bibinfo {year} {1981})}\BibitemShut
  {NoStop}%
\bibitem [{\citenamefont {Acharya}\ \emph
  {et~al.}(2022{\natexlab{b}})\citenamefont {Acharya} \emph
  {et~al.}}]{ALICE:2021edd}%
  \BibitemOpen
  \bibfield  {author} {\bibinfo {author} {\bibfnamefont {S.}~\bibnamefont
  {Acharya}} \emph {et~al.} (\bibinfo {collaboration} {ALICE}),\ }\href
  {\doibase 10.1007/JHEP03(2022)190} {\bibfield  {journal} {\bibinfo  {journal}
  {JHEP}\ }\textbf {\bibinfo {volume} {03}},\ \bibinfo {pages} {190} (\bibinfo
  {year} {2022}{\natexlab{b}})},\ \Eprint {http://arxiv.org/abs/2108.02523}
  {arXiv:2108.02523 [nucl-ex]} \BibitemShut {NoStop}%
\bibitem [{\citenamefont {Acharya}\ \emph {et~al.}(2019)\citenamefont {Acharya}
  \emph {et~al.}}]{ALICE:2019pid}%
  \BibitemOpen
  \bibfield  {author} {\bibinfo {author} {\bibfnamefont {S.}~\bibnamefont
  {Acharya}} \emph {et~al.} (\bibinfo {collaboration} {ALICE}),\ }\href
  {\doibase 10.1007/JHEP10(2019)084} {\bibfield  {journal} {\bibinfo  {journal}
  {JHEP}\ }\textbf {\bibinfo {volume} {10}},\ \bibinfo {pages} {084} (\bibinfo
  {year} {2019})},\ \Eprint {http://arxiv.org/abs/1905.07211} {arXiv:1905.07211
  [nucl-ex]} \BibitemShut {NoStop}%
\bibitem [{\citenamefont {Aaboud}\ \emph {et~al.}(2018)\citenamefont {Aaboud}
  \emph {et~al.}}]{ATLAS:2017prf}%
  \BibitemOpen
  \bibfield  {author} {\bibinfo {author} {\bibfnamefont {M.}~\bibnamefont
  {Aaboud}} \emph {et~al.} (\bibinfo {collaboration} {ATLAS}),\ }\href
  {\doibase 10.1140/epjc/s10052-018-5624-4} {\bibfield  {journal} {\bibinfo
  {journal} {Eur. Phys. J. C}\ }\textbf {\bibinfo {volume} {78}},\ \bibinfo
  {pages} {171} (\bibinfo {year} {2018})},\ \Eprint
  {http://arxiv.org/abs/1709.03089} {arXiv:1709.03089 [nucl-ex]} \BibitemShut
  {NoStop}%
\bibitem [{\citenamefont {Sirunyan}\ \emph
  {et~al.}(2019{\natexlab{a}})\citenamefont {Sirunyan} \emph
  {et~al.}}]{CMS:2018gbb}%
  \BibitemOpen
  \bibfield  {author} {\bibinfo {author} {\bibfnamefont {A.~M.}\ \bibnamefont
  {Sirunyan}} \emph {et~al.} (\bibinfo {collaboration} {CMS}),\ }\href
  {\doibase 10.1016/j.physletb.2019.01.058} {\bibfield  {journal} {\bibinfo
  {journal} {Phys. Lett. B}\ }\textbf {\bibinfo {volume} {790}},\ \bibinfo
  {pages} {509} (\bibinfo {year} {2019}{\natexlab{a}})},\ \Eprint
  {http://arxiv.org/abs/1805.02248} {arXiv:1805.02248 [hep-ex]} \BibitemShut
  {NoStop}%
\bibitem [{\citenamefont {Adare}\ \emph {et~al.}(2012)\citenamefont {Adare}
  \emph {et~al.}}]{PHENIX:2011gyb}%
  \BibitemOpen
  \bibfield  {author} {\bibinfo {author} {\bibfnamefont {A.}~\bibnamefont
  {Adare}} \emph {et~al.} (\bibinfo {collaboration} {PHENIX}),\ }\href
  {\doibase 10.1103/PhysRevD.85.092004} {\bibfield  {journal} {\bibinfo
  {journal} {Phys. Rev. D}\ }\textbf {\bibinfo {volume} {85}},\ \bibinfo
  {pages} {092004} (\bibinfo {year} {2012})},\ \Eprint
  {http://arxiv.org/abs/1105.1966} {arXiv:1105.1966 [hep-ex]} \BibitemShut
  {NoStop}%
\bibitem [{\citenamefont {Adam}\ \emph {et~al.}(2018)\citenamefont {Adam} \emph
  {et~al.}}]{STAR:2018smh}%
  \BibitemOpen
  \bibfield  {author} {\bibinfo {author} {\bibfnamefont {J.}~\bibnamefont
  {Adam}} \emph {et~al.} (\bibinfo {collaboration} {STAR}),\ }\href {\doibase
  10.1016/j.physletb.2018.09.029} {\bibfield  {journal} {\bibinfo  {journal}
  {Phys. Lett. B}\ }\textbf {\bibinfo {volume} {786}},\ \bibinfo {pages} {87}
  (\bibinfo {year} {2018})},\ \Eprint {http://arxiv.org/abs/1805.03745}
  {arXiv:1805.03745 [hep-ex]} \BibitemShut {NoStop}%
\bibitem [{\citenamefont {Sirunyan}\ \emph
  {et~al.}(2019{\natexlab{b}})\citenamefont {Sirunyan} \emph
  {et~al.}}]{CMS:2018zza}%
  \BibitemOpen
  \bibfield  {author} {\bibinfo {author} {\bibfnamefont {A.~M.}\ \bibnamefont
  {Sirunyan}} \emph {et~al.} (\bibinfo {collaboration} {CMS}),\ }\href
  {\doibase 10.1016/j.physletb.2019.01.006} {\bibfield  {journal} {\bibinfo
  {journal} {Phys. Lett. B}\ }\textbf {\bibinfo {volume} {790}},\ \bibinfo
  {pages} {270} (\bibinfo {year} {2019}{\natexlab{b}})},\ \Eprint
  {http://arxiv.org/abs/1805.09215} {arXiv:1805.09215 [hep-ex]} \BibitemShut
  {NoStop}%
\bibitem [{\citenamefont {Tumasyan}\ \emph {et~al.}(2022)\citenamefont
  {Tumasyan} \emph {et~al.}}]{CMS:2022sxl}%
  \BibitemOpen
  \bibfield  {author} {\bibinfo {author} {\bibfnamefont {A.}~\bibnamefont
  {Tumasyan}} \emph {et~al.} (\bibinfo {collaboration} {CMS}),\ }\href
  {\doibase 10.1103/PhysRevLett.128.252301} {\bibfield  {journal} {\bibinfo
  {journal} {Phys. Rev. Lett.}\ }\textbf {\bibinfo {volume} {128}},\ \bibinfo
  {pages} {252301} (\bibinfo {year} {2022})},\ \Eprint
  {http://arxiv.org/abs/2201.02659} {arXiv:2201.02659 [hep-ex]} \BibitemShut
  {NoStop}%
\bibitem [{\citenamefont {Ebert}\ \emph {et~al.}(2003)\citenamefont {Ebert},
  \citenamefont {Faustov},\ and\ \citenamefont {Galkin}}]{Ebert:2003cn}%
  \BibitemOpen
  \bibfield  {author} {\bibinfo {author} {\bibfnamefont {D.}~\bibnamefont
  {Ebert}}, \bibinfo {author} {\bibfnamefont {R.~N.}\ \bibnamefont {Faustov}},
  \ and\ \bibinfo {author} {\bibfnamefont {V.~O.}\ \bibnamefont {Galkin}},\
  }\href {\doibase 10.1103/PhysRevD.68.094020} {\bibfield  {journal} {\bibinfo
  {journal} {Phys. Rev. D}\ }\textbf {\bibinfo {volume} {68}},\ \bibinfo
  {pages} {094020} (\bibinfo {year} {2003})},\ \Eprint
  {http://arxiv.org/abs/hep-ph/0306306} {arXiv:hep-ph/0306306} \BibitemShut
  {NoStop}%
\bibitem [{\citenamefont {Hernandez}\ \emph {et~al.}(2006)\citenamefont
  {Hernandez}, \citenamefont {Nieves},\ and\ \citenamefont
  {Verde-Velasco}}]{Hernandez:2006gt}%
  \BibitemOpen
  \bibfield  {author} {\bibinfo {author} {\bibfnamefont {E.}~\bibnamefont
  {Hernandez}}, \bibinfo {author} {\bibfnamefont {J.}~\bibnamefont {Nieves}}, \
  and\ \bibinfo {author} {\bibfnamefont {J.~M.}\ \bibnamefont
  {Verde-Velasco}},\ }\href {\doibase 10.1103/PhysRevD.74.074008} {\bibfield
  {journal} {\bibinfo  {journal} {Phys. Rev. D}\ }\textbf {\bibinfo {volume}
  {74}},\ \bibinfo {pages} {074008} (\bibinfo {year} {2006})},\ \Eprint
  {http://arxiv.org/abs/hep-ph/0607150} {arXiv:hep-ph/0607150} \BibitemShut
  {NoStop}%
\bibitem [{\citenamefont {Qiao}\ and\ \citenamefont {Zhu}(2013)}]{Qiao:2012vt}%
  \BibitemOpen
  \bibfield  {author} {\bibinfo {author} {\bibfnamefont {C.-F.}\ \bibnamefont
  {Qiao}}\ and\ \bibinfo {author} {\bibfnamefont {R.-L.}\ \bibnamefont {Zhu}},\
  }\href {\doibase 10.1103/PhysRevD.87.014009} {\bibfield  {journal} {\bibinfo
  {journal} {Phys. Rev. D}\ }\textbf {\bibinfo {volume} {87}},\ \bibinfo
  {pages} {014009} (\bibinfo {year} {2013})},\ \Eprint
  {http://arxiv.org/abs/1208.5916} {arXiv:1208.5916 [hep-ph]} \BibitemShut
  {NoStop}%
\end{thebibliography}%

\end{document}